\documentclass[aps,reprint,superscriptaddress]{revtex4-2}

\usepackage{graphicx,latexsym} 
\usepackage{amssymb,amsthm,amsmath,amsfonts}
\usepackage{bm}
\usepackage{dcolumn}
\usepackage{gensymb}
\usepackage{braket}
\usepackage{siunitx}
\usepackage{hyperref}

\newcommand{\omd}{{\omega_{\text{d}}}}
\newcommand{\ome}{{\omega_{\text{e}}}}
\newcommand{\omn}{{\omega_{\text{n}}}}
\newcommand{\doh}{{\delta_{\mathrm{oh}}}}

\newcommand{\AffCam}{Cavendish Laboratory, University of Cambridge, Cambridge, UK}
\newcommand{\AffLinz}{Institute of Semiconductor and Solid State Physics, Johannes Kepler University, Linz, Austria}

\begin{document}
\preprint{APS/123-QED}

\title{Tuning the coherent interaction of an electron qubit and a nuclear magnon}

\author{Noah Shofer}
\affiliation{\AffCam}
\author{Leon Zaporski}
\affiliation{\AffCam}
\author{Martin Hayhurst Appel}
\affiliation{\AffCam}
\author{Santanu Manna}
\email[Current address: ]{Department of Electrical Engineering, Indian Institute of Technology Delhi, New Delhi, India}
\affiliation{\AffLinz}
\author{Saimon Covre da Silva}
\email[Current address: ]{Instituto de F\'{i}sica Gleb Wataghin, Universidade Estadual de Campinas, Campinas, Brazil}
\affiliation{\AffLinz}
\author{Alexander Ghorbal}
\affiliation{\AffCam}
\author{Urs Haeusler}
\affiliation{\AffCam}
\author{Armando Rastelli}
\affiliation{\AffLinz}
\author{Claire Le Gall}
\affiliation{\AffCam}
\author{Micha\l{} Gawe\l{}czyk}
\affiliation{Institute of Theoretical Physics, Wroc\l{}aw University of Science and Technology, Wroc\l{}aw, Poland}
\author{Mete Atatüre}
\email[Corresponding author: ]{ma424@cam.ac.uk}
\affiliation{\AffCam}
\author{Dorian A. Gangloff}
\email[Corresponding author: ]{dag50@cam.ac.uk}
\affiliation{\AffCam}


\begin{abstract}
A central spin qubit interacting coherently with an ensemble of proximal spins can be used to engineer entangled collective states or a multi-qubit register. Making full use of this many-body platform requires tuning the interaction between the central spin and its spin register. GaAs quantum dots offer a model realization of the central spin system where an electron qubit interacts with multiple ensembles of $\sim 10^{4}$ nuclear spins. In this work, we demonstrate tuning of the interaction between the electron qubit and the nuclear many-body system in a GaAs quantum dot. The homogeneity of the GaAs system allows us to perform high-precision and isotopically selective nuclear sideband spectroscopy, which reveals the single-nucleus electronic Knight field. Together with time-resolved spectroscopy of the nuclear field, this fully characterizes the electron-nuclear interaction for a priori control. An algorithmic feedback sequence selects the nuclear polarization precisely, which adjusts the electron-nuclear exchange interaction in situ via the electronic \emph{g}-factor anisotropy. This allows us to tune directly the activation rate of a collective nuclear excitation (magnon) and the coherence time of the electron qubit. Our method is applicable to similar central-spin systems and enables the programmable tuning of coherent interactions in the many-body regime.
\end{abstract}

\maketitle


\section{Introduction}

Quantum many-body systems offer an avenue for quantum simulation \cite{blochQuantumSimulationsUltracold2012} and computational tasks \cite{rotondoDickeSimulatorsEmergent2015,taylorControllingMesoscopicSpin2003,bluvsteinLogicalQuantumProcessor2024}, as well as for fundamental exploration of quantum matter, such as spin liquids \cite{semeghiniProbingTopologicalSpin2021}, quasi-crystals \cite{viebahnMatterWaveDiffractionQuasicrystalline2019}, and quantum phase transitions \cite{Zhang2017b}. Exploring and optimizing performance within this phase space relies on the ability to tune the coupling parameters between quantum particles, as pioneered in cold atom systems \cite{blochQuantumSimulationsUltracold2012,bluvsteinControllingQuantumManybody2021,Zhang2017b} where inter-particle distance and interaction cross-section can be varied continuously over a large dynamic range. Despite their fixed qubit coupling parameters, optically active defects in solids have also been used to demonstrate many-body effects such as time-crystalline behavior \cite{choiObservationDiscreteTimecrystalline2017,randallManybodyLocalizedDiscrete2021}. A particularly interesting subset of solid-state many-body systems is the central spin system, in which a spin qubit interacts coherently with an ensemble of proximal spins making up a register \cite{urbaszekNuclearSpinPhysics2013}. Following theoretical proposals for the development of a quantum memory \cite{taylorLongLivedMemoryMesoscopic2003}, physical realizations of the central spin system have been used to demonstrate quantum control of collective states \cite{ruskucNuclearSpinwaveQuantum2022} as well as techniques for quantum error correction using a many-spin register \cite{taminiauUniversalControlError2014}. The addition of in situ tuning of the coupling between the central spin qubit and its associated spin register would enable the dynamical selection of various operational regimes for the optimization of quantum information tasks \cite{denningCollectiveQuantumMemory2019}, the preparation of previously inaccessible entangled collective states \cite{zaporskiManyBodySingletPrepared2023}, or the observation of engineered phase transitions \cite{Kessler2012a,Goldman2023}.

In addition to excellent optical properties \cite{wangOptimalSinglephotonSources2019,liuSolidstateSourceStrongly2019,tommBrightFastSource2021,appelCoherentSpinPhotonInterface2021,thomasBrightPolarizedSinglePhoton2021,berezovskyPicosecondCoherentOptical2008,pressCompleteQuantumControl2008,schollResonanceFluorescenceGaAs2019,zhaiLownoiseGaAsQuantum2020}, quantum dots (QDs) offer an exemplary realization of the central spin model where an electronic qubit interacts with a dense mesoscopic ensemble of nuclear spins via a contact hyperfine interaction \cite{urbaszekNuclearSpinPhysics2013}. The QD central spin system benefits from both a one-to-all coupling of the qubit to the spin ensemble, allowing activation and measurement of collective excitations of the nuclear ensemble (magnons) \cite{gangloffQuantumInterfaceElectron2019,gangloffWitnessingQuantumCorrelations2021,jacksonQuantumSensingCoherent2021}, as well as an all-to-all coupling of the nuclear spins mediated by the qubit \cite{wustRoleElectronSpin2016} that could enable entangled-state engineering within the ensemble \cite{zaporskiManyBodySingletPrepared2023}. In this system, qubit-ensemble interactions can be tuned ex situ using an external magnetic field \cite{merkulovElectronSpinRelaxation2002} to suppress electron-nuclear spin-flips and improve electronic coherence times \cite{stockillQuantumDotSpin2016} or, owing to recent developments, Hamiltonian engineering techniques can be employed to controllably activate electron-nuclear dynamics \cite{jacksonOptimalPurificationSpin2022}. A tuneable interaction strength between the electron spin qubit and the nuclear ensemble would allow varying the rate of magnon injection, which forms the mechanism for quantum information storage in the QD nuclear register \cite{denningCollectiveQuantumMemory2019} and for introducing quantum correlations within the ensemble.\par

In this work, we demonstrate an in situ tuneable coherent interaction between a central electron spin qubit and an ensemble of nuclear spins in a GaAs quantum dot. We show that this interaction arises from an anisotropy of the electron \emph{g}-factor and enables strong driving of nuclear magnons. This enables us to implement an algorithmic feedback protocol \cite{jacksonOptimalPurificationSpin2022} that leads to nuclear polarization locking at a programmable setpoint and a hundredfold narrowing of the slow noise spectrum of the ensemble. We can thus perform precision nuclear spectroscopy to measure directly the single-nucleus hyperfine constants -- both collinear and non-collinear terms -- that fully characterize the qubit-register interaction. By shifting the mean-field polarization of the nuclear ensemble, we directly tune the spin-flip coupling strength, as observed through a modified electron qubit coherence time $T_2$. Finally, we use this in situ control to tune deterministically the injection rate of single nuclear magnons, as a first practical demonstration of direct control over the qubit-register interaction parameters.\par

\begin{figure*}
\includegraphics[scale = 0.86]{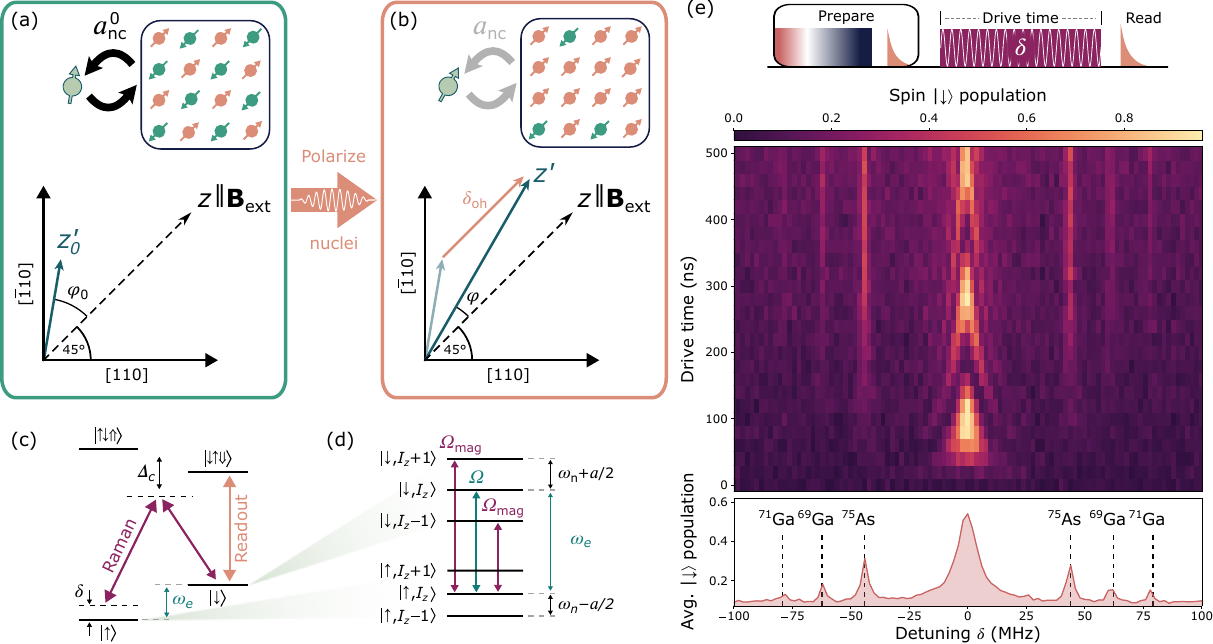}
\caption{(a) Top panel: the electron spin is in contact with an ensemble of $\sim\!\!10 ^5$ As and Ga nuclear spins, forming a central spin system. Bottom panel: A 6 T magnetic field $\mathbf{B}_{\mathrm{ext}}$ is applied perpendicularly to the [001] crystal axis (growth axis) of the QD, along [010] (i.e. at $45\degree$ relative to [110]). An electron \emph{g}-factor anisotropy leads to a different quantization axis for the electron and the nuclear spins, $z^{\prime}$ and $z$ respectively, separated by an angle $\varphi_{0}$, and to a non-collinear hyperfine interaction $a_{\mathrm{nc}}^0$ (b) As the mean-field Overhauser shift $\doh$ is increased, the angle $\varphi$ between the electron quantization axis $z^{\prime}$ and the external magnetic field $\mathbf{B}_{\mathrm{ext}}$ decreases, along with $a_{\mathrm{nc}}$. (c) The energy levels of a QD charged with a single electron, in a Voigt magnetic field geometry. The electron spin $\hat{S}_{z'}$-eigenstates $\ket{\downarrow}$ and $\ket{\uparrow}$ are split by $\omega_e$. The spin state is read out with a laser resonant with the trion transition $\ket{\downarrow}\rightarrow\ket{\downarrow\uparrow\Downarrow}$ at $782.9\,\mathrm{nm}$. Spin control is performed using a bichromatic laser detuned from the excited state by $\Delta_c \sim 400\,\mathrm{GHz}$, with a two-photon detuning of $\delta$ and a Rabi frequency $\Omega$.  (d) The hyperfine coupling to a nuclear ensemble with polarization $I_{z}$ (eigenstate of $\hat{I}_{z}$) results in an energy ladder (right panel) split by $\omega_{n}+a/2$ in the $\ket{\downarrow}$ manifold and $\omega_{n}-a/2$ in the $\ket{\uparrow}$ manifold, where $a$ is the single-nucleus hyperfine constant and $\omega_{n}$ is the nuclear Larmor frequency. (e) Top panel: the pulse sequence used to probe magnon dynamics. The preparation step locks the nuclear polarization $I_{z}$ using a algorithmic feedback \cite{jacksonOptimalPurificationSpin2022}. Middle panel: the electron spin resonance spectrum as function of Rabi drive time, with a measured Rabi frequency of $\Omega = 5.2\,\mathrm{MHz}$ at $\delta = 0$ and an electron splitting of $\omega_{e}^{0} = 3\,\mathrm{GHz}$. Bottom panel: the averaged spectrum over all Rabi drive times in the top panel.}
\label{fig:1}
\end{figure*}

\section{The electro-nuclear Hamiltonian}

Our central spin system consists of GaAs QDs (grown by droplet epitaxy) within an electrically gated device hosting a single electron together with the large number of spin-3/2 ${}^{75}\mathrm{As}$, $\mathrm{{}^{71}Ga}$, and $\mathrm{{}^{69}Ga}$ nuclei of the QD atoms. The nuclei are coupled to the single electron via a hyperfine contact interaction, realizing the illustration presented in the top panel of Fig.~\ref{fig:1}(a). The electron g-tensor has principal axes aligned with the $[110]$ and $[\bar{1}10]$ crystal axes, with values $g_{110}$ and $g_{\bar{1}10}$, and negligible off-diagonal elements in this basis \cite{botzemQuadrupolarAnisotropyEffects2016}. If an external magnetic field $\mathbf{B}_{\mathrm{ext}}$ is not oriented along one of these axes, a \emph{g}-factor anisotropy $g_{110} \neq g_{\bar{1}10}$ causes the electron spin to be quantized along an axis offset from $\mathbf{B}_{\mathrm{ext}}$ \cite{botzemQuadrupolarAnisotropyEffects2016,stanoFactorElectronsGatedefined2018,camenzindIsotropicAnisotropicFactor2021,SI}. The angle between the electron spin quantization axis and $\mathbf{B}_{\mathrm{ext}}$ is given by $\varphi_{0} = \arctan\left[(g_{110}-g_{\bar{1}10})/(g_{110}+g_{\bar{1}10})\right]$ (Fig.~\ref{fig:1}(a)). In the regime where the Zeeman effect dominates over the hyperfine interaction, the nuclei continue to be quantized along $\mathbf{B}_{\mathrm{ext}}$, thus causing a non-collinear interaction with the electron.\par

The Hamiltonian for the electronic spin $\hat{\bm{S}}$ in the frame defined by the external magnetic field, where $z$ is aligned with $\mathbf{B}_{\mathrm{ext}}$, is then:
\begin{equation}
\label{H_e}
\mathcal{H}_{e} =\omega_{e}^{0}\cos(\varphi_{0})\hat{S}_{z} + \omega_{e}^{0}\sin(\varphi_{0})\hat{S}_{x},
\end{equation}
where $\omega_{e}^{0}$ is the electronic Zeeman splitting due to $\mathbf{B}_{\mathrm{ext}}$ and we have set $h = 1$. The electron interacts with each nuclear species via an isotropic Fermi contact hyperfine interaction:
\begin{equation}
\label{H_en}
\mathcal{H}_{\mathrm{fc}} = a\!\left( \hat{S}_{x}\hat{I}_{x} + \hat{S}_{y}\hat{I}_{y} + \hat{S}_{z}\hat{I}_{z} \right),
\end{equation}
where $\hat{I}_\alpha \equiv \sum_j \hat{I}_\alpha^j$ ($\alpha = x,y,z$) are the collective nuclear spin operators summed over $j=1,..,N$ nuclei of that species and $a$ is the average single-nucleus hyperfine interaction strength \cite{SI}. We define $a = A/N$, where $A$ is the material hyperfine constant for the given nuclear species and $N$ is the number of those nuclei. In the presence of net nuclear polarization along $\mathbf{B}_{\mathrm{ext}}$ \cite{gangloffWitnessingQuantumCorrelations2021,jacksonQuantumSensingCoherent2021}, we can write the last term of Eq.~\ref{H_en} as:
\begin{equation}
\label{SzIz}
a\hat{S}_{z}\hat{I}_{z} = (\doh +  a\Delta\hat{I}_{z})\hat{S}_{z},
\end{equation}
where $\doh$ is the mean-field Overhauser shift \cite{overhauserPolarizationNucleiMetals1953} on the electron-spin splitting due to the effective magnetic field from polarized nuclei and $a\Delta\hat{I}_{z}$ captures the dynamic variation of polarization around its mean-field value that defines the inhomogeneous dephasing time for the electronic spin, $T_{2}^{*}$. Note that $a\Delta\hat{I}_{z}\ll\doh,\omega_{e}^{0}$ and as a result the electronic Hamiltonian in the presence of a net nuclear polarization is:
\begin{equation}
\label{H_e_OH}
\mathcal{H}_{e} \simeq \left( \omega_{e}^{0}\cos(\varphi_{0}) +\doh\right) \hat{S}_{z} + \omega_{e}^{0}\sin(\varphi_{0})\hat{S}_{x}. 
\end{equation}
We now diagonalize the electronic Hamiltonian, thus defining the electronic quantization axis $z^{\prime}$ and qubit energy splitting $\omega_{e}\equiv \sqrt{\left(\omega_{e}^{0}\cos(\varphi_{0}) + \doh \right)^{2} + \left( \omega_{e}^{0}\sin(\varphi_{0}) \right)^{2}}$ (Fig.~\ref{fig:1}(b),(c)). This dominant energy term suppresses (to first order in $\omega_{e}$) components of the hyperfine interaction not parallel to $z'$ \cite{SI}. With the addition of a coherent electronic drive with Rabi frequency $\Omega$ (\cite{bodeyOpticalSpinLocking2019}, Fig.\,1(c)), the resulting complete system Hamiltonian expressed in the drive rotating frame (frequency $\omd$) is: 
\begin{multline}
\label{H_diag}
\mathcal{H}^{\prime} = \delta\hat{S}_{z^{\prime}} + \Omega\hat{S}_{x^{\prime}} + \omega_{n}\hat{I}_{z}\\
+ a\hat{S}_{z^{\prime}}\!\left( \sin(\varphi)\hat{I}_{x} + \cos(\varphi)\Delta\hat{I}_{z}   \right),
\end{multline}
where $\delta = \omd - \omega_e$ is the drive detuning and the nuclear Zeeman splitting $\omega_{n} \gg a$ (i.e. nuclei remain oriented along $\mathbf{B}_{\mathrm{ext}}$). The angle $\varphi$ between the electron quantization axis $z'$ and $\mathbf{B}_{\mathrm{ext}}$ is thus a function of the Overhauser shift $\doh$ (Fig.~\ref{fig:1}(b)), and $\varphi = \varphi_{0}$ in the case of no net nuclear polarization ($\doh = 0$).\par

A consequence of the electronic spin \emph{g}-factor anisotropy is the emergence of the non-collinear hyperfine interaction captured by the term $a\sin(\varphi)\hat{S}_{z^{\prime}}\hat{I}_{x}$ \cite{vinkLockingElectronSpins2009,huangTheoreticalStudyNuclear2010,lattaHyperfineInteractionDominatedDynamics2011,hogeleDynamicNuclearSpin2012}, from which we define the constant $a_{\mathrm{nc}} = a\sin(\varphi)$. This interaction leads to a coherent spin exchange between the electron and the nuclear ensemble, where a collective nuclear mode $I_{z} \rightarrow I_{z}\pm 1$, referred to as a magnon \cite{gangloffQuantumInterfaceElectron2019}, is coherently excited in exchange for an electronic spin rotation (Fig.~\ref{fig:1}(d)). Due to the dependence of $\varphi$ on $\doh$, the strength of the non-collinear hyperfine constant $a_{\mathrm{nc}}$ is inversely proportional to $\doh$:
\begin{equation}
\label{eq:Anc_we}
    \frac{a_{\mathrm{nc}}}{a_{\mathrm{nc}}^{0}} = \frac{\sin(\varphi)}{\sin(\varphi_{0})} = \frac{\omega_{e}^{0}}{\omega_{e}}\simeq\frac{\omega_{e}^{0}}{\omega_{e}^0 + \doh}.
\end{equation}
 For example, as the Overhauser shift is increased by nuclear polarization -- increasing the total electron spin splitting $\omega_{e}$ -- the non-collinear coupling constant $a_{\mathrm{nc}}$ decreases proportionally. Selecting the mean-field polarization of the nuclear ensemble \cite{gangloffWitnessingQuantumCorrelations2021,jacksonQuantumSensingCoherent2021} via an algorithmic optical feedback protocol \cite{jacksonOptimalPurificationSpin2022} results in a controlled tuning of the non-collinear electron-nuclear interaction strength.\par

\begin{figure}
\includegraphics[scale = 0.85]{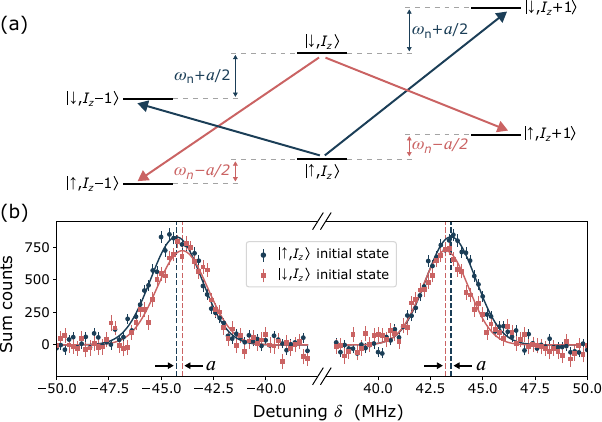}
\caption{(a) The ladder of nuclear states for a single species, highlighting the single-nucleus hyperfine constant $a$. The arrows correspond to the non-collinear enabled electron-nuclear spin flip transitions. (b) The negative and positive ${}^{75}\mathrm{As}$ sidebands at $t_{\mathrm{Rabi}}=400\,\mathrm{ns}$, for initial electron states $\ket{\downarrow}$ (red) and $\ket{\uparrow}$ (dark blue). The measured peaks correspond to the transitions of the same color in panel (a). The data are fit with a Gaussian distribution, with peak detuning differences between negative and positive sidebands of $0.26(7)\,\mathrm{MHz}$ and $0.29(7)\,\mathrm{MHz}$, respectively.} 
\label{fig:2}
\end{figure}



\section{Measuring the single-nucleus hyperfine constant}
We begin characterizing the electron-nuclear interface by performing electron-spin resonance (ESR) spectroscopy of the nuclear register for the condition $\Omega\ll \omega_{n}$, which will allow us to observe the manifold of nuclear states (Fig.~\ref{fig:1}(d)) and probe the longitudinal $a\cos(\varphi)\Delta\hat{I}_{z}$ term of our system Hamiltonian (Eq.~\ref{H_diag}). We perform coherent control of the electron qubit all optically via the QD trion state (Fig.~\ref{fig:1}(c)) following a stimulated Raman scheme \cite{bodeyOpticalSpinLocking2019}. We employ a pulse sequence during which the nuclear ensemble is prepared using an algorithmic feedback sequence \cite{jacksonOptimalPurificationSpin2022}, as was recently also realized in a GaAs QD system \cite{nguyenEnhancedElectronSpinCoherence2023}, locking the electron splitting to $\omega_{e}^{0}=3.0\,\mathrm{GHz}$ (its value at $\doh=0$) and extending its Ramsey coherence time a hundredfold to $T_{2}^{*} = 253(7)\,\mathrm{ns}$ \cite{SI}. To measure the ESR spectrum following preparation, the electron is driven at a fixed detuning $\delta$ and at a Rabi frequency of $\Omega = 5.2\,\mathrm{MHz}$, and the population of the $\ket{\downarrow}$ state is read-out optically. 

Figure~\ref{fig:1}(d) shows the measured ESR spectrum, providing a concrete visualization of the electro-nuclear states shown in Fig.~\ref{fig:1}(c). When the drive at effective Rabi frequency $\Omega' = \sqrt{\Omega^{2} + \delta^{2}}\approx \delta$ is resonant with one of the electro-nuclear transitions ($I_{z} \rightarrow I_{z}\pm 1$), which occurs at the nuclear Larmor frequencies of ${}^{75}\mathrm{As}$, ${}^{69}\mathrm{Ga}$, or ${}^{71}\mathrm{Ga}$, the measured $\ket{\downarrow}$ population increases as a result of driving the coherent electron-nuclear exchange. Unlike previous observations of nuclear sidebands in InGaAs QDs, where both $I_{z} \rightarrow I_{z}\pm 1$ and $I_{z} \rightarrow I_{z}\pm 2$ sidebands were allowed \cite{gangloffQuantumInterfaceElectron2019,jacksonQuantumSensingCoherent2021} due to the quadrupolar origin of the interaction, here we observe only single quanta spin flips $I_{z} \rightarrow I_{z}\pm 1$ due to the linear nature of the interaction derived from the \emph{g}-factor anisotropy (Eq.~\ref{H_diag}). The sidebands observed in this system are also strikingly narrower, reflecting the enhanced electron $T_{2}^{*}$ and reduced nuclear inhomogenous broadening due to the nearly strain-free growth of GaAs QDs \cite{dasilvaGaAsQuantumDots2021,chekhovichCrossCalibrationDeformation2018}. This allows for isotopically selective nuclear excitation and heralds a fully sideband-resolved spectroscopy regime for QD systems.\par


For full knowledge of the non-collinear interaction, which allows us to tune $a_{\mathrm{nc}}$ to the desired strength, we need to determine the single-nucleus hyperfine constant $a$ for a target species. To do so, we perform high-resolution nuclear-sideband spectroscopy of ${}^{75}\mathrm{As}$ to obtain $a\cos(\varphi)$ ($\sim\!a$ for small values of $\varphi$) directly from the Knight shift imparted on the nuclear eigenstates by the electron \cite{knightNuclearMagneticResonance1949,Lai2006,Sallen2014}. We prepare the electron into either the $\ket{\uparrow}$ state or into the $\ket{\downarrow}$ state -- for the latter using a coherent inversion pulse on the optically prepared state $\ket{\uparrow}$. As illustrated in Fig.~\ref{fig:2}(a), the two-photon detuning $|\delta| = \omn \pm a/2$ resonant with a nuclear sideband transition depends on the initial electron state $\ket{\uparrow}$ ($+$) or $\ket{\downarrow}$ ($-$). We thus probe the ${}^{75}\mathrm{As}$ sideband with a step size well below the sideband linewidth, at both negative and positive detunings, using a fixed Rabi drive time of $t_{\mathrm{Rabi}}=400\,\mathrm{ns}$. Figure~\ref{fig:2}(b) presents the measured spectra for the $\ket{\uparrow}$ and the $\ket{\downarrow}$ initial states, coded as dark blue and red circles, respectively. The shift in resonance condition for $\ket{\uparrow}$ and $\ket{\downarrow}$ data sets is visible by eye. We fit each sideband data set with a Gaussian distribution and obtain the frequency difference between $\ket{\uparrow}$ and $\ket{\downarrow}$ resonances for both negatively detuned and postively detuned sideband experiments. These frequency differences agree within one standard error and their average value is taken to be the single-nuclear hyperfine constant for the As species $a_{\mathrm{As}} = 0.28(7)\,\mathrm{MHz}$. This high-precision measurement of $a_{\mathrm{As}}$ additionally allows us to determine the effective number of As nuclei that interact with the electron as $N_{\mathrm{As}}=A_{\mathrm{As}}/a_{\mathrm{As}}=3.8(9)\times 10^{4}$, where we have used the known material constant $A_{\mathrm{As}}=65.3\,\mathrm{GHz}$ \cite{pagetLowFieldElectronnuclear1977,zaporskiIdealRefocusingOptically2023}. Given that ${}^{75}\mathrm{As}$ is present in all unit cells of GaAs, this also yields the total number of nuclei $N = 2N_{\mathrm{As}}= (7.6\pm1.8)\times 10^{4}$ interacting with the ground state electron in the QD. This agrees within a standard error with more indirect estimates of the number of nuclei based on Ramsey spin coherence measurements taken on similar QDs \cite{zaporskiIdealRefocusingOptically2023}.\par

\begin{figure}
\includegraphics[scale = 0.85]{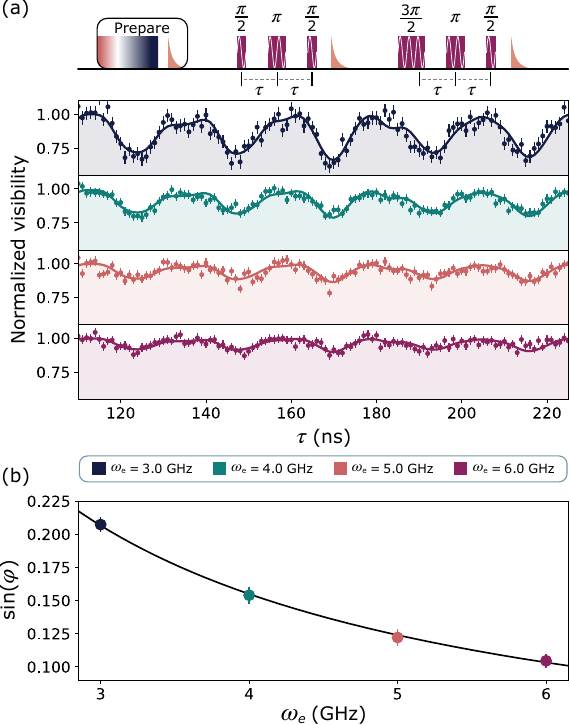}
\caption{(a) CP1 visibility as a function of time delay $\tau$ for the values of $\omega_e$ shown in the plot legend, corresponding to different degrees of nuclear polarization. The solid curve is a fit to the data based on a filter function model described in Eq.~\ref{eq:filterW}. Visibility is given by $v = (\mathrm{cts}_{\uparrow}-\mathrm{cts}_{\downarrow})/(\mathrm{cts}_{\uparrow}+\mathrm{cts}_{\downarrow})$, where are $\mathrm{cts}_{\uparrow}$ ($\mathrm{cts}_{\downarrow}$) are the background-corrected read-out counts from experiments initialized to the $\ket{\uparrow}$ ($\ket{\downarrow}$) state. (b) the angle $\varphi$ between $\hat{S}_{z^{\prime}}$ and $B_{\mathrm{ext}}$ as a function of $\omega_{e}$, extracted from global fits of Eq.~\ref{eq:filterW} to the CP1 and CP2 (see \cite{SI}) data. The solid curve is a fit $\sin(\varphi) = \sin(\varphi_{0})\times(\omega_{e}^{0}/\omega_{e})$ which gives  $\sin(\varphi_{0})=0.207(1)$.}
\label{fig:3}
\end{figure}

\begin{figure}
    \centering
    \includegraphics[scale = 0.8]{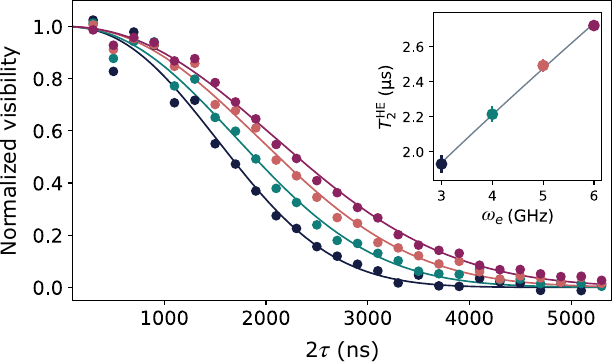}
    \caption{Qubit visibility under Hahn echo (CP1) sequence as a function of inter-pulse delay time $\tau$. The solid curves are stretched-exponential fits $\exp(-(\tau/T_{2})^{\alpha})$, with $T_{2}=\{1.93(5),2.21(5),2.49(4),2.72(3)\}\,\SI{}{\micro\second}$  and $\alpha = \{2.4(2),2.3(2),2.3(1),2.22(8)\}$ for $\omega_{e}=\{3,4,5,6\}\,\mathrm{GHz}$, respectively. Inset: $T_{2}$ as a function of $\omega_{e}$, along with a  fit $T_{2}(\omega_{e}) = 0.95(3)\,\SI{}{\micro\second}\times(\omega_{e}/\omega_{e}^{0})^{(2/\bar{\alpha})} + 0.99(4)\,\SI{}{\micro\second}$, where $\bar{\alpha} = 2.28$ is the mean of fitted $\alpha$ values.}
    \label{fig:HE}
\end{figure}

\begin{figure}
\includegraphics[scale = 0.85]{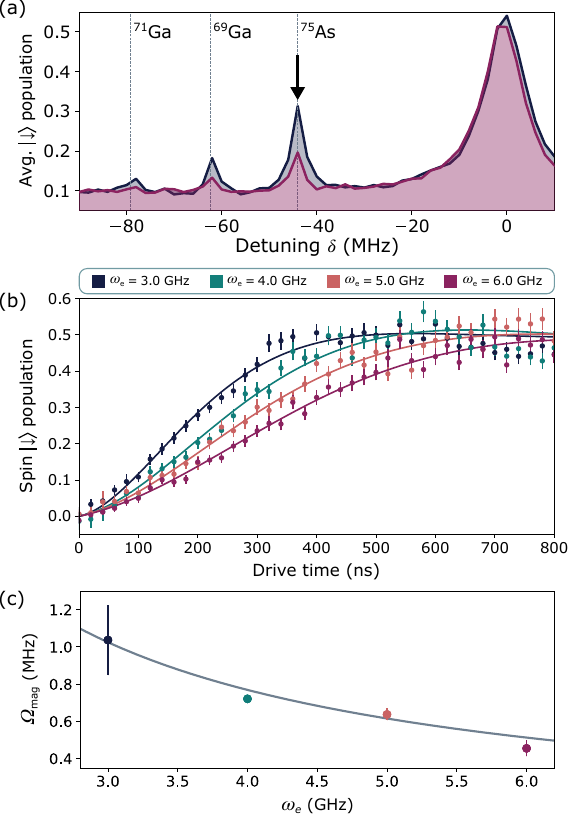}
\caption{(a) averaged QD sideband spectrum from $t_{\mathrm{Rabi}} = 0$ to 500 ns Rabi drive (the same range of Rabi drive times as in Fig.~\ref{fig:1}(f)) taken with a carrier Rabi frequency of $\Omega = 5.2\,\mathrm{MHz}$. The purple solid curve shows the spectrum at $\omega_{e}= 6\,\mathrm{GHz}$ while the dark blue curve is at $\omega_{e}^{0} = 3\,\mathrm{GHz}$. The arrow indicates the sideband transition driven in panel (b). (b) the $\ket{\downarrow}$ spin population of the negative ${}^{75}\mathrm{As}$ sideband as a function of Rabi drive time for several values of $\omega_{e}$. The solid curves are fits to a simulation of a damped two-level system. (c) the magnon Rabi frequency $\Omega_{\mathrm{mag}}$ versus the electron splitting $\omega_{e}$, averaged for the positive and negative As sidebands. The solid curve is an interpolated prediction of an ab initio calculation of $\Omega_{\mathrm{mag}}$ using Eq.\,\ref{eq:Omag}, where the one standard error from measured parameters falls under the width of the curve.}
\label{fig:4}
\end{figure}

\section{Measuring the transverse nuclear field}

Our second and final high-precision characterization yields the non-collinear hyperfine constant $a_{\mathrm{nc}}=a\sin(\varphi)$ by measuring the effect of the precessing transverse nuclear field $a_{\mathrm{nc}}\hat{I}_{x}$ on the electron spin coherence.
To probe this effect, we employ one-pulse and two-pulse Carr-Purcell sequences \cite{carrEffectsDiffusionFree1954} (CP1 and CP2, respectively) at several values of the qubit splitting $\ome$, and thus Overhauser shift $\doh$, chosen by our drive frequency $\omd$ \cite{jacksonOptimalPurificationSpin2022}. We obtain the electron spin visibility for every delay $\tau$ as the difference in $\ket{\downarrow}$ readout counts between two consecutive pulse sequences with initial state $\ket{\uparrow}$ or $\ket{\downarrow}$. Figure~\ref{fig:3}(a) shows the measured qubit visibility for the CP1 sequence (see \cite{SI} for CP2 measurements), where periodic modulation of this visibility occurs at the Larmor frequency of each nuclear species. Strikingly, as the mean-field nuclear polarization $\doh$ is increased by our choice of $\omega_e$, the magnitude of the dips in visibility decreases as we expect from a reduced non-collinear coupling to the transverse nuclear field (Eq.\,\ref{eq:Anc_we}). 

Indeed, shown in Fig.~\ref{fig:3}(a) are theory fits to the data with a well-established model based on filter function formalism \cite{cywinskiHowEnhanceDephasing2008,malinowskiNotchFilteringNuclear2017} (see \cite{SI} for details), where the qubit visibility $W(\tau)$ at pulse delay $\tau$ is given by:
\begin{equation}
\label{eq:filterW}
W(\tau) = \mathrm{exp} \left[ -\sum_{j}\frac{5}{4}\frac{1}{N}\left(\frac{A^{j}\sin(\varphi)\sqrt{c_{j}}}{\omega_{n}^{j}}\right)^{2} \mathcal{F}_{\mathrm{CP}}(\omega_{n}^{j}\tau)    \right],
\end{equation}
where $\omega_{n}^{j}$ are the nuclear Larmor frequencies of the nuclear species ${}^{75}\mathrm{As}$, ${}^{69}\mathrm{Ga}$, and ${}^{71}\mathrm{Ga}$, $A^{j}$ are the hyperfine material constants for each species \cite{pagetLowFieldElectronnuclear1977}, $c_{j}$ are the abundances of the nuclear isotopes \cite{zaporskiIdealRefocusingOptically2023}, $\mathcal{F}_{\mathrm{CP}}(\omega_{j}\tau)$ is the filter function of the given CP sequence at delay $\tau$ \cite{cywinskiHowEnhanceDephasing2008,malinowskiNotchFilteringNuclear2017}, and $N$ is the total number of nuclei in the quantum dot obtained from the Knight shift (Fig.~\ref{fig:2}). The nuclear Larmor frequency of the ${}^{75}\mathrm{As}$ is more prominent in the visibility modulation spectrum than that of the other species \cite{SI}, as straightforwardly explained by the $(\omega_\text{n}^j)^{-2}$ dependence in $W(\tau)$. We fit the electron spin visibility in Fig.~\ref{fig:3}(a) with $\sin(\varphi)$ as the only microscopic free parameter unconstrained in the model a priori. Fig.~\ref{fig:3}(b) shows the resulting $\sin(\varphi)$ values as a function of $\omega_{e}$. Fitting these values to Eq.~\ref{eq:Anc_we} yields $\sin(\varphi_{0}) = 0.207(1)$, or equivalently an anisotropy angle of $12\degree$ in the absence of nuclear polarization, in good agreement with numerical predictions based on AFM measurements of QDs on this wafer \cite{SI} -- the significant anisotropy observed here results from stronger electronic confinement within droplet QDs compared to bulk GaAs \cite{stanoFactorElectronsGatedefined2018,camenzindIsotropicAnisotropicFactor2021}. The excellent agreement with the $\omega_{e}^{-1}$ fit from Eq.~\ref{eq:Anc_we} further confirms the geometric tilt of the electron quantization axis $z'$ from the external magnetic field $\textbf{B}_{\mathrm{ext}}$ can be modified via a tuneable nuclear polarization.\par 

\section{Extension of $T_{2}$}
A direct consequence of our ability to increase the qubit splitting is the reduced interaction with the transverse nuclear field \cite{stockillQuantumDotSpin2016}. This is observed as an increase of the qubit's Hahn Echo time $T_{2}$. Figure~\ref{fig:HE} shows the qubit visibility under a Hahn Echo (CP1) sequence for the values of $\omega_{e}$ used in Fig.~\ref{fig:3}(a). The decay of qubit coherence over delay time $\tau$ is fitted with a stretched exponential characterized by a decay time $T_2$ and exponent $\alpha$. When decoherence is dominated by a decorrelating transverse nuclear field with amplitude $\propto \omega_{e}^{-1}$ \cite{stockillQuantumDotSpin2016, zaporskiIdealRefocusingOptically2023}, the electron $T_{2}$ is expected to increase with $\omega_{e}^{2/\alpha}$ \cite{SI}, in excellent agreement with our measurements (Fig.~\ref{fig:HE} inset). In contrast to previous works in QDs where an increase in $T_{2}$ was observed with increasing external magnetic field \cite{stockillQuantumDotSpin2016,zaporskiIdealRefocusingOptically2023}, here we are able to realize the same increase in $T_{2}$ via a programmable effective magnetic field induced by nuclear polarization. This highlights the unique nature of the non-collinear interaction mediated by the \emph{g}-factor anisotropy, which enables coupling and decoupling the qubit dynamically from its intrinsic interaction with its nuclear ensemble. \par

\section{Tuning the magnon activation rate}

Having shown tuning of $a_{\mathrm{nc}}$, we now demonstrate direct control of the resonantly driven electron-nuclear interaction, measured by the injection rate of single-spin excitations (magnons) in the nuclear ensemble. Figure~\ref{fig:4}(a) contrasts the negative-detuned sideband spectra for two different values of $\omega_{e}$, namely 3 GHz (as was shown in Fig.~\ref{fig:1}) and 6 GHz, averaged over the first 500 ns of sideband driving. The absence of a noticeable change in the $\delta = 0$ central electron spin resonance confirms that the bare electron Rabi rate remains identical. In contrast, we observe significant suppression of the sideband resonances in the $\omega_{e} = 6\,\mathrm{GHz}$ spectrum compared to the $\omega_{e} = 3\,\mathrm{GHz}$ spectrum. This suppression is a direct consequence of the reduced non-collinear interaction strength in the 6 GHz spectrum, indicating that magnons are created by the non-collinear interaction mediated by the electron \emph{g}-factor anisotropy and that the injection rate can be regulated by the tuneable Overhauser shift $\doh$.\par

Finally, we benchmark our control over the electro-nuclear interaction in the coherent regime by tuning the exchange (Rabi) frequency when on resonance with the ${}^{75}\mathrm{As}$ magnon sideband. For a fixed bare electron drive rate $\Omega$, this magnon Rabi frequency $\Omega_{\mathrm{mag}}$ will depend only on the strength of the coherent electron-nuclear exchange interaction, and is thus sensitive to tuning of $a_{\mathrm{nc}}$. Figure~\ref{fig:4}(b) shows the $\ket{\downarrow}$ population of the negative-detuned sideband as a function of Rabi drive time for the same range of $\omega_{e}$ values as used in Fig.~\ref{fig:3}. We fit the data with a damped two-level system model characterized by a Rabi frequency $\Omega_{\mathrm{mag}}$ and two dephasing parameters $\gamma_{1}$ and $\Gamma$ that account for spin-flip and dephasing channels (see \cite{SI}). Figure~\ref{fig:4}(c) shows the fitted Rabi frequency, averaged from fits to the negative- and positive-detuned sideband data, as a function of $\omega_{e}$. In the detuned driving regime we expect:

\begin{equation}
\label{eq:Omag}
    \Omega_{\mathrm{mag}} = \frac{a_\text{nc}\Omega}{2\omega_\text{n}}\sqrt{\frac{5N_\text{As}}{2}}\text{,}
\end{equation}
\noindent
where $\sqrt{5N_\text{As}/2}$ is the root-mean-square transverse angular momentum of an ensemble of $N_\text{As}$ $3/2$ spins at thermal equilibrium \cite{SI}. Figure~\ref{fig:4}(c) shows an interpolated prediction of the calculated values of $\Omega_{\mathrm{mag}}$ according to Eq.\,\ref{eq:Omag} and using our experimental values of $a_\text{nc}$ (Eq.~\ref{eq:Anc_we} and Fig.\,3) \cite{SI}. The excellent agreement to the predicted value shows that the magnon injection rate $\Omega_{\mathrm{mag}}$ can be tuned ab initio and highlights the predictive power of the polarization-based $a_{\mathrm{nc}}$ tuning approach.


\section{Conclusion and outlook}

In summary, we have demonstrated a method to tune the coherent interaction strength of a central spin qubit and a collective excitation of a mesoscopic spin ensemble. The electronic \emph{g}-factor anisotropy enables this in situ control of the mean-field nuclear polarization, which in turn changes the relative angle between the nuclear and electronic quantization axes. The \emph{g}-factor anisotropy is a result of the spin-orbit interaction in GaAs QDs and is a function of the size, shape, and composition of the quantum dot and surrounding barrier material \cite{SI}, offering a growth-based design mechanism on reaching the ideal dynamic tuning range for the electron-nuclear interaction. To extend the usefulness of our in situ tuning method, the use of few-microsecond direct rotation pulses \cite{chekhovichMeasurementSpinTemperature2017} on a fully polarized (or anti-polarized) nuclear ensemble could enable switching between the strongly and weakly interacting regimes, well within the hundred-microsecond coherence of the central spin \cite{zaporskiIdealRefocusingOptically2023}. Additionally, while in this work we have observed the nuclear ensemble in a classical regime $\Omega_{\mathrm{mag}} \sim \sqrt{N}$, the tuning methods demonstrated here would allow us to witness non-classical transverse field correlations \cite{gangloffWitnessingQuantumCorrelations2021,vitaglianoSpinSqueezingInequalities2011} as deviations from predicted magnon injection rates. This would indicate the formation of a Dicke state of the nuclear system \cite{dickeCoherenceSpontaneousRadiation1954}, such as a singlet state \cite{zaporskiManyBodySingletPrepared2023}. Our work demonstrates the realization of a tuneable solid-state central spin system for quantum information protocols and quantum many-body physics.\par


\begin{acknowledgements}
We acknowledge support from the US Office of Naval Research Global (N62909-19-1-2115;M.A.), the EU Horizon 2020 FET Open project QLUSTER (862035; M.A. and C.L.G), the EU Horizon 2020 research and innovation program under Marie Sklodowska-Curie grant QUDOT-TECH (861097; M.A.), the Royal Society (EA/181068; C.L.G), Qurope (899814; A.R.), ASCENT+ (871130; A.R.), the Austrian Science Fund (FWF; 10.55776/COE1; A.R.), the EU NextGenerationEU (10.55776/FG5; A.R.), SERB India (CRG/2023/007444; S.M.), and a collaborative research agreement with the Korean Electronics and Telecommunications Research Institute (D.A.G.). Part of the calculations have been carried out using resources provided by the Wroc\l{}aw Centre for Networking and Supercomputing, Grant No. 203 (M.G.). L.Z. acknowledges support from the EPSRC DTP (EP/R513180/1) and A.G. from a Harding scholarship and a Christ's College scholarship. D.A.G acknowledges a Royal Society University Research Fellowship. C.L.G. acknowledges a Dorothy Hodgkin Royal Society Fellowship.\par
We additionally would like to thank E. Chekhovich, C. Schimpf, and Z.X. Koong  for fruitful discussions, as well as M. Tribble for cleanroom advice. M.G. is grateful to K. Gawarecki for sharing his computational code. \par
\end{acknowledgements}



\bibliography{gaas_aniso}

\begin{thebibliography}{58}%
\makeatletter
\providecommand \@ifxundefined [1]{%
 \@ifx{#1\undefined}
}%
\providecommand \@ifnum [1]{%
 \ifnum #1\expandafter \@firstoftwo
 \else \expandafter \@secondoftwo
 \fi
}%
\providecommand \@ifx [1]{%
 \ifx #1\expandafter \@firstoftwo
 \else \expandafter \@secondoftwo
 \fi
}%
\providecommand \natexlab [1]{#1}%
\providecommand \enquote  [1]{``#1''}%
\providecommand \bibnamefont  [1]{#1}%
\providecommand \bibfnamefont [1]{#1}%
\providecommand \citenamefont [1]{#1}%
\providecommand \href@noop [0]{\@secondoftwo}%
\providecommand \href [0]{\begingroup \@sanitize@url \@href}%
\providecommand \@href[1]{\@@startlink{#1}\@@href}%
\providecommand \@@href[1]{\endgroup#1\@@endlink}%
\providecommand \@sanitize@url [0]{\catcode `\\12\catcode `\$12\catcode `\&12\catcode `\#12\catcode `\^12\catcode `\_12\catcode `\%12\relax}%
\providecommand \@@startlink[1]{}%
\providecommand \@@endlink[0]{}%
\providecommand \url  [0]{\begingroup\@sanitize@url \@url }%
\providecommand \@url [1]{\endgroup\@href {#1}{\urlprefix }}%
\providecommand \urlprefix  [0]{URL }%
\providecommand \Eprint [0]{\href }%
\providecommand \doibase [0]{https://doi.org/}%
\providecommand \selectlanguage [0]{\@gobble}%
\providecommand \bibinfo  [0]{\@secondoftwo}%
\providecommand \bibfield  [0]{\@secondoftwo}%
\providecommand \translation [1]{[#1]}%
\providecommand \BibitemOpen [0]{}%
\providecommand \bibitemStop [0]{}%
\providecommand \bibitemNoStop [0]{.\EOS\space}%
\providecommand \EOS [0]{\spacefactor3000\relax}%
\providecommand \BibitemShut  [1]{\csname bibitem#1\endcsname}%
\let\auto@bib@innerbib\@empty
\bibitem [{\citenamefont {Bloch}\ \emph {et~al.}(2012)\citenamefont {Bloch}, \citenamefont {Dalibard},\ and\ \citenamefont {Nascimb{\`e}ne}}]{blochQuantumSimulationsUltracold2012}%
  \BibitemOpen
  \bibfield  {author} {\bibinfo {author} {\bibfnamefont {I.}~\bibnamefont {Bloch}}, \bibinfo {author} {\bibfnamefont {J.}~\bibnamefont {Dalibard}},\ and\ \bibinfo {author} {\bibfnamefont {S.}~\bibnamefont {Nascimb{\`e}ne}},\ }\bibfield  {title} {\bibinfo {title} {Quantum simulations with ultracold quantum gases},\ }\href {https://doi.org/10.1038/nphys2259} {\bibfield  {journal} {\bibinfo  {journal} {Nature Phys}\ }\textbf {\bibinfo {volume} {8}},\ \bibinfo {pages} {267} (\bibinfo {year} {2012})}\BibitemShut {NoStop}%
\bibitem [{\citenamefont {Rotondo}\ \emph {et~al.}(2015)\citenamefont {Rotondo}, \citenamefont {Cosentino~Lagomarsino},\ and\ \citenamefont {Viola}}]{rotondoDickeSimulatorsEmergent2015}%
  \BibitemOpen
  \bibfield  {author} {\bibinfo {author} {\bibfnamefont {P.}~\bibnamefont {Rotondo}}, \bibinfo {author} {\bibfnamefont {M.}~\bibnamefont {Cosentino~Lagomarsino}},\ and\ \bibinfo {author} {\bibfnamefont {G.}~\bibnamefont {Viola}},\ }\bibfield  {title} {\bibinfo {title} {Dicke {{Simulators}} with {{Emergent Collective Quantum Computational Abilities}}},\ }\href {https://doi.org/10.1103/PhysRevLett.114.143601} {\bibfield  {journal} {\bibinfo  {journal} {Phys. Rev. Lett.}\ }\textbf {\bibinfo {volume} {114}},\ \bibinfo {pages} {143601} (\bibinfo {year} {2015})}\BibitemShut {NoStop}%
\bibitem [{\citenamefont {Taylor}\ \emph {et~al.}(2003{\natexlab{a}})\citenamefont {Taylor}, \citenamefont {Imamoglu},\ and\ \citenamefont {Lukin}}]{taylorControllingMesoscopicSpin2003}%
  \BibitemOpen
  \bibfield  {author} {\bibinfo {author} {\bibfnamefont {J.~M.}\ \bibnamefont {Taylor}}, \bibinfo {author} {\bibfnamefont {A.}~\bibnamefont {Imamoglu}},\ and\ \bibinfo {author} {\bibfnamefont {M.~D.}\ \bibnamefont {Lukin}},\ }\bibfield  {title} {\bibinfo {title} {Controlling a {{Mesoscopic Spin Environment}} by {{Quantum Bit Manipulation}}},\ }\href {https://doi.org/10.1103/PhysRevLett.91.246802} {\bibfield  {journal} {\bibinfo  {journal} {Phys. Rev. Lett.}\ }\textbf {\bibinfo {volume} {91}},\ \bibinfo {pages} {246802} (\bibinfo {year} {2003}{\natexlab{a}})}\BibitemShut {NoStop}%
\bibitem [{\citenamefont {Bluvstein}\ \emph {et~al.}(2024)\citenamefont {Bluvstein}, \citenamefont {Evered}, \citenamefont {Geim}, \citenamefont {Li}, \citenamefont {Zhou}, \citenamefont {Manovitz}, \citenamefont {Ebadi}, \citenamefont {Cain}, \citenamefont {Kalinowski}, \citenamefont {Hangleiter}, \citenamefont {Bonilla~Ataides}, \citenamefont {Maskara}, \citenamefont {Cong}, \citenamefont {Gao}, \citenamefont {Sales~Rodriguez}, \citenamefont {Karolyshyn}, \citenamefont {Semeghini}, \citenamefont {Gullans}, \citenamefont {Greiner}, \citenamefont {Vuleti{\'c}},\ and\ \citenamefont {Lukin}}]{bluvsteinLogicalQuantumProcessor2024}%
  \BibitemOpen
  \bibfield  {author} {\bibinfo {author} {\bibfnamefont {D.}~\bibnamefont {Bluvstein}}, \bibinfo {author} {\bibfnamefont {S.~J.}\ \bibnamefont {Evered}}, \bibinfo {author} {\bibfnamefont {A.~A.}\ \bibnamefont {Geim}}, \bibinfo {author} {\bibfnamefont {S.~H.}\ \bibnamefont {Li}}, \bibinfo {author} {\bibfnamefont {H.}~\bibnamefont {Zhou}}, \bibinfo {author} {\bibfnamefont {T.}~\bibnamefont {Manovitz}}, \bibinfo {author} {\bibfnamefont {S.}~\bibnamefont {Ebadi}}, \bibinfo {author} {\bibfnamefont {M.}~\bibnamefont {Cain}}, \bibinfo {author} {\bibfnamefont {M.}~\bibnamefont {Kalinowski}}, \bibinfo {author} {\bibfnamefont {D.}~\bibnamefont {Hangleiter}}, \bibinfo {author} {\bibfnamefont {J.~P.}\ \bibnamefont {Bonilla~Ataides}}, \bibinfo {author} {\bibfnamefont {N.}~\bibnamefont {Maskara}}, \bibinfo {author} {\bibfnamefont {I.}~\bibnamefont {Cong}}, \bibinfo {author} {\bibfnamefont {X.}~\bibnamefont {Gao}}, \bibinfo {author} {\bibfnamefont {P.}~\bibnamefont {Sales~Rodriguez}}, \bibinfo {author} {\bibfnamefont
  {T.}~\bibnamefont {Karolyshyn}}, \bibinfo {author} {\bibfnamefont {G.}~\bibnamefont {Semeghini}}, \bibinfo {author} {\bibfnamefont {M.~J.}\ \bibnamefont {Gullans}}, \bibinfo {author} {\bibfnamefont {M.}~\bibnamefont {Greiner}}, \bibinfo {author} {\bibfnamefont {V.}~\bibnamefont {Vuleti{\'c}}},\ and\ \bibinfo {author} {\bibfnamefont {M.~D.}\ \bibnamefont {Lukin}},\ }\bibfield  {title} {\bibinfo {title} {Logical quantum processor based on reconfigurable atom arrays},\ }\href {https://doi.org/10.1038/s41586-023-06927-3} {\bibfield  {journal} {\bibinfo  {journal} {Nature}\ }\textbf {\bibinfo {volume} {626}},\ \bibinfo {pages} {58} (\bibinfo {year} {2024})}\BibitemShut {NoStop}%
\bibitem [{\citenamefont {Semeghini}\ \emph {et~al.}(2021)\citenamefont {Semeghini}, \citenamefont {Levine}, \citenamefont {Keesling}, \citenamefont {Ebadi}, \citenamefont {Wang}, \citenamefont {Bluvstein}, \citenamefont {Verresen}, \citenamefont {Pichler}, \citenamefont {Kalinowski}, \citenamefont {Samajdar}, \citenamefont {Omran}, \citenamefont {Sachdev}, \citenamefont {Vishwanath}, \citenamefont {Greiner}, \citenamefont {Vuleti{\'c}},\ and\ \citenamefont {Lukin}}]{semeghiniProbingTopologicalSpin2021}%
  \BibitemOpen
  \bibfield  {author} {\bibinfo {author} {\bibfnamefont {G.}~\bibnamefont {Semeghini}}, \bibinfo {author} {\bibfnamefont {H.}~\bibnamefont {Levine}}, \bibinfo {author} {\bibfnamefont {A.}~\bibnamefont {Keesling}}, \bibinfo {author} {\bibfnamefont {S.}~\bibnamefont {Ebadi}}, \bibinfo {author} {\bibfnamefont {T.~T.}\ \bibnamefont {Wang}}, \bibinfo {author} {\bibfnamefont {D.}~\bibnamefont {Bluvstein}}, \bibinfo {author} {\bibfnamefont {R.}~\bibnamefont {Verresen}}, \bibinfo {author} {\bibfnamefont {H.}~\bibnamefont {Pichler}}, \bibinfo {author} {\bibfnamefont {M.}~\bibnamefont {Kalinowski}}, \bibinfo {author} {\bibfnamefont {R.}~\bibnamefont {Samajdar}}, \bibinfo {author} {\bibfnamefont {A.}~\bibnamefont {Omran}}, \bibinfo {author} {\bibfnamefont {S.}~\bibnamefont {Sachdev}}, \bibinfo {author} {\bibfnamefont {A.}~\bibnamefont {Vishwanath}}, \bibinfo {author} {\bibfnamefont {M.}~\bibnamefont {Greiner}}, \bibinfo {author} {\bibfnamefont {V.}~\bibnamefont {Vuleti{\'c}}},\ and\ \bibinfo {author} {\bibfnamefont
  {M.~D.}\ \bibnamefont {Lukin}},\ }\bibfield  {title} {\bibinfo {title} {Probing topological spin liquids on a programmable quantum simulator},\ }\href {https://doi.org/10.1126/science.abi8794} {\bibfield  {journal} {\bibinfo  {journal} {Science}\ }\textbf {\bibinfo {volume} {374}},\ \bibinfo {pages} {1242} (\bibinfo {year} {2021})}\BibitemShut {NoStop}%
\bibitem [{\citenamefont {Viebahn}\ \emph {et~al.}(2019)\citenamefont {Viebahn}, \citenamefont {Sbroscia}, \citenamefont {Carter}, \citenamefont {Yu},\ and\ \citenamefont {Schneider}}]{viebahnMatterWaveDiffractionQuasicrystalline2019}%
  \BibitemOpen
  \bibfield  {author} {\bibinfo {author} {\bibfnamefont {K.}~\bibnamefont {Viebahn}}, \bibinfo {author} {\bibfnamefont {M.}~\bibnamefont {Sbroscia}}, \bibinfo {author} {\bibfnamefont {E.}~\bibnamefont {Carter}}, \bibinfo {author} {\bibfnamefont {J.-C.}\ \bibnamefont {Yu}},\ and\ \bibinfo {author} {\bibfnamefont {U.}~\bibnamefont {Schneider}},\ }\bibfield  {title} {\bibinfo {title} {Matter-{{Wave Diffraction}} from a {{Quasicrystalline Optical Lattice}}},\ }\href {https://doi.org/10.1103/PhysRevLett.122.110404} {\bibfield  {journal} {\bibinfo  {journal} {Phys. Rev. Lett.}\ }\textbf {\bibinfo {volume} {122}},\ \bibinfo {pages} {110404} (\bibinfo {year} {2019})}\BibitemShut {NoStop}%
\bibitem [{\citenamefont {Zhang}\ \emph {et~al.}(2017)\citenamefont {Zhang}, \citenamefont {Pagano}, \citenamefont {Hess}, \citenamefont {Kyprianidis}, \citenamefont {Becker}, \citenamefont {Kaplan}, \citenamefont {Gorshkov}, \citenamefont {Gong},\ and\ \citenamefont {Monroe}}]{Zhang2017b}%
  \BibitemOpen
  \bibfield  {author} {\bibinfo {author} {\bibfnamefont {J.}~\bibnamefont {Zhang}}, \bibinfo {author} {\bibfnamefont {G.}~\bibnamefont {Pagano}}, \bibinfo {author} {\bibfnamefont {P.~W.}\ \bibnamefont {Hess}}, \bibinfo {author} {\bibfnamefont {A.}~\bibnamefont {Kyprianidis}}, \bibinfo {author} {\bibfnamefont {P.}~\bibnamefont {Becker}}, \bibinfo {author} {\bibfnamefont {H.}~\bibnamefont {Kaplan}}, \bibinfo {author} {\bibfnamefont {A.~V.}\ \bibnamefont {Gorshkov}}, \bibinfo {author} {\bibfnamefont {Z.-X.}\ \bibnamefont {Gong}},\ and\ \bibinfo {author} {\bibfnamefont {C.}~\bibnamefont {Monroe}},\ }\bibfield  {title} {\bibinfo {title} {{Observation of a many-body dynamical phase transition with a 53-qubit quantum simulator}},\ }\href {https://doi.org/10.1038/nature24654} {\bibfield  {journal} {\bibinfo  {journal} {Nature}\ }\textbf {\bibinfo {volume} {551}},\ \bibinfo {pages} {601} (\bibinfo {year} {2017})}\BibitemShut {NoStop}%
\bibitem [{\citenamefont {Bluvstein}\ \emph {et~al.}(2021)\citenamefont {Bluvstein}, \citenamefont {Omran}, \citenamefont {Levine}, \citenamefont {Keesling}, \citenamefont {Semeghini}, \citenamefont {Ebadi}, \citenamefont {Wang}, \citenamefont {Michailidis}, \citenamefont {Maskara}, \citenamefont {Ho}, \citenamefont {Choi}, \citenamefont {Serbyn}, \citenamefont {Greiner}, \citenamefont {Vuleti{\'c}},\ and\ \citenamefont {Lukin}}]{bluvsteinControllingQuantumManybody2021}%
  \BibitemOpen
  \bibfield  {author} {\bibinfo {author} {\bibfnamefont {D.}~\bibnamefont {Bluvstein}}, \bibinfo {author} {\bibfnamefont {A.}~\bibnamefont {Omran}}, \bibinfo {author} {\bibfnamefont {H.}~\bibnamefont {Levine}}, \bibinfo {author} {\bibfnamefont {A.}~\bibnamefont {Keesling}}, \bibinfo {author} {\bibfnamefont {G.}~\bibnamefont {Semeghini}}, \bibinfo {author} {\bibfnamefont {S.}~\bibnamefont {Ebadi}}, \bibinfo {author} {\bibfnamefont {T.~T.}\ \bibnamefont {Wang}}, \bibinfo {author} {\bibfnamefont {A.~A.}\ \bibnamefont {Michailidis}}, \bibinfo {author} {\bibfnamefont {N.}~\bibnamefont {Maskara}}, \bibinfo {author} {\bibfnamefont {W.~W.}\ \bibnamefont {Ho}}, \bibinfo {author} {\bibfnamefont {S.}~\bibnamefont {Choi}}, \bibinfo {author} {\bibfnamefont {M.}~\bibnamefont {Serbyn}}, \bibinfo {author} {\bibfnamefont {M.}~\bibnamefont {Greiner}}, \bibinfo {author} {\bibfnamefont {V.}~\bibnamefont {Vuleti{\'c}}},\ and\ \bibinfo {author} {\bibfnamefont {M.~D.}\ \bibnamefont {Lukin}},\ }\bibfield  {title} {\bibinfo {title}
  {Controlling quantum many-body dynamics in driven {{Rydberg}} atom arrays},\ }\href {https://doi.org/10.1126/science.abg2530} {\bibfield  {journal} {\bibinfo  {journal} {Science}\ }\textbf {\bibinfo {volume} {371}},\ \bibinfo {pages} {1355} (\bibinfo {year} {2021})}\BibitemShut {NoStop}%
\bibitem [{\citenamefont {Choi}\ \emph {et~al.}(2017)\citenamefont {Choi}, \citenamefont {Choi}, \citenamefont {Landig}, \citenamefont {Kucsko}, \citenamefont {Zhou}, \citenamefont {Isoya}, \citenamefont {Jelezko}, \citenamefont {Onoda}, \citenamefont {Sumiya}, \citenamefont {Khemani}, \citenamefont {Von~Keyserlingk}, \citenamefont {Yao}, \citenamefont {Demler},\ and\ \citenamefont {Lukin}}]{choiObservationDiscreteTimecrystalline2017}%
  \BibitemOpen
  \bibfield  {author} {\bibinfo {author} {\bibfnamefont {S.}~\bibnamefont {Choi}}, \bibinfo {author} {\bibfnamefont {J.}~\bibnamefont {Choi}}, \bibinfo {author} {\bibfnamefont {R.}~\bibnamefont {Landig}}, \bibinfo {author} {\bibfnamefont {G.}~\bibnamefont {Kucsko}}, \bibinfo {author} {\bibfnamefont {H.}~\bibnamefont {Zhou}}, \bibinfo {author} {\bibfnamefont {J.}~\bibnamefont {Isoya}}, \bibinfo {author} {\bibfnamefont {F.}~\bibnamefont {Jelezko}}, \bibinfo {author} {\bibfnamefont {S.}~\bibnamefont {Onoda}}, \bibinfo {author} {\bibfnamefont {H.}~\bibnamefont {Sumiya}}, \bibinfo {author} {\bibfnamefont {V.}~\bibnamefont {Khemani}}, \bibinfo {author} {\bibfnamefont {C.}~\bibnamefont {Von~Keyserlingk}}, \bibinfo {author} {\bibfnamefont {N.~Y.}\ \bibnamefont {Yao}}, \bibinfo {author} {\bibfnamefont {E.}~\bibnamefont {Demler}},\ and\ \bibinfo {author} {\bibfnamefont {M.~D.}\ \bibnamefont {Lukin}},\ }\bibfield  {title} {\bibinfo {title} {Observation of discrete time-crystalline order in a disordered dipolar many-body
  system},\ }\href {https://doi.org/10.1038/nature21426} {\bibfield  {journal} {\bibinfo  {journal} {Nature}\ }\textbf {\bibinfo {volume} {543}},\ \bibinfo {pages} {221} (\bibinfo {year} {2017})}\BibitemShut {NoStop}%
\bibitem [{\citenamefont {Randall}\ \emph {et~al.}(2021)\citenamefont {Randall}, \citenamefont {Bradley}, \citenamefont {Van Der~Gronden}, \citenamefont {Galicia}, \citenamefont {Abobeih}, \citenamefont {Markham}, \citenamefont {Twitchen}, \citenamefont {Machado}, \citenamefont {Yao},\ and\ \citenamefont {Taminiau}}]{randallManybodyLocalizedDiscrete2021}%
  \BibitemOpen
  \bibfield  {author} {\bibinfo {author} {\bibfnamefont {J.}~\bibnamefont {Randall}}, \bibinfo {author} {\bibfnamefont {C.~E.}\ \bibnamefont {Bradley}}, \bibinfo {author} {\bibfnamefont {F.~V.}\ \bibnamefont {Van Der~Gronden}}, \bibinfo {author} {\bibfnamefont {A.}~\bibnamefont {Galicia}}, \bibinfo {author} {\bibfnamefont {M.~H.}\ \bibnamefont {Abobeih}}, \bibinfo {author} {\bibfnamefont {M.}~\bibnamefont {Markham}}, \bibinfo {author} {\bibfnamefont {D.~J.}\ \bibnamefont {Twitchen}}, \bibinfo {author} {\bibfnamefont {F.}~\bibnamefont {Machado}}, \bibinfo {author} {\bibfnamefont {N.~Y.}\ \bibnamefont {Yao}},\ and\ \bibinfo {author} {\bibfnamefont {T.~H.}\ \bibnamefont {Taminiau}},\ }\bibfield  {title} {\bibinfo {title} {Many-body--localized discrete time crystal with a programmable spin-based quantum simulator},\ }\href {https://doi.org/10.1126/science.abk0603} {\bibfield  {journal} {\bibinfo  {journal} {Science}\ }\textbf {\bibinfo {volume} {374}},\ \bibinfo {pages} {1474} (\bibinfo {year}
  {2021})}\BibitemShut {NoStop}%
\bibitem [{\citenamefont {Urbaszek}\ \emph {et~al.}(2013)\citenamefont {Urbaszek}, \citenamefont {Marie}, \citenamefont {Amand}, \citenamefont {Krebs}, \citenamefont {Voisin}, \citenamefont {Maletinsky}, \citenamefont {H{\"o}gele},\ and\ \citenamefont {Imamoglu}}]{urbaszekNuclearSpinPhysics2013}%
  \BibitemOpen
  \bibfield  {author} {\bibinfo {author} {\bibfnamefont {B.}~\bibnamefont {Urbaszek}}, \bibinfo {author} {\bibfnamefont {X.}~\bibnamefont {Marie}}, \bibinfo {author} {\bibfnamefont {T.}~\bibnamefont {Amand}}, \bibinfo {author} {\bibfnamefont {O.}~\bibnamefont {Krebs}}, \bibinfo {author} {\bibfnamefont {P.}~\bibnamefont {Voisin}}, \bibinfo {author} {\bibfnamefont {P.}~\bibnamefont {Maletinsky}}, \bibinfo {author} {\bibfnamefont {A.}~\bibnamefont {H{\"o}gele}},\ and\ \bibinfo {author} {\bibfnamefont {A.}~\bibnamefont {Imamoglu}},\ }\bibfield  {title} {\bibinfo {title} {Nuclear spin physics in quantum dots: {{An}} optical investigation},\ }\href {https://doi.org/10.1103/RevModPhys.85.79} {\bibfield  {journal} {\bibinfo  {journal} {Rev. Mod. Phys.}\ }\textbf {\bibinfo {volume} {85}},\ \bibinfo {pages} {79} (\bibinfo {year} {2013})}\BibitemShut {NoStop}%
\bibitem [{\citenamefont {Taylor}\ \emph {et~al.}(2003{\natexlab{b}})\citenamefont {Taylor}, \citenamefont {Marcus},\ and\ \citenamefont {Lukin}}]{taylorLongLivedMemoryMesoscopic2003}%
  \BibitemOpen
  \bibfield  {author} {\bibinfo {author} {\bibfnamefont {J.~M.}\ \bibnamefont {Taylor}}, \bibinfo {author} {\bibfnamefont {C.~M.}\ \bibnamefont {Marcus}},\ and\ \bibinfo {author} {\bibfnamefont {M.~D.}\ \bibnamefont {Lukin}},\ }\bibfield  {title} {\bibinfo {title} {Long-{{Lived Memory}} for {{Mesoscopic Quantum Bits}}},\ }\href@noop {} {\bibfield  {journal} {\bibinfo  {journal} {Physical Review Letters}\ }\textbf {\bibinfo {volume} {90}} (\bibinfo {year} {2003}{\natexlab{b}})}\BibitemShut {NoStop}%
\bibitem [{\citenamefont {Ruskuc}\ \emph {et~al.}(2022)\citenamefont {Ruskuc}, \citenamefont {Wu}, \citenamefont {Rochman}, \citenamefont {Choi},\ and\ \citenamefont {Faraon}}]{ruskucNuclearSpinwaveQuantum2022}%
  \BibitemOpen
  \bibfield  {author} {\bibinfo {author} {\bibfnamefont {A.}~\bibnamefont {Ruskuc}}, \bibinfo {author} {\bibfnamefont {C.-J.}\ \bibnamefont {Wu}}, \bibinfo {author} {\bibfnamefont {J.}~\bibnamefont {Rochman}}, \bibinfo {author} {\bibfnamefont {J.}~\bibnamefont {Choi}},\ and\ \bibinfo {author} {\bibfnamefont {A.}~\bibnamefont {Faraon}},\ }\bibfield  {title} {\bibinfo {title} {Nuclear spin-wave quantum register for a solid-state qubit},\ }\href {https://doi.org/10.1038/s41586-021-04293-6} {\bibfield  {journal} {\bibinfo  {journal} {Nature}\ }\textbf {\bibinfo {volume} {602}},\ \bibinfo {pages} {408} (\bibinfo {year} {2022})}\BibitemShut {NoStop}%
\bibitem [{\citenamefont {Taminiau}\ \emph {et~al.}(2014)\citenamefont {Taminiau}, \citenamefont {Cramer}, \citenamefont {Van Der~Sar}, \citenamefont {Dobrovitski},\ and\ \citenamefont {Hanson}}]{taminiauUniversalControlError2014}%
  \BibitemOpen
  \bibfield  {author} {\bibinfo {author} {\bibfnamefont {T.~H.}\ \bibnamefont {Taminiau}}, \bibinfo {author} {\bibfnamefont {J.}~\bibnamefont {Cramer}}, \bibinfo {author} {\bibfnamefont {T.}~\bibnamefont {Van Der~Sar}}, \bibinfo {author} {\bibfnamefont {V.~V.}\ \bibnamefont {Dobrovitski}},\ and\ \bibinfo {author} {\bibfnamefont {R.}~\bibnamefont {Hanson}},\ }\bibfield  {title} {\bibinfo {title} {Universal control and error correction in multi-qubit spin registers in diamond},\ }\href {https://doi.org/10.1038/nnano.2014.2} {\bibfield  {journal} {\bibinfo  {journal} {Nature Nanotech}\ }\textbf {\bibinfo {volume} {9}},\ \bibinfo {pages} {171} (\bibinfo {year} {2014})}\BibitemShut {NoStop}%
\bibitem [{\citenamefont {Denning}\ \emph {et~al.}(2019)\citenamefont {Denning}, \citenamefont {Gangloff}, \citenamefont {Atat{\"u}re}, \citenamefont {M{\o}rk},\ and\ \citenamefont {Le~Gall}}]{denningCollectiveQuantumMemory2019}%
  \BibitemOpen
  \bibfield  {author} {\bibinfo {author} {\bibfnamefont {E.~V.}\ \bibnamefont {Denning}}, \bibinfo {author} {\bibfnamefont {D.~A.}\ \bibnamefont {Gangloff}}, \bibinfo {author} {\bibfnamefont {M.}~\bibnamefont {Atat{\"u}re}}, \bibinfo {author} {\bibfnamefont {J.}~\bibnamefont {M{\o}rk}},\ and\ \bibinfo {author} {\bibfnamefont {C.}~\bibnamefont {Le~Gall}},\ }\bibfield  {title} {\bibinfo {title} {Collective {{Quantum Memory Activated}} by a {{Driven Central Spin}}},\ }\href {https://doi.org/10.1103/PhysRevLett.123.140502} {\bibfield  {journal} {\bibinfo  {journal} {Phys. Rev. Lett.}\ }\textbf {\bibinfo {volume} {123}},\ \bibinfo {pages} {140502} (\bibinfo {year} {2019})}\BibitemShut {NoStop}%
\bibitem [{\citenamefont {Zaporski}\ \emph {et~al.}(2023{\natexlab{a}})\citenamefont {Zaporski}, \citenamefont {De~Wit}, \citenamefont {Isogawa}, \citenamefont {Hayhurst~Appel}, \citenamefont {Le~Gall}, \citenamefont {Atat{\"u}re},\ and\ \citenamefont {Gangloff}}]{zaporskiManyBodySingletPrepared2023}%
  \BibitemOpen
  \bibfield  {author} {\bibinfo {author} {\bibfnamefont {L.}~\bibnamefont {Zaporski}}, \bibinfo {author} {\bibfnamefont {S.~R.}\ \bibnamefont {De~Wit}}, \bibinfo {author} {\bibfnamefont {T.}~\bibnamefont {Isogawa}}, \bibinfo {author} {\bibfnamefont {M.}~\bibnamefont {Hayhurst~Appel}}, \bibinfo {author} {\bibfnamefont {C.}~\bibnamefont {Le~Gall}}, \bibinfo {author} {\bibfnamefont {M.}~\bibnamefont {Atat{\"u}re}},\ and\ \bibinfo {author} {\bibfnamefont {D.~A.}\ \bibnamefont {Gangloff}},\ }\bibfield  {title} {\bibinfo {title} {Many-{{Body Singlet Prepared}} by a {{Central-Spin Qubit}}},\ }\href {https://doi.org/10.1103/PRXQuantum.4.040343} {\bibfield  {journal} {\bibinfo  {journal} {PRX Quantum}\ }\textbf {\bibinfo {volume} {4}},\ \bibinfo {pages} {040343} (\bibinfo {year} {2023}{\natexlab{a}})}\BibitemShut {NoStop}%
\bibitem [{\citenamefont {Kessler}\ \emph {et~al.}(2012)\citenamefont {Kessler}, \citenamefont {Giedke}, \citenamefont {Imamoglu}, \citenamefont {Yelin}, \citenamefont {Lukin},\ and\ \citenamefont {Cirac}}]{Kessler2012a}%
  \BibitemOpen
  \bibfield  {author} {\bibinfo {author} {\bibfnamefont {E.~M.}\ \bibnamefont {Kessler}}, \bibinfo {author} {\bibfnamefont {G.}~\bibnamefont {Giedke}}, \bibinfo {author} {\bibfnamefont {A.}~\bibnamefont {Imamoglu}}, \bibinfo {author} {\bibfnamefont {S.~F.}\ \bibnamefont {Yelin}}, \bibinfo {author} {\bibfnamefont {M.~D.}\ \bibnamefont {Lukin}},\ and\ \bibinfo {author} {\bibfnamefont {J.~I.}\ \bibnamefont {Cirac}},\ }\bibfield  {title} {\bibinfo {title} {{Dissipative phase transition in a central spin system}},\ }\href {https://doi.org/10.1103/PhysRevA.86.012116} {\bibfield  {journal} {\bibinfo  {journal} {Phys. Rev. A}\ }\textbf {\bibinfo {volume} {86}},\ \bibinfo {pages} {012116} (\bibinfo {year} {2012})}\BibitemShut {NoStop}%
\bibitem [{\citenamefont {Goldman}\ \emph {et~al.}(2023)\citenamefont {Goldman}, \citenamefont {Diessel}, \citenamefont {Barbiero}, \citenamefont {Pr{\"{u}}fer}, \citenamefont {{Di Liberto}},\ and\ \citenamefont {{Peralta Gavensky}}}]{Goldman2023}%
  \BibitemOpen
  \bibfield  {author} {\bibinfo {author} {\bibfnamefont {N.}~\bibnamefont {Goldman}}, \bibinfo {author} {\bibfnamefont {O.}~\bibnamefont {Diessel}}, \bibinfo {author} {\bibfnamefont {L.}~\bibnamefont {Barbiero}}, \bibinfo {author} {\bibfnamefont {M.}~\bibnamefont {Pr{\"{u}}fer}}, \bibinfo {author} {\bibfnamefont {M.}~\bibnamefont {{Di Liberto}}},\ and\ \bibinfo {author} {\bibfnamefont {L.}~\bibnamefont {{Peralta Gavensky}}},\ }\bibfield  {title} {\bibinfo {title} {{Floquet-Engineered Nonlinearities and Controllable Pair-Hopping Processes: From Optical Kerr Cavities to Correlated Quantum Matter}},\ }\href {https://doi.org/10.1103/PRXQuantum.4.040327} {\bibfield  {journal} {\bibinfo  {journal} {PRX Quantum}\ }\textbf {\bibinfo {volume} {4}},\ \bibinfo {pages} {040327} (\bibinfo {year} {2023})}\BibitemShut {NoStop}%
\bibitem [{\citenamefont {Wang}\ \emph {et~al.}(2019)\citenamefont {Wang}, \citenamefont {He}, \citenamefont {Chung}, \citenamefont {Hu}, \citenamefont {Yu}, \citenamefont {Chen}, \citenamefont {Ding}, \citenamefont {Chen}, \citenamefont {Qin}, \citenamefont {Yang}, \citenamefont {Liu}, \citenamefont {Duan}, \citenamefont {Li}, \citenamefont {Gerhardt}, \citenamefont {Winkler}, \citenamefont {Jurkat}, \citenamefont {Wang}, \citenamefont {Gregersen}, \citenamefont {Huo}, \citenamefont {Dai}, \citenamefont {Yu}, \citenamefont {H{\"o}fling}, \citenamefont {Lu},\ and\ \citenamefont {Pan}}]{wangOptimalSinglephotonSources2019}%
  \BibitemOpen
  \bibfield  {author} {\bibinfo {author} {\bibfnamefont {H.}~\bibnamefont {Wang}}, \bibinfo {author} {\bibfnamefont {Y.-M.}\ \bibnamefont {He}}, \bibinfo {author} {\bibfnamefont {T.-H.}\ \bibnamefont {Chung}}, \bibinfo {author} {\bibfnamefont {H.}~\bibnamefont {Hu}}, \bibinfo {author} {\bibfnamefont {Y.}~\bibnamefont {Yu}}, \bibinfo {author} {\bibfnamefont {S.}~\bibnamefont {Chen}}, \bibinfo {author} {\bibfnamefont {X.}~\bibnamefont {Ding}}, \bibinfo {author} {\bibfnamefont {M.-C.}\ \bibnamefont {Chen}}, \bibinfo {author} {\bibfnamefont {J.}~\bibnamefont {Qin}}, \bibinfo {author} {\bibfnamefont {X.}~\bibnamefont {Yang}}, \bibinfo {author} {\bibfnamefont {R.-Z.}\ \bibnamefont {Liu}}, \bibinfo {author} {\bibfnamefont {Z.-C.}\ \bibnamefont {Duan}}, \bibinfo {author} {\bibfnamefont {J.-P.}\ \bibnamefont {Li}}, \bibinfo {author} {\bibfnamefont {S.}~\bibnamefont {Gerhardt}}, \bibinfo {author} {\bibfnamefont {K.}~\bibnamefont {Winkler}}, \bibinfo {author} {\bibfnamefont {J.}~\bibnamefont {Jurkat}}, \bibinfo {author}
  {\bibfnamefont {L.-J.}\ \bibnamefont {Wang}}, \bibinfo {author} {\bibfnamefont {N.}~\bibnamefont {Gregersen}}, \bibinfo {author} {\bibfnamefont {Y.-H.}\ \bibnamefont {Huo}}, \bibinfo {author} {\bibfnamefont {Q.}~\bibnamefont {Dai}}, \bibinfo {author} {\bibfnamefont {S.}~\bibnamefont {Yu}}, \bibinfo {author} {\bibfnamefont {S.}~\bibnamefont {H{\"o}fling}}, \bibinfo {author} {\bibfnamefont {C.-Y.}\ \bibnamefont {Lu}},\ and\ \bibinfo {author} {\bibfnamefont {J.-W.}\ \bibnamefont {Pan}},\ }\bibfield  {title} {\bibinfo {title} {Towards optimal single-photon sources from polarized microcavities},\ }\href {https://doi.org/10.1038/s41566-019-0494-3} {\bibfield  {journal} {\bibinfo  {journal} {Nat. Photonics}\ }\textbf {\bibinfo {volume} {13}},\ \bibinfo {pages} {770} (\bibinfo {year} {2019})}\BibitemShut {NoStop}%
\bibitem [{\citenamefont {Liu}\ \emph {et~al.}(2019)\citenamefont {Liu}, \citenamefont {Su}, \citenamefont {Wei}, \citenamefont {Yao}, \citenamefont {Silva}, \citenamefont {Yu}, \citenamefont {{Iles-Smith}}, \citenamefont {Srinivasan}, \citenamefont {Rastelli}, \citenamefont {Li},\ and\ \citenamefont {Wang}}]{liuSolidstateSourceStrongly2019}%
  \BibitemOpen
  \bibfield  {author} {\bibinfo {author} {\bibfnamefont {J.}~\bibnamefont {Liu}}, \bibinfo {author} {\bibfnamefont {R.}~\bibnamefont {Su}}, \bibinfo {author} {\bibfnamefont {Y.}~\bibnamefont {Wei}}, \bibinfo {author} {\bibfnamefont {B.}~\bibnamefont {Yao}}, \bibinfo {author} {\bibfnamefont {S.~F. C.~D.}\ \bibnamefont {Silva}}, \bibinfo {author} {\bibfnamefont {Y.}~\bibnamefont {Yu}}, \bibinfo {author} {\bibfnamefont {J.}~\bibnamefont {{Iles-Smith}}}, \bibinfo {author} {\bibfnamefont {K.}~\bibnamefont {Srinivasan}}, \bibinfo {author} {\bibfnamefont {A.}~\bibnamefont {Rastelli}}, \bibinfo {author} {\bibfnamefont {J.}~\bibnamefont {Li}},\ and\ \bibinfo {author} {\bibfnamefont {X.}~\bibnamefont {Wang}},\ }\bibfield  {title} {\bibinfo {title} {A solid-state source of strongly entangled photon pairs with high brightness and indistinguishability},\ }\href {https://doi.org/10.1038/s41565-019-0435-9} {\bibfield  {journal} {\bibinfo  {journal} {Nat. Nanotechnol.}\ }\textbf {\bibinfo {volume} {14}},\ \bibinfo {pages}
  {586} (\bibinfo {year} {2019})}\BibitemShut {NoStop}%
\bibitem [{\citenamefont {Tomm}\ \emph {et~al.}(2021)\citenamefont {Tomm}, \citenamefont {Javadi}, \citenamefont {Antoniadis}, \citenamefont {Najer}, \citenamefont {L{\"o}bl}, \citenamefont {Korsch}, \citenamefont {Schott}, \citenamefont {Valentin}, \citenamefont {Wieck}, \citenamefont {Ludwig},\ and\ \citenamefont {Warburton}}]{tommBrightFastSource2021}%
  \BibitemOpen
  \bibfield  {author} {\bibinfo {author} {\bibfnamefont {N.}~\bibnamefont {Tomm}}, \bibinfo {author} {\bibfnamefont {A.}~\bibnamefont {Javadi}}, \bibinfo {author} {\bibfnamefont {N.~O.}\ \bibnamefont {Antoniadis}}, \bibinfo {author} {\bibfnamefont {D.}~\bibnamefont {Najer}}, \bibinfo {author} {\bibfnamefont {M.~C.}\ \bibnamefont {L{\"o}bl}}, \bibinfo {author} {\bibfnamefont {A.~R.}\ \bibnamefont {Korsch}}, \bibinfo {author} {\bibfnamefont {R.}~\bibnamefont {Schott}}, \bibinfo {author} {\bibfnamefont {S.~R.}\ \bibnamefont {Valentin}}, \bibinfo {author} {\bibfnamefont {A.~D.}\ \bibnamefont {Wieck}}, \bibinfo {author} {\bibfnamefont {A.}~\bibnamefont {Ludwig}},\ and\ \bibinfo {author} {\bibfnamefont {R.~J.}\ \bibnamefont {Warburton}},\ }\bibfield  {title} {\bibinfo {title} {A bright and fast source of coherent single photons},\ }\href {https://doi.org/10.1038/s41565-020-00831-x} {\bibfield  {journal} {\bibinfo  {journal} {Nat. Nanotechnol.}\ }\textbf {\bibinfo {volume} {16}},\ \bibinfo {pages} {399} (\bibinfo
  {year} {2021})}\BibitemShut {NoStop}%
\bibitem [{\citenamefont {Appel}\ \emph {et~al.}(2021)\citenamefont {Appel}, \citenamefont {Tiranov}, \citenamefont {Javadi}, \citenamefont {L{\"o}bl}, \citenamefont {Wang}, \citenamefont {Scholz}, \citenamefont {Wieck}, \citenamefont {Ludwig}, \citenamefont {Warburton},\ and\ \citenamefont {Lodahl}}]{appelCoherentSpinPhotonInterface2021}%
  \BibitemOpen
  \bibfield  {author} {\bibinfo {author} {\bibfnamefont {M.~H.}\ \bibnamefont {Appel}}, \bibinfo {author} {\bibfnamefont {A.}~\bibnamefont {Tiranov}}, \bibinfo {author} {\bibfnamefont {A.}~\bibnamefont {Javadi}}, \bibinfo {author} {\bibfnamefont {M.~C.}\ \bibnamefont {L{\"o}bl}}, \bibinfo {author} {\bibfnamefont {Y.}~\bibnamefont {Wang}}, \bibinfo {author} {\bibfnamefont {S.}~\bibnamefont {Scholz}}, \bibinfo {author} {\bibfnamefont {A.~D.}\ \bibnamefont {Wieck}}, \bibinfo {author} {\bibfnamefont {A.}~\bibnamefont {Ludwig}}, \bibinfo {author} {\bibfnamefont {R.~J.}\ \bibnamefont {Warburton}},\ and\ \bibinfo {author} {\bibfnamefont {P.}~\bibnamefont {Lodahl}},\ }\bibfield  {title} {\bibinfo {title} {Coherent {{Spin-Photon Interface}} with {{Waveguide Induced Cycling Transitions}}},\ }\href {https://doi.org/10.1103/PhysRevLett.126.013602} {\bibfield  {journal} {\bibinfo  {journal} {Phys. Rev. Lett.}\ }\textbf {\bibinfo {volume} {126}},\ \bibinfo {pages} {013602} (\bibinfo {year} {2021})}\BibitemShut {NoStop}%
\bibitem [{\citenamefont {Thomas}\ \emph {et~al.}(2021)\citenamefont {Thomas}, \citenamefont {Billard}, \citenamefont {Coste}, \citenamefont {Wein}, \citenamefont {{Priya}}, \citenamefont {Ollivier}, \citenamefont {Krebs}, \citenamefont {Taza\"{\i}rt}, \citenamefont {Harouri}, \citenamefont {Lemaitre}, \citenamefont {Sagnes}, \citenamefont {Anton}, \citenamefont {Lanco}, \citenamefont {Somaschi}, \citenamefont {Loredo},\ and\ \citenamefont {Senellart}}]{thomasBrightPolarizedSinglePhoton2021}%
  \BibitemOpen
  \bibfield  {author} {\bibinfo {author} {\bibfnamefont {S.~E.}\ \bibnamefont {Thomas}}, \bibinfo {author} {\bibfnamefont {M.}~\bibnamefont {Billard}}, \bibinfo {author} {\bibfnamefont {N.}~\bibnamefont {Coste}}, \bibinfo {author} {\bibfnamefont {S.~C.}\ \bibnamefont {Wein}}, \bibinfo {author} {\bibnamefont {{Priya}}}, \bibinfo {author} {\bibfnamefont {H.}~\bibnamefont {Ollivier}}, \bibinfo {author} {\bibfnamefont {O.}~\bibnamefont {Krebs}}, \bibinfo {author} {\bibfnamefont {L.}~\bibnamefont {Taza\"{\i}rt}}, \bibinfo {author} {\bibfnamefont {A.}~\bibnamefont {Harouri}}, \bibinfo {author} {\bibfnamefont {A.}~\bibnamefont {Lemaitre}}, \bibinfo {author} {\bibfnamefont {I.}~\bibnamefont {Sagnes}}, \bibinfo {author} {\bibfnamefont {C.}~\bibnamefont {Anton}}, \bibinfo {author} {\bibfnamefont {L.}~\bibnamefont {Lanco}}, \bibinfo {author} {\bibfnamefont {N.}~\bibnamefont {Somaschi}}, \bibinfo {author} {\bibfnamefont {J.~C.}\ \bibnamefont {Loredo}},\ and\ \bibinfo {author} {\bibfnamefont {P.}~\bibnamefont
  {Senellart}},\ }\bibfield  {title} {\bibinfo {title} {Bright {{Polarized Single-Photon Source Based}} on a {{Linear Dipole}}},\ }\href {https://doi.org/10.1103/PhysRevLett.126.233601} {\bibfield  {journal} {\bibinfo  {journal} {Phys. Rev. Lett.}\ }\textbf {\bibinfo {volume} {126}},\ \bibinfo {pages} {233601} (\bibinfo {year} {2021})}\BibitemShut {NoStop}%
\bibitem [{\citenamefont {Berezovsky}\ \emph {et~al.}(2008)\citenamefont {Berezovsky}, \citenamefont {Mikkelsen}, \citenamefont {Stoltz}, \citenamefont {Coldren},\ and\ \citenamefont {Awschalom}}]{berezovskyPicosecondCoherentOptical2008}%
  \BibitemOpen
  \bibfield  {author} {\bibinfo {author} {\bibfnamefont {J.}~\bibnamefont {Berezovsky}}, \bibinfo {author} {\bibfnamefont {M.~H.}\ \bibnamefont {Mikkelsen}}, \bibinfo {author} {\bibfnamefont {N.~G.}\ \bibnamefont {Stoltz}}, \bibinfo {author} {\bibfnamefont {L.~A.}\ \bibnamefont {Coldren}},\ and\ \bibinfo {author} {\bibfnamefont {D.~D.}\ \bibnamefont {Awschalom}},\ }\bibfield  {title} {\bibinfo {title} {Picosecond {{Coherent Optical Manipulation}} of a {{Single Electron Spin}} in a {{Quantum Dot}}},\ }\href {https://doi.org/10.1126/science.1154798} {\bibfield  {journal} {\bibinfo  {journal} {Science}\ }\textbf {\bibinfo {volume} {320}},\ \bibinfo {pages} {349} (\bibinfo {year} {2008})}\BibitemShut {NoStop}%
\bibitem [{\citenamefont {Press}\ \emph {et~al.}(2008)\citenamefont {Press}, \citenamefont {Ladd}, \citenamefont {Zhang},\ and\ \citenamefont {Yamamoto}}]{pressCompleteQuantumControl2008}%
  \BibitemOpen
  \bibfield  {author} {\bibinfo {author} {\bibfnamefont {D.}~\bibnamefont {Press}}, \bibinfo {author} {\bibfnamefont {T.~D.}\ \bibnamefont {Ladd}}, \bibinfo {author} {\bibfnamefont {B.}~\bibnamefont {Zhang}},\ and\ \bibinfo {author} {\bibfnamefont {Y.}~\bibnamefont {Yamamoto}},\ }\bibfield  {title} {\bibinfo {title} {Complete quantum control of a single quantum dot spin using ultrafast optical pulses},\ }\href {https://doi.org/10.1038/nature07530} {\bibfield  {journal} {\bibinfo  {journal} {Nature}\ }\textbf {\bibinfo {volume} {456}},\ \bibinfo {pages} {218} (\bibinfo {year} {2008})}\BibitemShut {NoStop}%
\bibitem [{\citenamefont {Sch{\"o}ll}\ \emph {et~al.}(2019)\citenamefont {Sch{\"o}ll}, \citenamefont {Hanschke}, \citenamefont {Schweickert}, \citenamefont {Zeuner}, \citenamefont {Reindl}, \citenamefont {{Covre da Silva}}, \citenamefont {Lettner}, \citenamefont {Trotta}, \citenamefont {Finley}, \citenamefont {M{\"u}ller}, \citenamefont {Rastelli}, \citenamefont {Zwiller},\ and\ \citenamefont {J{\"o}ns}}]{schollResonanceFluorescenceGaAs2019}%
  \BibitemOpen
  \bibfield  {author} {\bibinfo {author} {\bibfnamefont {E.}~\bibnamefont {Sch{\"o}ll}}, \bibinfo {author} {\bibfnamefont {L.}~\bibnamefont {Hanschke}}, \bibinfo {author} {\bibfnamefont {L.}~\bibnamefont {Schweickert}}, \bibinfo {author} {\bibfnamefont {K.~D.}\ \bibnamefont {Zeuner}}, \bibinfo {author} {\bibfnamefont {M.}~\bibnamefont {Reindl}}, \bibinfo {author} {\bibfnamefont {S.~F.}\ \bibnamefont {{Covre da Silva}}}, \bibinfo {author} {\bibfnamefont {T.}~\bibnamefont {Lettner}}, \bibinfo {author} {\bibfnamefont {R.}~\bibnamefont {Trotta}}, \bibinfo {author} {\bibfnamefont {J.~J.}\ \bibnamefont {Finley}}, \bibinfo {author} {\bibfnamefont {K.}~\bibnamefont {M{\"u}ller}}, \bibinfo {author} {\bibfnamefont {A.}~\bibnamefont {Rastelli}}, \bibinfo {author} {\bibfnamefont {V.}~\bibnamefont {Zwiller}},\ and\ \bibinfo {author} {\bibfnamefont {K.~D.}\ \bibnamefont {J{\"o}ns}},\ }\bibfield  {title} {\bibinfo {title} {Resonance {{Fluorescence}} of {{GaAs Quantum Dots}} with {{Near-Unity Photon Indistinguishability}}},\
  }\href {https://doi.org/10.1021/acs.nanolett.8b05132} {\bibfield  {journal} {\bibinfo  {journal} {Nano Lett.}\ }\textbf {\bibinfo {volume} {19}},\ \bibinfo {pages} {2404} (\bibinfo {year} {2019})}\BibitemShut {NoStop}%
\bibitem [{\citenamefont {Zhai}\ \emph {et~al.}(2020)\citenamefont {Zhai}, \citenamefont {L{\"o}bl}, \citenamefont {Nguyen}, \citenamefont {Ritzmann}, \citenamefont {Javadi}, \citenamefont {Spinnler}, \citenamefont {Wieck}, \citenamefont {Ludwig},\ and\ \citenamefont {Warburton}}]{zhaiLownoiseGaAsQuantum2020}%
  \BibitemOpen
  \bibfield  {author} {\bibinfo {author} {\bibfnamefont {L.}~\bibnamefont {Zhai}}, \bibinfo {author} {\bibfnamefont {M.~C.}\ \bibnamefont {L{\"o}bl}}, \bibinfo {author} {\bibfnamefont {G.~N.}\ \bibnamefont {Nguyen}}, \bibinfo {author} {\bibfnamefont {J.}~\bibnamefont {Ritzmann}}, \bibinfo {author} {\bibfnamefont {A.}~\bibnamefont {Javadi}}, \bibinfo {author} {\bibfnamefont {C.}~\bibnamefont {Spinnler}}, \bibinfo {author} {\bibfnamefont {A.~D.}\ \bibnamefont {Wieck}}, \bibinfo {author} {\bibfnamefont {A.}~\bibnamefont {Ludwig}},\ and\ \bibinfo {author} {\bibfnamefont {R.~J.}\ \bibnamefont {Warburton}},\ }\bibfield  {title} {\bibinfo {title} {Low-noise {{GaAs}} quantum dots for quantum photonics},\ }\href {https://doi.org/10.1038/s41467-020-18625-z} {\bibfield  {journal} {\bibinfo  {journal} {Nat Commun}\ }\textbf {\bibinfo {volume} {11}},\ \bibinfo {pages} {4745} (\bibinfo {year} {2020})}\BibitemShut {NoStop}%
\bibitem [{\citenamefont {Gangloff}\ \emph {et~al.}(2019)\citenamefont {Gangloff}, \citenamefont {{\'E}thier-Majcher}, \citenamefont {Lang}, \citenamefont {Denning}, \citenamefont {Bodey}, \citenamefont {Jackson}, \citenamefont {Clarke}, \citenamefont {Hugues}, \citenamefont {Le~Gall},\ and\ \citenamefont {Atat{\"u}re}}]{gangloffQuantumInterfaceElectron2019}%
  \BibitemOpen
  \bibfield  {author} {\bibinfo {author} {\bibfnamefont {D.~A.}\ \bibnamefont {Gangloff}}, \bibinfo {author} {\bibfnamefont {G.}~\bibnamefont {{\'E}thier-Majcher}}, \bibinfo {author} {\bibfnamefont {C.}~\bibnamefont {Lang}}, \bibinfo {author} {\bibfnamefont {E.~V.}\ \bibnamefont {Denning}}, \bibinfo {author} {\bibfnamefont {J.~H.}\ \bibnamefont {Bodey}}, \bibinfo {author} {\bibfnamefont {D.~M.}\ \bibnamefont {Jackson}}, \bibinfo {author} {\bibfnamefont {E.}~\bibnamefont {Clarke}}, \bibinfo {author} {\bibfnamefont {M.}~\bibnamefont {Hugues}}, \bibinfo {author} {\bibfnamefont {C.}~\bibnamefont {Le~Gall}},\ and\ \bibinfo {author} {\bibfnamefont {M.}~\bibnamefont {Atat{\"u}re}},\ }\bibfield  {title} {\bibinfo {title} {Quantum interface of an electron and a nuclear ensemble},\ }\href {https://doi.org/10.1126/science.aaw2906} {\bibfield  {journal} {\bibinfo  {journal} {Science}\ }\textbf {\bibinfo {volume} {364}},\ \bibinfo {pages} {62} (\bibinfo {year} {2019})}\BibitemShut {NoStop}%
\bibitem [{\citenamefont {Gangloff}\ \emph {et~al.}(2021)\citenamefont {Gangloff}, \citenamefont {Zaporski}, \citenamefont {Bodey}, \citenamefont {Bachorz}, \citenamefont {Jackson}, \citenamefont {{\'E}thier-Majcher}, \citenamefont {Lang}, \citenamefont {Clarke}, \citenamefont {Hugues}, \citenamefont {Le~Gall},\ and\ \citenamefont {Atat{\"u}re}}]{gangloffWitnessingQuantumCorrelations2021}%
  \BibitemOpen
  \bibfield  {author} {\bibinfo {author} {\bibfnamefont {D.~A.}\ \bibnamefont {Gangloff}}, \bibinfo {author} {\bibfnamefont {L.}~\bibnamefont {Zaporski}}, \bibinfo {author} {\bibfnamefont {J.~H.}\ \bibnamefont {Bodey}}, \bibinfo {author} {\bibfnamefont {C.}~\bibnamefont {Bachorz}}, \bibinfo {author} {\bibfnamefont {D.~M.}\ \bibnamefont {Jackson}}, \bibinfo {author} {\bibfnamefont {G.}~\bibnamefont {{\'E}thier-Majcher}}, \bibinfo {author} {\bibfnamefont {C.}~\bibnamefont {Lang}}, \bibinfo {author} {\bibfnamefont {E.}~\bibnamefont {Clarke}}, \bibinfo {author} {\bibfnamefont {M.}~\bibnamefont {Hugues}}, \bibinfo {author} {\bibfnamefont {C.}~\bibnamefont {Le~Gall}},\ and\ \bibinfo {author} {\bibfnamefont {M.}~\bibnamefont {Atat{\"u}re}},\ }\bibfield  {title} {\bibinfo {title} {Witnessing quantum correlations in a nuclear ensemble via an electron spin qubit},\ }\href {https://doi.org/10.1038/s41567-021-01344-7} {\bibfield  {journal} {\bibinfo  {journal} {Nat. Phys.}\ }\textbf {\bibinfo {volume} {17}},\ \bibinfo
  {pages} {1247} (\bibinfo {year} {2021})}\BibitemShut {NoStop}%
\bibitem [{\citenamefont {Jackson}\ \emph {et~al.}(2021)\citenamefont {Jackson}, \citenamefont {Gangloff}, \citenamefont {Bodey}, \citenamefont {Zaporski}, \citenamefont {Bachorz}, \citenamefont {Clarke}, \citenamefont {Hugues}, \citenamefont {Le~Gall},\ and\ \citenamefont {Atat{\"u}re}}]{jacksonQuantumSensingCoherent2021}%
  \BibitemOpen
  \bibfield  {author} {\bibinfo {author} {\bibfnamefont {D.~M.}\ \bibnamefont {Jackson}}, \bibinfo {author} {\bibfnamefont {D.~A.}\ \bibnamefont {Gangloff}}, \bibinfo {author} {\bibfnamefont {J.~H.}\ \bibnamefont {Bodey}}, \bibinfo {author} {\bibfnamefont {L.}~\bibnamefont {Zaporski}}, \bibinfo {author} {\bibfnamefont {C.}~\bibnamefont {Bachorz}}, \bibinfo {author} {\bibfnamefont {E.}~\bibnamefont {Clarke}}, \bibinfo {author} {\bibfnamefont {M.}~\bibnamefont {Hugues}}, \bibinfo {author} {\bibfnamefont {C.}~\bibnamefont {Le~Gall}},\ and\ \bibinfo {author} {\bibfnamefont {M.}~\bibnamefont {Atat{\"u}re}},\ }\bibfield  {title} {\bibinfo {title} {Quantum sensing of a coherent single spin excitation in a nuclear ensemble},\ }\href {https://doi.org/10.1038/s41567-020-01161-4} {\bibfield  {journal} {\bibinfo  {journal} {Nat. Phys.}\ }\textbf {\bibinfo {volume} {17}},\ \bibinfo {pages} {585} (\bibinfo {year} {2021})}\BibitemShut {NoStop}%
\bibitem [{\citenamefont {W{\"u}st}\ \emph {et~al.}(2016)\citenamefont {W{\"u}st}, \citenamefont {Munsch}, \citenamefont {Maier}, \citenamefont {Kuhlmann}, \citenamefont {Ludwig}, \citenamefont {Wieck}, \citenamefont {Loss}, \citenamefont {Poggio},\ and\ \citenamefont {Warburton}}]{wustRoleElectronSpin2016}%
  \BibitemOpen
  \bibfield  {author} {\bibinfo {author} {\bibfnamefont {G.}~\bibnamefont {W{\"u}st}}, \bibinfo {author} {\bibfnamefont {M.}~\bibnamefont {Munsch}}, \bibinfo {author} {\bibfnamefont {F.}~\bibnamefont {Maier}}, \bibinfo {author} {\bibfnamefont {A.~V.}\ \bibnamefont {Kuhlmann}}, \bibinfo {author} {\bibfnamefont {A.}~\bibnamefont {Ludwig}}, \bibinfo {author} {\bibfnamefont {A.~D.}\ \bibnamefont {Wieck}}, \bibinfo {author} {\bibfnamefont {D.}~\bibnamefont {Loss}}, \bibinfo {author} {\bibfnamefont {M.}~\bibnamefont {Poggio}},\ and\ \bibinfo {author} {\bibfnamefont {R.~J.}\ \bibnamefont {Warburton}},\ }\bibfield  {title} {\bibinfo {title} {Role of the electron spin in determining the coherence of the nuclear spins in a quantum dot},\ }\href {https://doi.org/10.1038/nnano.2016.114} {\bibfield  {journal} {\bibinfo  {journal} {Nature Nanotech}\ }\textbf {\bibinfo {volume} {11}},\ \bibinfo {pages} {885} (\bibinfo {year} {2016})}\BibitemShut {NoStop}%
\bibitem [{\citenamefont {Merkulov}\ \emph {et~al.}(2002)\citenamefont {Merkulov}, \citenamefont {Efros},\ and\ \citenamefont {Rosen}}]{merkulovElectronSpinRelaxation2002}%
  \BibitemOpen
  \bibfield  {author} {\bibinfo {author} {\bibfnamefont {I.~A.}\ \bibnamefont {Merkulov}}, \bibinfo {author} {\bibfnamefont {{\relax Al}.~L.}\ \bibnamefont {Efros}},\ and\ \bibinfo {author} {\bibfnamefont {M.}~\bibnamefont {Rosen}},\ }\bibfield  {title} {\bibinfo {title} {Electron spin relaxation by nuclei in semiconductor quantum dots},\ }\href {https://doi.org/10.1103/PhysRevB.65.205309} {\bibfield  {journal} {\bibinfo  {journal} {Phys. Rev. B}\ }\textbf {\bibinfo {volume} {65}},\ \bibinfo {pages} {205309} (\bibinfo {year} {2002})}\BibitemShut {NoStop}%
\bibitem [{\citenamefont {Stockill}\ \emph {et~al.}(2016)\citenamefont {Stockill}, \citenamefont {Le~Gall}, \citenamefont {Matthiesen}, \citenamefont {Huthmacher}, \citenamefont {Clarke}, \citenamefont {Hugues},\ and\ \citenamefont {Atat{\"u}re}}]{stockillQuantumDotSpin2016}%
  \BibitemOpen
  \bibfield  {author} {\bibinfo {author} {\bibfnamefont {R.}~\bibnamefont {Stockill}}, \bibinfo {author} {\bibfnamefont {C.}~\bibnamefont {Le~Gall}}, \bibinfo {author} {\bibfnamefont {C.}~\bibnamefont {Matthiesen}}, \bibinfo {author} {\bibfnamefont {L.}~\bibnamefont {Huthmacher}}, \bibinfo {author} {\bibfnamefont {E.}~\bibnamefont {Clarke}}, \bibinfo {author} {\bibfnamefont {M.}~\bibnamefont {Hugues}},\ and\ \bibinfo {author} {\bibfnamefont {M.}~\bibnamefont {Atat{\"u}re}},\ }\bibfield  {title} {\bibinfo {title} {Quantum dot spin coherence governed by a strained nuclear environment},\ }\href {https://doi.org/10.1038/ncomms12745} {\bibfield  {journal} {\bibinfo  {journal} {Nat Commun}\ }\textbf {\bibinfo {volume} {7}},\ \bibinfo {pages} {12745} (\bibinfo {year} {2016})}\BibitemShut {NoStop}%
\bibitem [{\citenamefont {Jackson}\ \emph {et~al.}(2022)\citenamefont {Jackson}, \citenamefont {Haeusler}, \citenamefont {Zaporski}, \citenamefont {Bodey}, \citenamefont {Shofer}, \citenamefont {Clarke}, \citenamefont {Hugues}, \citenamefont {Atat{\"u}re}, \citenamefont {Le~Gall},\ and\ \citenamefont {Gangloff}}]{jacksonOptimalPurificationSpin2022}%
  \BibitemOpen
  \bibfield  {author} {\bibinfo {author} {\bibfnamefont {D.~M.}\ \bibnamefont {Jackson}}, \bibinfo {author} {\bibfnamefont {U.}~\bibnamefont {Haeusler}}, \bibinfo {author} {\bibfnamefont {L.}~\bibnamefont {Zaporski}}, \bibinfo {author} {\bibfnamefont {J.~H.}\ \bibnamefont {Bodey}}, \bibinfo {author} {\bibfnamefont {N.}~\bibnamefont {Shofer}}, \bibinfo {author} {\bibfnamefont {E.}~\bibnamefont {Clarke}}, \bibinfo {author} {\bibfnamefont {M.}~\bibnamefont {Hugues}}, \bibinfo {author} {\bibfnamefont {M.}~\bibnamefont {Atat{\"u}re}}, \bibinfo {author} {\bibfnamefont {C.}~\bibnamefont {Le~Gall}},\ and\ \bibinfo {author} {\bibfnamefont {D.~A.}\ \bibnamefont {Gangloff}},\ }\bibfield  {title} {\bibinfo {title} {Optimal {{Purification}} of a {{Spin Ensemble}} by {{Quantum-Algorithmic Feedback}}},\ }\href {https://doi.org/10.1103/PhysRevX.12.031014} {\bibfield  {journal} {\bibinfo  {journal} {Phys. Rev. X}\ }\textbf {\bibinfo {volume} {12}},\ \bibinfo {pages} {031014} (\bibinfo {year} {2022})}\BibitemShut {NoStop}%
\bibitem [{\citenamefont {Botzem}\ \emph {et~al.}(2016)\citenamefont {Botzem}, \citenamefont {McNeil}, \citenamefont {Mol}, \citenamefont {Schuh}, \citenamefont {Bougeard},\ and\ \citenamefont {Bluhm}}]{botzemQuadrupolarAnisotropyEffects2016}%
  \BibitemOpen
  \bibfield  {author} {\bibinfo {author} {\bibfnamefont {T.}~\bibnamefont {Botzem}}, \bibinfo {author} {\bibfnamefont {R.~P.~G.}\ \bibnamefont {McNeil}}, \bibinfo {author} {\bibfnamefont {J.-M.}\ \bibnamefont {Mol}}, \bibinfo {author} {\bibfnamefont {D.}~\bibnamefont {Schuh}}, \bibinfo {author} {\bibfnamefont {D.}~\bibnamefont {Bougeard}},\ and\ \bibinfo {author} {\bibfnamefont {H.}~\bibnamefont {Bluhm}},\ }\bibfield  {title} {\bibinfo {title} {Quadrupolar and anisotropy effects on dephasing in two-electron spin qubits in {{GaAs}}},\ }\href {https://doi.org/10.1038/ncomms11170} {\bibfield  {journal} {\bibinfo  {journal} {Nat Commun}\ }\textbf {\bibinfo {volume} {7}},\ \bibinfo {pages} {11170} (\bibinfo {year} {2016})}\BibitemShut {NoStop}%
\bibitem [{\citenamefont {Stano}\ \emph {et~al.}(2018)\citenamefont {Stano}, \citenamefont {Hsu}, \citenamefont {Serina}, \citenamefont {Camenzind}, \citenamefont {Zumb{\"u}hl},\ and\ \citenamefont {Loss}}]{stanoFactorElectronsGatedefined2018}%
  \BibitemOpen
  \bibfield  {author} {\bibinfo {author} {\bibfnamefont {P.}~\bibnamefont {Stano}}, \bibinfo {author} {\bibfnamefont {C.-H.}\ \bibnamefont {Hsu}}, \bibinfo {author} {\bibfnamefont {M.}~\bibnamefont {Serina}}, \bibinfo {author} {\bibfnamefont {L.~C.}\ \bibnamefont {Camenzind}}, \bibinfo {author} {\bibfnamefont {D.~M.}\ \bibnamefont {Zumb{\"u}hl}},\ and\ \bibinfo {author} {\bibfnamefont {D.}~\bibnamefont {Loss}},\ }\bibfield  {title} {\bibinfo {title} {G -factor of electrons in gate-defined quantum dots in a strong in-plane magnetic field},\ }\href {https://doi.org/10.1103/PhysRevB.98.195314} {\bibfield  {journal} {\bibinfo  {journal} {Phys. Rev. B}\ }\textbf {\bibinfo {volume} {98}},\ \bibinfo {pages} {195314} (\bibinfo {year} {2018})}\BibitemShut {NoStop}%
\bibitem [{\citenamefont {Camenzind}\ \emph {et~al.}(2021)\citenamefont {Camenzind}, \citenamefont {Svab}, \citenamefont {Stano}, \citenamefont {Yu}, \citenamefont {Zimmerman}, \citenamefont {Gossard}, \citenamefont {Loss},\ and\ \citenamefont {Zumb{\"u}hl}}]{camenzindIsotropicAnisotropicFactor2021}%
  \BibitemOpen
  \bibfield  {author} {\bibinfo {author} {\bibfnamefont {L.~C.}\ \bibnamefont {Camenzind}}, \bibinfo {author} {\bibfnamefont {S.}~\bibnamefont {Svab}}, \bibinfo {author} {\bibfnamefont {P.}~\bibnamefont {Stano}}, \bibinfo {author} {\bibfnamefont {L.}~\bibnamefont {Yu}}, \bibinfo {author} {\bibfnamefont {J.~D.}\ \bibnamefont {Zimmerman}}, \bibinfo {author} {\bibfnamefont {A.~C.}\ \bibnamefont {Gossard}}, \bibinfo {author} {\bibfnamefont {D.}~\bibnamefont {Loss}},\ and\ \bibinfo {author} {\bibfnamefont {D.~M.}\ \bibnamefont {Zumb{\"u}hl}},\ }\bibfield  {title} {\bibinfo {title} {Isotropic and {{Anisotropic}} g -{{Factor Corrections}} in {{GaAs Quantum Dots}}},\ }\href {https://doi.org/10.1103/PhysRevLett.127.057701} {\bibfield  {journal} {\bibinfo  {journal} {Phys. Rev. Lett.}\ }\textbf {\bibinfo {volume} {127}},\ \bibinfo {pages} {057701} (\bibinfo {year} {2021})}\BibitemShut {NoStop}%
\bibitem [{SI()}]{SI}%
  \BibitemOpen
  \href@noop {} {\ }\bibinfo {note} {See Supplemental Material}\BibitemShut {NoStop}%
\bibitem [{\citenamefont {Overhauser}(1953)}]{overhauserPolarizationNucleiMetals1953}%
  \BibitemOpen
  \bibfield  {author} {\bibinfo {author} {\bibfnamefont {A.~W.}\ \bibnamefont {Overhauser}},\ }\bibfield  {title} {\bibinfo {title} {Polarization of {{Nuclei}} in {{Metals}}},\ }\href {https://doi.org/10.1103/PhysRev.92.411} {\bibfield  {journal} {\bibinfo  {journal} {Phys. Rev.}\ }\textbf {\bibinfo {volume} {92}},\ \bibinfo {pages} {411} (\bibinfo {year} {1953})}\BibitemShut {NoStop}%
\bibitem [{\citenamefont {Bodey}\ \emph {et~al.}(2019)\citenamefont {Bodey}, \citenamefont {Stockill}, \citenamefont {Denning}, \citenamefont {Gangloff}, \citenamefont {{Ethier-Majcher}}, \citenamefont {Jackson}, \citenamefont {Clarke}, \citenamefont {Hugues}, \citenamefont {Gall},\ and\ \citenamefont {Atature}}]{bodeyOpticalSpinLocking2019}%
  \BibitemOpen
  \bibfield  {author} {\bibinfo {author} {\bibfnamefont {J.~H.}\ \bibnamefont {Bodey}}, \bibinfo {author} {\bibfnamefont {R.}~\bibnamefont {Stockill}}, \bibinfo {author} {\bibfnamefont {E.~V.}\ \bibnamefont {Denning}}, \bibinfo {author} {\bibfnamefont {D.~A.}\ \bibnamefont {Gangloff}}, \bibinfo {author} {\bibfnamefont {G.}~\bibnamefont {{Ethier-Majcher}}}, \bibinfo {author} {\bibfnamefont {D.~M.}\ \bibnamefont {Jackson}}, \bibinfo {author} {\bibfnamefont {E.}~\bibnamefont {Clarke}}, \bibinfo {author} {\bibfnamefont {M.}~\bibnamefont {Hugues}}, \bibinfo {author} {\bibfnamefont {C.~L.}\ \bibnamefont {Gall}},\ and\ \bibinfo {author} {\bibfnamefont {M.}~\bibnamefont {Atature}},\ }\bibfield  {title} {\bibinfo {title} {Optical spin locking of a solid-state qubit},\ }\href {https://doi.org/10.1038/s41534-019-0206-3} {\bibfield  {journal} {\bibinfo  {journal} {npj Quantum Inf}\ }\textbf {\bibinfo {volume} {5}},\ \bibinfo {pages} {95} (\bibinfo {year} {2019})},\ \Eprint {https://arxiv.org/abs/1906.00427}
  {arxiv:1906.00427} \BibitemShut {NoStop}%
\bibitem [{\citenamefont {Vink}\ \emph {et~al.}(2009)\citenamefont {Vink}, \citenamefont {Nowack}, \citenamefont {Koppens}, \citenamefont {Danon}, \citenamefont {Nazarov},\ and\ \citenamefont {Vandersypen}}]{vinkLockingElectronSpins2009}%
  \BibitemOpen
  \bibfield  {author} {\bibinfo {author} {\bibfnamefont {I.~T.}\ \bibnamefont {Vink}}, \bibinfo {author} {\bibfnamefont {K.~C.}\ \bibnamefont {Nowack}}, \bibinfo {author} {\bibfnamefont {F.~H.~L.}\ \bibnamefont {Koppens}}, \bibinfo {author} {\bibfnamefont {J.}~\bibnamefont {Danon}}, \bibinfo {author} {\bibfnamefont {Y.~V.}\ \bibnamefont {Nazarov}},\ and\ \bibinfo {author} {\bibfnamefont {L.~M.~K.}\ \bibnamefont {Vandersypen}},\ }\bibfield  {title} {\bibinfo {title} {Locking electron spins into magnetic resonance by electron--nuclear feedback},\ }\href {https://doi.org/10.1038/nphys1366} {\bibfield  {journal} {\bibinfo  {journal} {Nature Phys}\ }\textbf {\bibinfo {volume} {5}},\ \bibinfo {pages} {764} (\bibinfo {year} {2009})}\BibitemShut {NoStop}%
\bibitem [{\citenamefont {Huang}\ and\ \citenamefont {Hu}(2010)}]{huangTheoreticalStudyNuclear2010}%
  \BibitemOpen
  \bibfield  {author} {\bibinfo {author} {\bibfnamefont {C.-W.}\ \bibnamefont {Huang}}\ and\ \bibinfo {author} {\bibfnamefont {X.}~\bibnamefont {Hu}},\ }\bibfield  {title} {\bibinfo {title} {Theoretical study of nuclear spin polarization and depolarization in self-assembled quantum dots},\ }\href {https://doi.org/10.1103/PhysRevB.81.205304} {\bibfield  {journal} {\bibinfo  {journal} {Phys. Rev. B}\ }\textbf {\bibinfo {volume} {81}},\ \bibinfo {pages} {205304} (\bibinfo {year} {2010})}\BibitemShut {NoStop}%
\bibitem [{\citenamefont {Latta}\ \emph {et~al.}(2011)\citenamefont {Latta}, \citenamefont {Srivastava},\ and\ \citenamefont {Imamoglu}}]{lattaHyperfineInteractionDominatedDynamics2011}%
  \BibitemOpen
  \bibfield  {author} {\bibinfo {author} {\bibfnamefont {C.}~\bibnamefont {Latta}}, \bibinfo {author} {\bibfnamefont {A.}~\bibnamefont {Srivastava}},\ and\ \bibinfo {author} {\bibfnamefont {A.}~\bibnamefont {Imamoglu}},\ }\bibfield  {title} {\bibinfo {title} {Hyperfine {{Interaction-Dominated Dynamics}} of {{Nuclear Spins}} in {{Self-Assembled InGaAs Quantum Dots}}},\ }\href@noop {} {\bibfield  {journal} {\bibinfo  {journal} {Physical Review Letters}\ }\textbf {\bibinfo {volume} {107}},\ \bibinfo {pages} {167401} (\bibinfo {year} {2011})}\BibitemShut {NoStop}%
\bibitem [{\citenamefont {H{\"o}gele}\ \emph {et~al.}(2012)\citenamefont {H{\"o}gele}, \citenamefont {Kroner}, \citenamefont {Latta}, \citenamefont {Claassen}, \citenamefont {Carusotto}, \citenamefont {Bulutay},\ and\ \citenamefont {Imamoglu}}]{hogeleDynamicNuclearSpin2012}%
  \BibitemOpen
  \bibfield  {author} {\bibinfo {author} {\bibfnamefont {A.}~\bibnamefont {H{\"o}gele}}, \bibinfo {author} {\bibfnamefont {M.}~\bibnamefont {Kroner}}, \bibinfo {author} {\bibfnamefont {C.}~\bibnamefont {Latta}}, \bibinfo {author} {\bibfnamefont {M.}~\bibnamefont {Claassen}}, \bibinfo {author} {\bibfnamefont {I.}~\bibnamefont {Carusotto}}, \bibinfo {author} {\bibfnamefont {C.}~\bibnamefont {Bulutay}},\ and\ \bibinfo {author} {\bibfnamefont {A.}~\bibnamefont {Imamoglu}},\ }\bibfield  {title} {\bibinfo {title} {Dynamic {{Nuclear Spin Polarization}} in the {{Resonant Laser Excitation}} of an {{InGaAs Quantum Dot}}},\ }\href {https://doi.org/10.1103/PhysRevLett.108.197403} {\bibfield  {journal} {\bibinfo  {journal} {Phys. Rev. Lett.}\ }\textbf {\bibinfo {volume} {108}},\ \bibinfo {pages} {197403} (\bibinfo {year} {2012})}\BibitemShut {NoStop}%
\bibitem [{\citenamefont {Nguyen}\ \emph {et~al.}(2023)\citenamefont {Nguyen}, \citenamefont {Spinnler}, \citenamefont {Hogg}, \citenamefont {Zhai}, \citenamefont {Javadi}, \citenamefont {Schrader}, \citenamefont {Erbe}, \citenamefont {Wyss}, \citenamefont {Ritzmann}, \citenamefont {Babin}, \citenamefont {Wieck}, \citenamefont {Ludwig},\ and\ \citenamefont {Warburton}}]{nguyenEnhancedElectronSpinCoherence2023}%
  \BibitemOpen
  \bibfield  {author} {\bibinfo {author} {\bibfnamefont {G.~N.}\ \bibnamefont {Nguyen}}, \bibinfo {author} {\bibfnamefont {C.}~\bibnamefont {Spinnler}}, \bibinfo {author} {\bibfnamefont {M.~R.}\ \bibnamefont {Hogg}}, \bibinfo {author} {\bibfnamefont {L.}~\bibnamefont {Zhai}}, \bibinfo {author} {\bibfnamefont {A.}~\bibnamefont {Javadi}}, \bibinfo {author} {\bibfnamefont {C.~A.}\ \bibnamefont {Schrader}}, \bibinfo {author} {\bibfnamefont {M.}~\bibnamefont {Erbe}}, \bibinfo {author} {\bibfnamefont {M.}~\bibnamefont {Wyss}}, \bibinfo {author} {\bibfnamefont {J.}~\bibnamefont {Ritzmann}}, \bibinfo {author} {\bibfnamefont {H.-G.}\ \bibnamefont {Babin}}, \bibinfo {author} {\bibfnamefont {A.~D.}\ \bibnamefont {Wieck}}, \bibinfo {author} {\bibfnamefont {A.}~\bibnamefont {Ludwig}},\ and\ \bibinfo {author} {\bibfnamefont {R.~J.}\ \bibnamefont {Warburton}},\ }\bibfield  {title} {\bibinfo {title} {Enhanced {{Electron-Spin Coherence}} in a {{GaAs Quantum Emitter}}},\ }\href {https://doi.org/10.1103/PhysRevLett.131.210805}
  {\bibfield  {journal} {\bibinfo  {journal} {Phys. Rev. Lett.}\ }\textbf {\bibinfo {volume} {131}},\ \bibinfo {pages} {210805} (\bibinfo {year} {2023})}\BibitemShut {NoStop}%
\bibitem [{\citenamefont {{da Silva}}\ \emph {et~al.}(2021)\citenamefont {{da Silva}}, \citenamefont {Undeutsch}, \citenamefont {Lehner}, \citenamefont {Manna}, \citenamefont {Krieger}, \citenamefont {Reindl}, \citenamefont {Schimpf}, \citenamefont {Trotta},\ and\ \citenamefont {Rastelli}}]{dasilvaGaAsQuantumDots2021}%
  \BibitemOpen
  \bibfield  {author} {\bibinfo {author} {\bibfnamefont {S.~F.~C.}\ \bibnamefont {{da Silva}}}, \bibinfo {author} {\bibfnamefont {G.}~\bibnamefont {Undeutsch}}, \bibinfo {author} {\bibfnamefont {B.}~\bibnamefont {Lehner}}, \bibinfo {author} {\bibfnamefont {S.}~\bibnamefont {Manna}}, \bibinfo {author} {\bibfnamefont {T.~M.}\ \bibnamefont {Krieger}}, \bibinfo {author} {\bibfnamefont {M.}~\bibnamefont {Reindl}}, \bibinfo {author} {\bibfnamefont {C.}~\bibnamefont {Schimpf}}, \bibinfo {author} {\bibfnamefont {R.}~\bibnamefont {Trotta}},\ and\ \bibinfo {author} {\bibfnamefont {A.}~\bibnamefont {Rastelli}},\ }\bibfield  {title} {\bibinfo {title} {{{GaAs}} quantum dots grown by droplet etching epitaxy as quantum light sources},\ }\href {https://doi.org/10.1063/5.0057070} {\bibfield  {journal} {\bibinfo  {journal} {Appl. Phys. Lett.}\ }\textbf {\bibinfo {volume} {119}},\ \bibinfo {pages} {120502} (\bibinfo {year} {2021})}\BibitemShut {NoStop}%
\bibitem [{\citenamefont {Chekhovich}\ \emph {et~al.}(2018)\citenamefont {Chekhovich}, \citenamefont {Griffiths}, \citenamefont {Skolnick}, \citenamefont {Huang}, \citenamefont {{Covre da Silva}}, \citenamefont {Yuan},\ and\ \citenamefont {Rastelli}}]{chekhovichCrossCalibrationDeformation2018}%
  \BibitemOpen
  \bibfield  {author} {\bibinfo {author} {\bibfnamefont {E.~A.}\ \bibnamefont {Chekhovich}}, \bibinfo {author} {\bibfnamefont {I.~M.}\ \bibnamefont {Griffiths}}, \bibinfo {author} {\bibfnamefont {M.~S.}\ \bibnamefont {Skolnick}}, \bibinfo {author} {\bibfnamefont {H.}~\bibnamefont {Huang}}, \bibinfo {author} {\bibfnamefont {S.~F.}\ \bibnamefont {{Covre da Silva}}}, \bibinfo {author} {\bibfnamefont {X.}~\bibnamefont {Yuan}},\ and\ \bibinfo {author} {\bibfnamefont {A.}~\bibnamefont {Rastelli}},\ }\bibfield  {title} {\bibinfo {title} {Cross calibration of deformation potentials and gradient-elastic tensors of {{GaAs}} using photoluminescence and nuclear mangetic resonance spectroscopy in {{GaAs}}/{{AlGaAs}} quantum dot structures},\ }\href@noop {} {\bibfield  {journal} {\bibinfo  {journal} {Physical Review B}\ }\textbf {\bibinfo {volume} {97}},\ \bibinfo {pages} {235311} (\bibinfo {year} {2018})}\BibitemShut {NoStop}%
\bibitem [{\citenamefont {Knight}(1949)}]{knightNuclearMagneticResonance1949}%
  \BibitemOpen
  \bibfield  {author} {\bibinfo {author} {\bibfnamefont {W.~D.}\ \bibnamefont {Knight}},\ }\bibfield  {title} {\bibinfo {title} {Nuclear {{Magnetic Resonance Shift}} in {{Metals}}},\ }\href {https://doi.org/10.1103/PhysRev.76.1259.2} {\bibfield  {journal} {\bibinfo  {journal} {Phys. Rev.}\ }\textbf {\bibinfo {volume} {76}},\ \bibinfo {pages} {1259} (\bibinfo {year} {1949})}\BibitemShut {NoStop}%
\bibitem [{\citenamefont {Lai}\ \emph {et~al.}(2006)\citenamefont {Lai}, \citenamefont {Maletinsky}, \citenamefont {Badolato},\ and\ \citenamefont {Imamoglu}}]{Lai2006}%
  \BibitemOpen
  \bibfield  {author} {\bibinfo {author} {\bibfnamefont {C.~W.}\ \bibnamefont {Lai}}, \bibinfo {author} {\bibfnamefont {P.}~\bibnamefont {Maletinsky}}, \bibinfo {author} {\bibfnamefont {A.}~\bibnamefont {Badolato}},\ and\ \bibinfo {author} {\bibfnamefont {A.}~\bibnamefont {Imamoglu}},\ }\bibfield  {title} {\bibinfo {title} {{Knight-Field-Enabled Nuclear Spin Polarization in Single Quantum Dots}},\ }\href {https://doi.org/10.1103/PhysRevLett.96.167403} {\bibfield  {journal} {\bibinfo  {journal} {Phys. Rev. Lett.}\ }\textbf {\bibinfo {volume} {96}},\ \bibinfo {pages} {167403} (\bibinfo {year} {2006})}\BibitemShut {NoStop}%
\bibitem [{\citenamefont {Sallen}\ \emph {et~al.}(2014)\citenamefont {Sallen}, \citenamefont {Kunz}, \citenamefont {Amand}, \citenamefont {Bouet}, \citenamefont {Kuroda}, \citenamefont {Mano}, \citenamefont {Paget}, \citenamefont {Krebs}, \citenamefont {Marie}, \citenamefont {Sakoda},\ and\ \citenamefont {Urbaszek}}]{Sallen2014}%
  \BibitemOpen
  \bibfield  {author} {\bibinfo {author} {\bibfnamefont {G.}~\bibnamefont {Sallen}}, \bibinfo {author} {\bibfnamefont {S.}~\bibnamefont {Kunz}}, \bibinfo {author} {\bibfnamefont {T.}~\bibnamefont {Amand}}, \bibinfo {author} {\bibfnamefont {L.}~\bibnamefont {Bouet}}, \bibinfo {author} {\bibfnamefont {T.}~\bibnamefont {Kuroda}}, \bibinfo {author} {\bibfnamefont {T.}~\bibnamefont {Mano}}, \bibinfo {author} {\bibfnamefont {D.}~\bibnamefont {Paget}}, \bibinfo {author} {\bibfnamefont {O.}~\bibnamefont {Krebs}}, \bibinfo {author} {\bibfnamefont {X.}~\bibnamefont {Marie}}, \bibinfo {author} {\bibfnamefont {K.}~\bibnamefont {Sakoda}},\ and\ \bibinfo {author} {\bibfnamefont {B.}~\bibnamefont {Urbaszek}},\ }\bibfield  {title} {\bibinfo {title} {{Nuclear magnetization in gallium arsenide quantum dots at zero magnetic field}},\ }\href {https://doi.org/10.1038/ncomms4268} {\bibfield  {journal} {\bibinfo  {journal} {Nat. Commun.}\ }\textbf {\bibinfo {volume} {5}},\ \bibinfo {pages} {3268} (\bibinfo {year}
  {2014})}\BibitemShut {NoStop}%
\bibitem [{\citenamefont {Paget}\ \emph {et~al.}(1977)\citenamefont {Paget}, \citenamefont {Lampel}, \citenamefont {Sapoval},\ and\ \citenamefont {Safarov}}]{pagetLowFieldElectronnuclear1977}%
  \BibitemOpen
  \bibfield  {author} {\bibinfo {author} {\bibfnamefont {D.}~\bibnamefont {Paget}}, \bibinfo {author} {\bibfnamefont {G.}~\bibnamefont {Lampel}}, \bibinfo {author} {\bibfnamefont {B.}~\bibnamefont {Sapoval}},\ and\ \bibinfo {author} {\bibfnamefont {V.~I.}\ \bibnamefont {Safarov}},\ }\bibfield  {title} {\bibinfo {title} {Low field electron-nuclear spin coupling in gallium arsenide under optical pumping conditions},\ }\href {https://doi.org/10.1103/PhysRevB.15.5780} {\bibfield  {journal} {\bibinfo  {journal} {Phys. Rev. B}\ }\textbf {\bibinfo {volume} {15}},\ \bibinfo {pages} {5780} (\bibinfo {year} {1977})}\BibitemShut {NoStop}%
\bibitem [{\citenamefont {Zaporski}\ \emph {et~al.}(2023{\natexlab{b}})\citenamefont {Zaporski}, \citenamefont {Shofer}, \citenamefont {Bodey}, \citenamefont {Manna}, \citenamefont {Gillard}, \citenamefont {Appel}, \citenamefont {Schimpf}, \citenamefont {Covre Da~Silva}, \citenamefont {Jarman}, \citenamefont {Delamare}, \citenamefont {Park}, \citenamefont {Haeusler}, \citenamefont {Chekhovich}, \citenamefont {Rastelli}, \citenamefont {Gangloff}, \citenamefont {Atat{\"u}re},\ and\ \citenamefont {Le~Gall}}]{zaporskiIdealRefocusingOptically2023}%
  \BibitemOpen
  \bibfield  {author} {\bibinfo {author} {\bibfnamefont {L.}~\bibnamefont {Zaporski}}, \bibinfo {author} {\bibfnamefont {N.}~\bibnamefont {Shofer}}, \bibinfo {author} {\bibfnamefont {J.~H.}\ \bibnamefont {Bodey}}, \bibinfo {author} {\bibfnamefont {S.}~\bibnamefont {Manna}}, \bibinfo {author} {\bibfnamefont {G.}~\bibnamefont {Gillard}}, \bibinfo {author} {\bibfnamefont {M.~H.}\ \bibnamefont {Appel}}, \bibinfo {author} {\bibfnamefont {C.}~\bibnamefont {Schimpf}}, \bibinfo {author} {\bibfnamefont {S.~F.}\ \bibnamefont {Covre Da~Silva}}, \bibinfo {author} {\bibfnamefont {J.}~\bibnamefont {Jarman}}, \bibinfo {author} {\bibfnamefont {G.}~\bibnamefont {Delamare}}, \bibinfo {author} {\bibfnamefont {G.}~\bibnamefont {Park}}, \bibinfo {author} {\bibfnamefont {U.}~\bibnamefont {Haeusler}}, \bibinfo {author} {\bibfnamefont {E.~A.}\ \bibnamefont {Chekhovich}}, \bibinfo {author} {\bibfnamefont {A.}~\bibnamefont {Rastelli}}, \bibinfo {author} {\bibfnamefont {D.~A.}\ \bibnamefont {Gangloff}}, \bibinfo {author} {\bibfnamefont
  {M.}~\bibnamefont {Atat{\"u}re}},\ and\ \bibinfo {author} {\bibfnamefont {C.}~\bibnamefont {Le~Gall}},\ }\bibfield  {title} {\bibinfo {title} {Ideal refocusing of an optically active spin qubit under strong hyperfine interactions},\ }\href {https://doi.org/10.1038/s41565-022-01282-2} {\bibfield  {journal} {\bibinfo  {journal} {Nat. Nanotechnol.}\ }\textbf {\bibinfo {volume} {18}},\ \bibinfo {pages} {257} (\bibinfo {year} {2023}{\natexlab{b}})}\BibitemShut {NoStop}%
\bibitem [{\citenamefont {Carr}\ and\ \citenamefont {Purcell}(1954)}]{carrEffectsDiffusionFree1954}%
  \BibitemOpen
  \bibfield  {author} {\bibinfo {author} {\bibfnamefont {H.~Y.}\ \bibnamefont {Carr}}\ and\ \bibinfo {author} {\bibfnamefont {E.~M.}\ \bibnamefont {Purcell}},\ }\bibfield  {title} {\bibinfo {title} {Effects of {{Diffusion}} on {{Free Precession}} in {{Nuclear Magnetic Resonance Experiments}}},\ }\href {https://doi.org/10.1103/PhysRev.94.630} {\bibfield  {journal} {\bibinfo  {journal} {Phys. Rev.}\ }\textbf {\bibinfo {volume} {94}},\ \bibinfo {pages} {630} (\bibinfo {year} {1954})}\BibitemShut {NoStop}%
\bibitem [{\citenamefont {Cywi{\'n}ski}\ \emph {et~al.}(2008)\citenamefont {Cywi{\'n}ski}, \citenamefont {Lutchyn}, \citenamefont {Nave},\ and\ \citenamefont {Das~Sarma}}]{cywinskiHowEnhanceDephasing2008}%
  \BibitemOpen
  \bibfield  {author} {\bibinfo {author} {\bibfnamefont {{\L}.}~\bibnamefont {Cywi{\'n}ski}}, \bibinfo {author} {\bibfnamefont {R.~M.}\ \bibnamefont {Lutchyn}}, \bibinfo {author} {\bibfnamefont {C.~P.}\ \bibnamefont {Nave}},\ and\ \bibinfo {author} {\bibfnamefont {S.}~\bibnamefont {Das~Sarma}},\ }\bibfield  {title} {\bibinfo {title} {How to enhance dephasing time in superconducting qubits},\ }\href {https://doi.org/10.1103/PhysRevB.77.174509} {\bibfield  {journal} {\bibinfo  {journal} {Phys. Rev. B}\ }\textbf {\bibinfo {volume} {77}},\ \bibinfo {pages} {174509} (\bibinfo {year} {2008})}\BibitemShut {NoStop}%
\bibitem [{\citenamefont {Malinowski}\ \emph {et~al.}(2017)\citenamefont {Malinowski}, \citenamefont {Martins}, \citenamefont {Nissen}, \citenamefont {Barnes}, \citenamefont {Cywi{\'n}ski}, \citenamefont {Rudner}, \citenamefont {Fallahi}, \citenamefont {Gardner}, \citenamefont {Manfra}, \citenamefont {Marcus},\ and\ \citenamefont {Kuemmeth}}]{malinowskiNotchFilteringNuclear2017}%
  \BibitemOpen
  \bibfield  {author} {\bibinfo {author} {\bibfnamefont {F.~K.}\ \bibnamefont {Malinowski}}, \bibinfo {author} {\bibfnamefont {F.}~\bibnamefont {Martins}}, \bibinfo {author} {\bibfnamefont {P.~D.}\ \bibnamefont {Nissen}}, \bibinfo {author} {\bibfnamefont {E.}~\bibnamefont {Barnes}}, \bibinfo {author} {\bibfnamefont {{\L}.}~\bibnamefont {Cywi{\'n}ski}}, \bibinfo {author} {\bibfnamefont {M.~S.}\ \bibnamefont {Rudner}}, \bibinfo {author} {\bibfnamefont {S.}~\bibnamefont {Fallahi}}, \bibinfo {author} {\bibfnamefont {G.~C.}\ \bibnamefont {Gardner}}, \bibinfo {author} {\bibfnamefont {M.~J.}\ \bibnamefont {Manfra}}, \bibinfo {author} {\bibfnamefont {C.~M.}\ \bibnamefont {Marcus}},\ and\ \bibinfo {author} {\bibfnamefont {F.}~\bibnamefont {Kuemmeth}},\ }\bibfield  {title} {\bibinfo {title} {Notch filtering the nuclear environment of a spin qubit},\ }\href {https://doi.org/10.1038/nnano.2016.170} {\bibfield  {journal} {\bibinfo  {journal} {Nature Nanotech}\ }\textbf {\bibinfo {volume} {12}},\ \bibinfo {pages} {16}
  (\bibinfo {year} {2017})}\BibitemShut {NoStop}%
\bibitem [{\citenamefont {Chekhovich}\ \emph {et~al.}(2017)\citenamefont {Chekhovich}, \citenamefont {Ulhaq}, \citenamefont {Zallo}, \citenamefont {Ding}, \citenamefont {Schmidt},\ and\ \citenamefont {Skolnick}}]{chekhovichMeasurementSpinTemperature2017}%
  \BibitemOpen
  \bibfield  {author} {\bibinfo {author} {\bibfnamefont {E.~A.}\ \bibnamefont {Chekhovich}}, \bibinfo {author} {\bibfnamefont {A.}~\bibnamefont {Ulhaq}}, \bibinfo {author} {\bibfnamefont {E.}~\bibnamefont {Zallo}}, \bibinfo {author} {\bibfnamefont {F.}~\bibnamefont {Ding}}, \bibinfo {author} {\bibfnamefont {O.~G.}\ \bibnamefont {Schmidt}},\ and\ \bibinfo {author} {\bibfnamefont {M.~S.}\ \bibnamefont {Skolnick}},\ }\bibfield  {title} {\bibinfo {title} {Measurement of the spin temperature of optically cooled nuclei and {{GaAs}} hyperfine constants in {{GaAs}}/{{AlGaAs}} quantum dots},\ }\href {https://doi.org/10.1038/nmat4959} {\bibfield  {journal} {\bibinfo  {journal} {Nature Mater}\ }\textbf {\bibinfo {volume} {16}},\ \bibinfo {pages} {982} (\bibinfo {year} {2017})}\BibitemShut {NoStop}%
\bibitem [{\citenamefont {Vitagliano}\ \emph {et~al.}(2011)\citenamefont {Vitagliano}, \citenamefont {Hyllus}, \citenamefont {Egusquiza},\ and\ \citenamefont {T{\'o}th}}]{vitaglianoSpinSqueezingInequalities2011}%
  \BibitemOpen
  \bibfield  {author} {\bibinfo {author} {\bibfnamefont {G.}~\bibnamefont {Vitagliano}}, \bibinfo {author} {\bibfnamefont {P.}~\bibnamefont {Hyllus}}, \bibinfo {author} {\bibfnamefont {I.~L.}\ \bibnamefont {Egusquiza}},\ and\ \bibinfo {author} {\bibfnamefont {G.}~\bibnamefont {T{\'o}th}},\ }\bibfield  {title} {\bibinfo {title} {Spin {{Squeezing Inequalities}} for {{Arbitrary Spin}}},\ }\href {https://doi.org/10.1103/PhysRevLett.107.240502} {\bibfield  {journal} {\bibinfo  {journal} {Phys. Rev. Lett.}\ }\textbf {\bibinfo {volume} {107}},\ \bibinfo {pages} {240502} (\bibinfo {year} {2011})}\BibitemShut {NoStop}%
\bibitem [{\citenamefont {Dicke}(1954)}]{dickeCoherenceSpontaneousRadiation1954}%
  \BibitemOpen
  \bibfield  {author} {\bibinfo {author} {\bibfnamefont {R.~H.}\ \bibnamefont {Dicke}},\ }\bibfield  {title} {\bibinfo {title} {Coherence in {{Spontaneous Radiation Processes}}},\ }\href {https://doi.org/10.1103/PhysRev.93.99} {\bibfield  {journal} {\bibinfo  {journal} {Phys. Rev.}\ }\textbf {\bibinfo {volume} {93}},\ \bibinfo {pages} {99} (\bibinfo {year} {1954})}\BibitemShut {NoStop}%
\end{thebibliography}%


\begin{thebibliography}{19}%
\makeatletter
\providecommand \@ifxundefined [1]{%
 \@ifx{#1\undefined}
}%
\providecommand \@ifnum [1]{%
 \ifnum #1\expandafter \@firstoftwo
 \else \expandafter \@secondoftwo
 \fi
}%
\providecommand \@ifx [1]{%
 \ifx #1\expandafter \@firstoftwo
 \else \expandafter \@secondoftwo
 \fi
}%
\providecommand \natexlab [1]{#1}%
\providecommand \enquote  [1]{``#1''}%
\providecommand \bibnamefont  [1]{#1}%
\providecommand \bibfnamefont [1]{#1}%
\providecommand \citenamefont [1]{#1}%
\providecommand \href@noop [0]{\@secondoftwo}%
\providecommand \href [0]{\begingroup \@sanitize@url \@href}%
\providecommand \@href[1]{\@@startlink{#1}\@@href}%
\providecommand \@@href[1]{\endgroup#1\@@endlink}%
\providecommand \@sanitize@url [0]{\catcode `\\12\catcode `\$12\catcode `\&12\catcode `\#12\catcode `\^12\catcode `\_12\catcode `\%12\relax}%
\providecommand \@@startlink[1]{}%
\providecommand \@@endlink[0]{}%
\providecommand \url  [0]{\begingroup\@sanitize@url \@url }%
\providecommand \@url [1]{\endgroup\@href {#1}{\urlprefix }}%
\providecommand \urlprefix  [0]{URL }%
\providecommand \Eprint [0]{\href }%
\providecommand \doibase [0]{https://doi.org/}%
\providecommand \selectlanguage [0]{\@gobble}%
\providecommand \bibinfo  [0]{\@secondoftwo}%
\providecommand \bibfield  [0]{\@secondoftwo}%
\providecommand \translation [1]{[#1]}%
\providecommand \BibitemOpen [0]{}%
\providecommand \bibitemStop [0]{}%
\providecommand \bibitemNoStop [0]{.\EOS\space}%
\providecommand \EOS [0]{\spacefactor3000\relax}%
\providecommand \BibitemShut  [1]{\csname bibitem#1\endcsname}%
\let\auto@bib@innerbib\@empty
\bibitem [{\citenamefont {Zaporski}\ \emph {et~al.}(2023)\citenamefont {Zaporski}, \citenamefont {Shofer}, \citenamefont {Bodey}, \citenamefont {Manna}, \citenamefont {Gillard}, \citenamefont {Appel}, \citenamefont {Schimpf}, \citenamefont {Covre Da~Silva}, \citenamefont {Jarman}, \citenamefont {Delamare}, \citenamefont {Park}, \citenamefont {Haeusler}, \citenamefont {Chekhovich}, \citenamefont {Rastelli}, \citenamefont {Gangloff}, \citenamefont {Atat{\"u}re},\ and\ \citenamefont {Le~Gall}}]{zaporskiIdealRefocusingOptically2023}%
  \BibitemOpen
  \bibfield  {author} {\bibinfo {author} {\bibfnamefont {L.}~\bibnamefont {Zaporski}}, \bibinfo {author} {\bibfnamefont {N.}~\bibnamefont {Shofer}}, \bibinfo {author} {\bibfnamefont {J.~H.}\ \bibnamefont {Bodey}}, \bibinfo {author} {\bibfnamefont {S.}~\bibnamefont {Manna}}, \bibinfo {author} {\bibfnamefont {G.}~\bibnamefont {Gillard}}, \bibinfo {author} {\bibfnamefont {M.~H.}\ \bibnamefont {Appel}}, \bibinfo {author} {\bibfnamefont {C.}~\bibnamefont {Schimpf}}, \bibinfo {author} {\bibfnamefont {S.~F.}\ \bibnamefont {Covre Da~Silva}}, \bibinfo {author} {\bibfnamefont {J.}~\bibnamefont {Jarman}}, \bibinfo {author} {\bibfnamefont {G.}~\bibnamefont {Delamare}}, \bibinfo {author} {\bibfnamefont {G.}~\bibnamefont {Park}}, \bibinfo {author} {\bibfnamefont {U.}~\bibnamefont {Haeusler}}, \bibinfo {author} {\bibfnamefont {E.~A.}\ \bibnamefont {Chekhovich}}, \bibinfo {author} {\bibfnamefont {A.}~\bibnamefont {Rastelli}}, \bibinfo {author} {\bibfnamefont {D.~A.}\ \bibnamefont {Gangloff}}, \bibinfo {author} {\bibfnamefont
  {M.}~\bibnamefont {Atat{\"u}re}},\ and\ \bibinfo {author} {\bibfnamefont {C.}~\bibnamefont {Le~Gall}},\ }\bibfield  {title} {\bibinfo {title} {Ideal refocusing of an optically active spin qubit under strong hyperfine interactions},\ }\href {https://doi.org/10.1038/s41565-022-01282-2} {\bibfield  {journal} {\bibinfo  {journal} {Nat. Nanotechnol.}\ }\textbf {\bibinfo {volume} {18}},\ \bibinfo {pages} {257} (\bibinfo {year} {2023})}\BibitemShut {NoStop}%
\bibitem [{\citenamefont {{da Silva}}\ \emph {et~al.}(2021)\citenamefont {{da Silva}}, \citenamefont {Undeutsch}, \citenamefont {Lehner}, \citenamefont {Manna}, \citenamefont {Krieger}, \citenamefont {Reindl}, \citenamefont {Schimpf}, \citenamefont {Trotta},\ and\ \citenamefont {Rastelli}}]{dasilvaGaAsQuantumDots2021}%
  \BibitemOpen
  \bibfield  {author} {\bibinfo {author} {\bibfnamefont {S.~F.~C.}\ \bibnamefont {{da Silva}}}, \bibinfo {author} {\bibfnamefont {G.}~\bibnamefont {Undeutsch}}, \bibinfo {author} {\bibfnamefont {B.}~\bibnamefont {Lehner}}, \bibinfo {author} {\bibfnamefont {S.}~\bibnamefont {Manna}}, \bibinfo {author} {\bibfnamefont {T.~M.}\ \bibnamefont {Krieger}}, \bibinfo {author} {\bibfnamefont {M.}~\bibnamefont {Reindl}}, \bibinfo {author} {\bibfnamefont {C.}~\bibnamefont {Schimpf}}, \bibinfo {author} {\bibfnamefont {R.}~\bibnamefont {Trotta}},\ and\ \bibinfo {author} {\bibfnamefont {A.}~\bibnamefont {Rastelli}},\ }\bibfield  {title} {\bibinfo {title} {{{GaAs}} quantum dots grown by droplet etching epitaxy as quantum light sources},\ }\href {https://doi.org/10.1063/5.0057070} {\bibfield  {journal} {\bibinfo  {journal} {Appl. Phys. Lett.}\ }\textbf {\bibinfo {volume} {119}},\ \bibinfo {pages} {120502} (\bibinfo {year} {2021})}\BibitemShut {NoStop}%
\bibitem [{\citenamefont {Zhai}\ \emph {et~al.}(2020)\citenamefont {Zhai}, \citenamefont {L{\"o}bl}, \citenamefont {Nguyen}, \citenamefont {Ritzmann}, \citenamefont {Javadi}, \citenamefont {Spinnler}, \citenamefont {Wieck}, \citenamefont {Ludwig},\ and\ \citenamefont {Warburton}}]{zhaiLownoiseGaAsQuantum2020}%
  \BibitemOpen
  \bibfield  {author} {\bibinfo {author} {\bibfnamefont {L.}~\bibnamefont {Zhai}}, \bibinfo {author} {\bibfnamefont {M.~C.}\ \bibnamefont {L{\"o}bl}}, \bibinfo {author} {\bibfnamefont {G.~N.}\ \bibnamefont {Nguyen}}, \bibinfo {author} {\bibfnamefont {J.}~\bibnamefont {Ritzmann}}, \bibinfo {author} {\bibfnamefont {A.}~\bibnamefont {Javadi}}, \bibinfo {author} {\bibfnamefont {C.}~\bibnamefont {Spinnler}}, \bibinfo {author} {\bibfnamefont {A.~D.}\ \bibnamefont {Wieck}}, \bibinfo {author} {\bibfnamefont {A.}~\bibnamefont {Ludwig}},\ and\ \bibinfo {author} {\bibfnamefont {R.~J.}\ \bibnamefont {Warburton}},\ }\bibfield  {title} {\bibinfo {title} {Low-noise {{GaAs}} quantum dots for quantum photonics},\ }\href {https://doi.org/10.1038/s41467-020-18625-z} {\bibfield  {journal} {\bibinfo  {journal} {Nat Commun}\ }\textbf {\bibinfo {volume} {11}},\ \bibinfo {pages} {4745} (\bibinfo {year} {2020})}\BibitemShut {NoStop}%
\bibitem [{\citenamefont {Schimpf}\ \emph {et~al.}(2021)\citenamefont {Schimpf}, \citenamefont {Manna}, \citenamefont {Covre Da~Silva}, \citenamefont {Aigner},\ and\ \citenamefont {Rastelli}}]{schimpfEntanglementbasedQuantumKey2021}%
  \BibitemOpen
  \bibfield  {author} {\bibinfo {author} {\bibfnamefont {C.}~\bibnamefont {Schimpf}}, \bibinfo {author} {\bibfnamefont {S.}~\bibnamefont {Manna}}, \bibinfo {author} {\bibfnamefont {S.~F.}\ \bibnamefont {Covre Da~Silva}}, \bibinfo {author} {\bibfnamefont {M.}~\bibnamefont {Aigner}},\ and\ \bibinfo {author} {\bibfnamefont {A.}~\bibnamefont {Rastelli}},\ }\bibfield  {title} {\bibinfo {title} {Entanglement-based quantum key distribution with a blinking-free quantum dot operated at a temperature up to 20 {{K}}},\ }\bibfield  {journal} {\bibinfo  {journal} {Adv. Photon.}\ }\textbf {\bibinfo {volume} {3}},\ \href {https://doi.org/10.1117/1.AP.3.6.065001} {10.1117/1.AP.3.6.065001} (\bibinfo {year} {2021})\BibitemShut {NoStop}%
\bibitem [{\citenamefont {Chekhovich}\ \emph {et~al.}(2018)\citenamefont {Chekhovich}, \citenamefont {Griffiths}, \citenamefont {Skolnick}, \citenamefont {Huang}, \citenamefont {{Covre da Silva}}, \citenamefont {Yuan},\ and\ \citenamefont {Rastelli}}]{chekhovichCrossCalibrationDeformation2018}%
  \BibitemOpen
  \bibfield  {author} {\bibinfo {author} {\bibfnamefont {E.~A.}\ \bibnamefont {Chekhovich}}, \bibinfo {author} {\bibfnamefont {I.~M.}\ \bibnamefont {Griffiths}}, \bibinfo {author} {\bibfnamefont {M.~S.}\ \bibnamefont {Skolnick}}, \bibinfo {author} {\bibfnamefont {H.}~\bibnamefont {Huang}}, \bibinfo {author} {\bibfnamefont {S.~F.}\ \bibnamefont {{Covre da Silva}}}, \bibinfo {author} {\bibfnamefont {X.}~\bibnamefont {Yuan}},\ and\ \bibinfo {author} {\bibfnamefont {A.}~\bibnamefont {Rastelli}},\ }\bibfield  {title} {\bibinfo {title} {Cross calibration of deformation potentials and gradient-elastic tensors of {{GaAs}} using photoluminescence and nuclear mangetic resonance spectroscopy in {{GaAs}}/{{AlGaAs}} quantum dot structures},\ }\href@noop {} {\bibfield  {journal} {\bibinfo  {journal} {Physical Review B}\ }\textbf {\bibinfo {volume} {97}},\ \bibinfo {pages} {235311} (\bibinfo {year} {2018})}\BibitemShut {NoStop}%
\bibitem [{\citenamefont {Gangloff}\ \emph {et~al.}(2019)\citenamefont {Gangloff}, \citenamefont {{\'E}thier-Majcher}, \citenamefont {Lang}, \citenamefont {Denning}, \citenamefont {Bodey}, \citenamefont {Jackson}, \citenamefont {Clarke}, \citenamefont {Hugues}, \citenamefont {Le~Gall},\ and\ \citenamefont {Atat{\"u}re}}]{gangloffQuantumInterfaceElectron2019}%
  \BibitemOpen
  \bibfield  {author} {\bibinfo {author} {\bibfnamefont {D.~A.}\ \bibnamefont {Gangloff}}, \bibinfo {author} {\bibfnamefont {G.}~\bibnamefont {{\'E}thier-Majcher}}, \bibinfo {author} {\bibfnamefont {C.}~\bibnamefont {Lang}}, \bibinfo {author} {\bibfnamefont {E.~V.}\ \bibnamefont {Denning}}, \bibinfo {author} {\bibfnamefont {J.~H.}\ \bibnamefont {Bodey}}, \bibinfo {author} {\bibfnamefont {D.~M.}\ \bibnamefont {Jackson}}, \bibinfo {author} {\bibfnamefont {E.}~\bibnamefont {Clarke}}, \bibinfo {author} {\bibfnamefont {M.}~\bibnamefont {Hugues}}, \bibinfo {author} {\bibfnamefont {C.}~\bibnamefont {Le~Gall}},\ and\ \bibinfo {author} {\bibfnamefont {M.}~\bibnamefont {Atat{\"u}re}},\ }\bibfield  {title} {\bibinfo {title} {Quantum interface of an electron and a nuclear ensemble},\ }\href {https://doi.org/10.1126/science.aaw2906} {\bibfield  {journal} {\bibinfo  {journal} {Science}\ }\textbf {\bibinfo {volume} {364}},\ \bibinfo {pages} {62} (\bibinfo {year} {2019})}\BibitemShut {NoStop}%
\bibitem [{\citenamefont {Bahder}(1990)}]{bahderEightbandModelStrained1990}%
  \BibitemOpen
  \bibfield  {author} {\bibinfo {author} {\bibfnamefont {T.~B.}\ \bibnamefont {Bahder}},\ }\bibfield  {title} {\bibinfo {title} {Eight-band {\textbf{k}} {$\cdot$} {\textbf{p}} model of strained zinc-blende crystals},\ }\href {https://doi.org/10.1103/PhysRevB.41.11992} {\bibfield  {journal} {\bibinfo  {journal} {Phys. Rev. B}\ }\textbf {\bibinfo {volume} {41}},\ \bibinfo {pages} {11992} (\bibinfo {year} {1990})}\BibitemShut {NoStop}%
\bibitem [{\citenamefont {Gawarecki}\ \emph {et~al.}(2014)\citenamefont {Gawarecki}, \citenamefont {Machnikowski},\ and\ \citenamefont {Kuhn}}]{gawareckiElectronStatesDouble2014}%
  \BibitemOpen
  \bibfield  {author} {\bibinfo {author} {\bibfnamefont {K.}~\bibnamefont {Gawarecki}}, \bibinfo {author} {\bibfnamefont {P.}~\bibnamefont {Machnikowski}},\ and\ \bibinfo {author} {\bibfnamefont {T.}~\bibnamefont {Kuhn}},\ }\bibfield  {title} {\bibinfo {title} {Electron states in a double quantum dot with broken axial symmetry},\ }\href {https://doi.org/10.1103/PhysRevB.90.085437} {\bibfield  {journal} {\bibinfo  {journal} {Phys. Rev. B}\ }\textbf {\bibinfo {volume} {90}},\ \bibinfo {pages} {085437} (\bibinfo {year} {2014})}\BibitemShut {NoStop}%
\bibitem [{\citenamefont {{Mielnik-Pyszczorski}}\ \emph {et~al.}(2018)\citenamefont {{Mielnik-Pyszczorski}}, \citenamefont {Gawarecki}, \citenamefont {Gawe{\l}czyk},\ and\ \citenamefont {Machnikowski}}]{mielnik-pyszczorskiDominantRoleShear2018}%
  \BibitemOpen
  \bibfield  {author} {\bibinfo {author} {\bibfnamefont {A.}~\bibnamefont {{Mielnik-Pyszczorski}}}, \bibinfo {author} {\bibfnamefont {K.}~\bibnamefont {Gawarecki}}, \bibinfo {author} {\bibfnamefont {M.}~\bibnamefont {Gawe{\l}czyk}},\ and\ \bibinfo {author} {\bibfnamefont {P.}~\bibnamefont {Machnikowski}},\ }\bibfield  {title} {\bibinfo {title} {Dominant role of the shear strain induced admixture in spin-flip processes in self-assembled quantum dots},\ }\href {https://doi.org/10.1103/PhysRevB.97.245313} {\bibfield  {journal} {\bibinfo  {journal} {Phys. Rev. B}\ }\textbf {\bibinfo {volume} {97}},\ \bibinfo {pages} {245313} (\bibinfo {year} {2018})}\BibitemShut {NoStop}%
\bibitem [{\citenamefont {Gawe{\l}czyk}\ \emph {et~al.}(2017)\citenamefont {Gawe{\l}czyk}, \citenamefont {Syperek}, \citenamefont {Mary{\'n}ski}, \citenamefont {Mrowi{\'n}ski}, \citenamefont {Dusanowski}, \citenamefont {Gawarecki}, \citenamefont {Misiewicz}, \citenamefont {Somers}, \citenamefont {Reithmaier}, \citenamefont {H{\"o}fling},\ and\ \citenamefont {Sek}}]{gawelczykExcitonLifetimeEmission2017}%
  \BibitemOpen
  \bibfield  {author} {\bibinfo {author} {\bibfnamefont {M.}~\bibnamefont {Gawe{\l}czyk}}, \bibinfo {author} {\bibfnamefont {M.}~\bibnamefont {Syperek}}, \bibinfo {author} {\bibfnamefont {A.}~\bibnamefont {Mary{\'n}ski}}, \bibinfo {author} {\bibfnamefont {P.}~\bibnamefont {Mrowi{\'n}ski}}, \bibinfo {author} {\bibfnamefont {{\L}.}~\bibnamefont {Dusanowski}}, \bibinfo {author} {\bibfnamefont {K.}~\bibnamefont {Gawarecki}}, \bibinfo {author} {\bibfnamefont {J.}~\bibnamefont {Misiewicz}}, \bibinfo {author} {\bibfnamefont {A.}~\bibnamefont {Somers}}, \bibinfo {author} {\bibfnamefont {J.~P.}\ \bibnamefont {Reithmaier}}, \bibinfo {author} {\bibfnamefont {S.}~\bibnamefont {H{\"o}fling}},\ and\ \bibinfo {author} {\bibfnamefont {G.}~\bibnamefont {Sek}},\ }\bibfield  {title} {\bibinfo {title} {Exciton lifetime and emission polarization dispersion in strongly in-plane asymmetric nanostructures},\ }\href {https://doi.org/10.1103/PhysRevB.96.245425} {\bibfield  {journal} {\bibinfo  {journal} {Phys. Rev. B}\ }\textbf
  {\bibinfo {volume} {96}},\ \bibinfo {pages} {245425} (\bibinfo {year} {2017})}\BibitemShut {NoStop}%
\bibitem [{\citenamefont {Xu}\ \emph {et~al.}(2007)\citenamefont {Xu}, \citenamefont {Wu}, \citenamefont {Sun}, \citenamefont {Huang}, \citenamefont {Cheng}, \citenamefont {Steel}, \citenamefont {Bracker}, \citenamefont {Gammon}, \citenamefont {Emary},\ and\ \citenamefont {Sham}}]{xuFastSpinState2007}%
  \BibitemOpen
  \bibfield  {author} {\bibinfo {author} {\bibfnamefont {X.}~\bibnamefont {Xu}}, \bibinfo {author} {\bibfnamefont {Y.}~\bibnamefont {Wu}}, \bibinfo {author} {\bibfnamefont {B.}~\bibnamefont {Sun}}, \bibinfo {author} {\bibfnamefont {Q.}~\bibnamefont {Huang}}, \bibinfo {author} {\bibfnamefont {J.}~\bibnamefont {Cheng}}, \bibinfo {author} {\bibfnamefont {D.~G.}\ \bibnamefont {Steel}}, \bibinfo {author} {\bibfnamefont {A.~S.}\ \bibnamefont {Bracker}}, \bibinfo {author} {\bibfnamefont {D.}~\bibnamefont {Gammon}}, \bibinfo {author} {\bibfnamefont {C.}~\bibnamefont {Emary}},\ and\ \bibinfo {author} {\bibfnamefont {L.~J.}\ \bibnamefont {Sham}},\ }\bibfield  {title} {\bibinfo {title} {Fast {{Spin State Initialization}} in a {{Singly Charged InAs-GaAs Quantum Dot}} by {{Optical Cooling}}},\ }\href@noop {} {\bibfield  {journal} {\bibinfo  {journal} {Physical Review Letters}\ }\textbf {\bibinfo {volume} {99}} (\bibinfo {year} {2007})}\BibitemShut {NoStop}%
\bibitem [{\citenamefont {Bodey}\ \emph {et~al.}(2019)\citenamefont {Bodey}, \citenamefont {Stockill}, \citenamefont {Denning}, \citenamefont {Gangloff}, \citenamefont {{Ethier-Majcher}}, \citenamefont {Jackson}, \citenamefont {Clarke}, \citenamefont {Hugues}, \citenamefont {Gall},\ and\ \citenamefont {Atature}}]{bodeyOpticalSpinLocking2019}%
  \BibitemOpen
  \bibfield  {author} {\bibinfo {author} {\bibfnamefont {J.~H.}\ \bibnamefont {Bodey}}, \bibinfo {author} {\bibfnamefont {R.}~\bibnamefont {Stockill}}, \bibinfo {author} {\bibfnamefont {E.~V.}\ \bibnamefont {Denning}}, \bibinfo {author} {\bibfnamefont {D.~A.}\ \bibnamefont {Gangloff}}, \bibinfo {author} {\bibfnamefont {G.}~\bibnamefont {{Ethier-Majcher}}}, \bibinfo {author} {\bibfnamefont {D.~M.}\ \bibnamefont {Jackson}}, \bibinfo {author} {\bibfnamefont {E.}~\bibnamefont {Clarke}}, \bibinfo {author} {\bibfnamefont {M.}~\bibnamefont {Hugues}}, \bibinfo {author} {\bibfnamefont {C.~L.}\ \bibnamefont {Gall}},\ and\ \bibinfo {author} {\bibfnamefont {M.}~\bibnamefont {Atature}},\ }\bibfield  {title} {\bibinfo {title} {Optical spin locking of a solid-state qubit},\ }\href {https://doi.org/10.1038/s41534-019-0206-3} {\bibfield  {journal} {\bibinfo  {journal} {npj Quantum Inf}\ }\textbf {\bibinfo {volume} {5}},\ \bibinfo {pages} {95} (\bibinfo {year} {2019})},\ \Eprint {https://arxiv.org/abs/1906.00427}
  {arxiv:1906.00427} \BibitemShut {NoStop}%
\bibitem [{\citenamefont {Press}\ \emph {et~al.}(2008)\citenamefont {Press}, \citenamefont {Ladd}, \citenamefont {Zhang},\ and\ \citenamefont {Yamamoto}}]{pressCompleteQuantumControl2008}%
  \BibitemOpen
  \bibfield  {author} {\bibinfo {author} {\bibfnamefont {D.}~\bibnamefont {Press}}, \bibinfo {author} {\bibfnamefont {T.~D.}\ \bibnamefont {Ladd}}, \bibinfo {author} {\bibfnamefont {B.}~\bibnamefont {Zhang}},\ and\ \bibinfo {author} {\bibfnamefont {Y.}~\bibnamefont {Yamamoto}},\ }\bibfield  {title} {\bibinfo {title} {Complete quantum control of a single quantum dot spin using ultrafast optical pulses},\ }\href {https://doi.org/10.1038/nature07530} {\bibfield  {journal} {\bibinfo  {journal} {Nature}\ }\textbf {\bibinfo {volume} {456}},\ \bibinfo {pages} {218} (\bibinfo {year} {2008})}\BibitemShut {NoStop}%
\bibitem [{\citenamefont {Cywi{\'n}ski}\ \emph {et~al.}(2008)\citenamefont {Cywi{\'n}ski}, \citenamefont {Lutchyn}, \citenamefont {Nave},\ and\ \citenamefont {Das~Sarma}}]{cywinskiHowEnhanceDephasing2008}%
  \BibitemOpen
  \bibfield  {author} {\bibinfo {author} {\bibfnamefont {{\L}.}~\bibnamefont {Cywi{\'n}ski}}, \bibinfo {author} {\bibfnamefont {R.~M.}\ \bibnamefont {Lutchyn}}, \bibinfo {author} {\bibfnamefont {C.~P.}\ \bibnamefont {Nave}},\ and\ \bibinfo {author} {\bibfnamefont {S.}~\bibnamefont {Das~Sarma}},\ }\bibfield  {title} {\bibinfo {title} {How to enhance dephasing time in superconducting qubits},\ }\href {https://doi.org/10.1103/PhysRevB.77.174509} {\bibfield  {journal} {\bibinfo  {journal} {Phys. Rev. B}\ }\textbf {\bibinfo {volume} {77}},\ \bibinfo {pages} {174509} (\bibinfo {year} {2008})}\BibitemShut {NoStop}%
\bibitem [{\citenamefont {Jackson}\ \emph {et~al.}(2022)\citenamefont {Jackson}, \citenamefont {Haeusler}, \citenamefont {Zaporski}, \citenamefont {Bodey}, \citenamefont {Shofer}, \citenamefont {Clarke}, \citenamefont {Hugues}, \citenamefont {Atat{\"u}re}, \citenamefont {Le~Gall},\ and\ \citenamefont {Gangloff}}]{jacksonOptimalPurificationSpin2022}%
  \BibitemOpen
  \bibfield  {author} {\bibinfo {author} {\bibfnamefont {D.~M.}\ \bibnamefont {Jackson}}, \bibinfo {author} {\bibfnamefont {U.}~\bibnamefont {Haeusler}}, \bibinfo {author} {\bibfnamefont {L.}~\bibnamefont {Zaporski}}, \bibinfo {author} {\bibfnamefont {J.~H.}\ \bibnamefont {Bodey}}, \bibinfo {author} {\bibfnamefont {N.}~\bibnamefont {Shofer}}, \bibinfo {author} {\bibfnamefont {E.}~\bibnamefont {Clarke}}, \bibinfo {author} {\bibfnamefont {M.}~\bibnamefont {Hugues}}, \bibinfo {author} {\bibfnamefont {M.}~\bibnamefont {Atat{\"u}re}}, \bibinfo {author} {\bibfnamefont {C.}~\bibnamefont {Le~Gall}},\ and\ \bibinfo {author} {\bibfnamefont {D.~A.}\ \bibnamefont {Gangloff}},\ }\bibfield  {title} {\bibinfo {title} {Optimal {{Purification}} of a {{Spin Ensemble}} by {{Quantum-Algorithmic Feedback}}},\ }\href {https://doi.org/10.1103/PhysRevX.12.031014} {\bibfield  {journal} {\bibinfo  {journal} {Phys. Rev. X}\ }\textbf {\bibinfo {volume} {12}},\ \bibinfo {pages} {031014} (\bibinfo {year} {2022})}\BibitemShut {NoStop}%
\bibitem [{\citenamefont {Malinowski}\ \emph {et~al.}(2017)\citenamefont {Malinowski}, \citenamefont {Martins}, \citenamefont {Nissen}, \citenamefont {Barnes}, \citenamefont {Cywi{\'n}ski}, \citenamefont {Rudner}, \citenamefont {Fallahi}, \citenamefont {Gardner}, \citenamefont {Manfra}, \citenamefont {Marcus},\ and\ \citenamefont {Kuemmeth}}]{malinowskiNotchFilteringNuclear2017}%
  \BibitemOpen
  \bibfield  {author} {\bibinfo {author} {\bibfnamefont {F.~K.}\ \bibnamefont {Malinowski}}, \bibinfo {author} {\bibfnamefont {F.}~\bibnamefont {Martins}}, \bibinfo {author} {\bibfnamefont {P.~D.}\ \bibnamefont {Nissen}}, \bibinfo {author} {\bibfnamefont {E.}~\bibnamefont {Barnes}}, \bibinfo {author} {\bibfnamefont {{\L}.}~\bibnamefont {Cywi{\'n}ski}}, \bibinfo {author} {\bibfnamefont {M.~S.}\ \bibnamefont {Rudner}}, \bibinfo {author} {\bibfnamefont {S.}~\bibnamefont {Fallahi}}, \bibinfo {author} {\bibfnamefont {G.~C.}\ \bibnamefont {Gardner}}, \bibinfo {author} {\bibfnamefont {M.~J.}\ \bibnamefont {Manfra}}, \bibinfo {author} {\bibfnamefont {C.~M.}\ \bibnamefont {Marcus}},\ and\ \bibinfo {author} {\bibfnamefont {F.}~\bibnamefont {Kuemmeth}},\ }\bibfield  {title} {\bibinfo {title} {Notch filtering the nuclear environment of a spin qubit},\ }\href {https://doi.org/10.1038/nnano.2016.170} {\bibfield  {journal} {\bibinfo  {journal} {Nature Nanotech}\ }\textbf {\bibinfo {volume} {12}},\ \bibinfo {pages} {16}
  (\bibinfo {year} {2017})}\BibitemShut {NoStop}%
\bibitem [{\citenamefont {Stockill}\ \emph {et~al.}(2016)\citenamefont {Stockill}, \citenamefont {Le~Gall}, \citenamefont {Matthiesen}, \citenamefont {Huthmacher}, \citenamefont {Clarke}, \citenamefont {Hugues},\ and\ \citenamefont {Atat{\"u}re}}]{stockillQuantumDotSpin2016}%
  \BibitemOpen
  \bibfield  {author} {\bibinfo {author} {\bibfnamefont {R.}~\bibnamefont {Stockill}}, \bibinfo {author} {\bibfnamefont {C.}~\bibnamefont {Le~Gall}}, \bibinfo {author} {\bibfnamefont {C.}~\bibnamefont {Matthiesen}}, \bibinfo {author} {\bibfnamefont {L.}~\bibnamefont {Huthmacher}}, \bibinfo {author} {\bibfnamefont {E.}~\bibnamefont {Clarke}}, \bibinfo {author} {\bibfnamefont {M.}~\bibnamefont {Hugues}},\ and\ \bibinfo {author} {\bibfnamefont {M.}~\bibnamefont {Atat{\"u}re}},\ }\bibfield  {title} {\bibinfo {title} {Quantum dot spin coherence governed by a strained nuclear environment},\ }\href {https://doi.org/10.1038/ncomms12745} {\bibfield  {journal} {\bibinfo  {journal} {Nat Commun}\ }\textbf {\bibinfo {volume} {7}},\ \bibinfo {pages} {12745} (\bibinfo {year} {2016})}\BibitemShut {NoStop}%
\bibitem [{\citenamefont {Steck}(2019)}]{steckQuantumAtomOptics2019}%
  \BibitemOpen
  \bibfield  {author} {\bibinfo {author} {\bibfnamefont {D.}~\bibnamefont {Steck}},\ }\href@noop {} {\emph {\bibinfo {title} {Quantum and {{Atom Optics}}}}},\ Available Online at {{http://steck.us.teaching}}\ (\bibinfo  {publisher} {{revision 0.12.6}},\ \bibinfo {year} {2019})\BibitemShut {NoStop}%
\bibitem [{\citenamefont {Johansson}\ \emph {et~al.}(2013)\citenamefont {Johansson}, \citenamefont {Nation},\ and\ \citenamefont {Nori}}]{johanssonQuTiPPythonFramework2013}%
  \BibitemOpen
  \bibfield  {author} {\bibinfo {author} {\bibfnamefont {J.}~\bibnamefont {Johansson}}, \bibinfo {author} {\bibfnamefont {P.}~\bibnamefont {Nation}},\ and\ \bibinfo {author} {\bibfnamefont {F.}~\bibnamefont {Nori}},\ }\bibfield  {title} {\bibinfo {title} {{{QuTiP}} 2: {{A Python}} framework for the dynamics of open quantum systems},\ }\href {https://doi.org/10.1016/j.cpc.2012.11.019} {\bibfield  {journal} {\bibinfo  {journal} {Computer Physics Communications}\ }\textbf {\bibinfo {volume} {184}},\ \bibinfo {pages} {1234} (\bibinfo {year} {2013})}\BibitemShut {NoStop}%
\end{thebibliography}%

\end{document}


\preprint{APS/123-QED}

\title{Supplemental Material for: Tuning the coherent interaction of an electron qubit and a nuclear magnon}

\author{Noah Shofer}
\affiliation{\AffCam}
\author{Leon Zaporski}
\affiliation{\AffCam}
\author{Martin Hayhurst Appel}
\affiliation{\AffCam}
\author{Santanu Manna}
\email[Current address: ]{Department of Electrical Engineering, Indian Institute of Technology Delhi, New Delhi, India}
\affiliation{\AffLinz}
\author{Saimon Covre da Silva}
\email[Current address: ]{Instituto de F\'{i}sica Gleb Wataghin, Universidade Estadual de Campinas, Campinas, Brazil}
\affiliation{\AffLinz}
\author{Alexander Ghorbal}
\affiliation{\AffCam}
\author{Urs Haeusler}
\affiliation{\AffCam}
\author{Armando Rastelli}
\affiliation{\AffLinz}
\author{Claire Le Gall}
\affiliation{\AffCam}
\author{Micha\l{} Gawe\l{}czyk}
\affiliation{Institute of Theoretical Physics, Wroc\l{}aw University of Science and Technology, Wroc\l{}aw, Poland}
\author{Mete Atatüre}
\email[Corresponding author: ]{ma424@cam.ac.uk}
\affiliation{\AffCam}
\author{Dorian A. Gangloff}
\email[Corresponding author: ]{dag50@cam.ac.uk}
\affiliation{\AffCam}

\maketitle

\tableofcontents

\section{Quantum dot devices}
\begin{figure*}
    \includegraphics[scale=0.9]{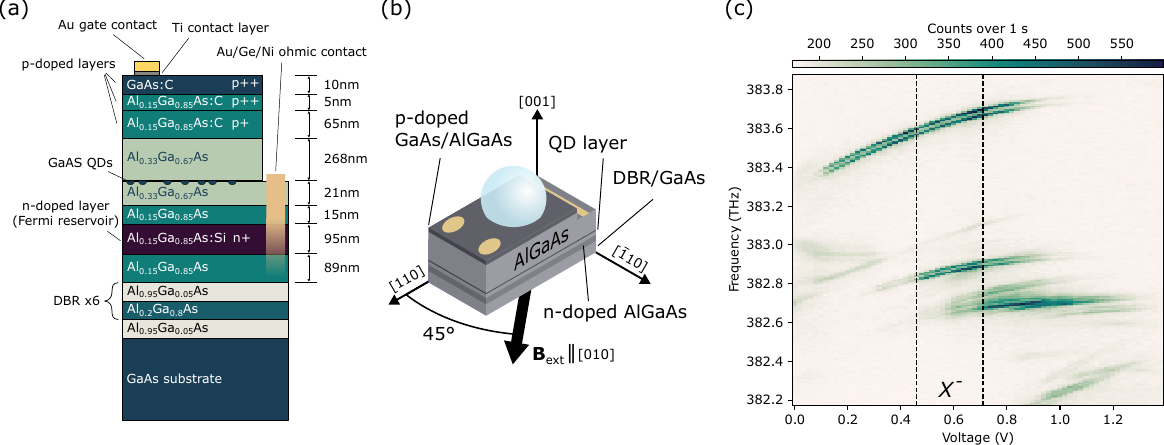}
    \caption{(a) The heterostructure that comprises the QD device used in this work. (b) A depiction of the sample containing the QD. A cubic zirconia superhemispherical solid immersion lens (SIL) is mounted on top of the sample to improve excitation and collection efficiency. A 6T magnetic field $B_{\mathrm{ext}}$ is applied perpendicularly to the [001] axis (growth axis) of the QD, at $45\degree$ relative to the cleaved [110] crystal axis of the wafer (i.e. along the the [010] axis). (c) A detail of the photoluminescence spectrum of the studied QD as a function of the gate voltage applied to the \textit{pin} diode. The gate voltage range that corresponds to the negatively charged exciton (centered at 382.9 THz) is highlighted with dashed lines. This PL spectrum was measured using an $\sim\! \SI{650}{\nano\meter}$ excitation laser.}
    \label{fig:SI_diode}
\end{figure*}

The GaAs/AlGaAs heterostructure (Fig.~\ref{fig:SI_diode}(a)) containing the QDs is grown via molecular beam epitaxy (MBE), in a manner similar to that used in Ref.~\cite{zaporskiIdealRefocusingOptically2023}. The GaAs quantum dots used in this work are grown by infilling nanoholes etched in an AlGaAs barrier \cite{dasilvaGaAsQuantumDots2021}. The QD-containing layer is sandwiched between a heavily $\textit{n}$-doped and a heavily $p$-doped layer, which enables a \textit{pin} diode to be fabricated around the QDs. The diode structure enables single electrons to be deterministically loaded into the QD via Coulomb blockade by applying a voltage to the contacts of the diode \cite{zhaiLownoiseGaAsQuantum2020,schimpfEntanglementbasedQuantumKey2021,zaporskiIdealRefocusingOptically2023}. The voltage-dependent photoluminescence of the QD is shown in Fig.~\ref{fig:SI_diode}(c). The negatively charged exciton $X^{-}$ used in this work can be seen centered at roughly 382.9 THz, along with other voltage-dependent charge states. A solid immersion lens (SIL) mounted on top of the sample enhances the collection efficiency of photons emitted from the QD, as well as the power delivered to the QD by the laser. The sample is held so that the $[110]$ and $[1\bar{1}0]$ crystal axes of the sample are at $45\degree$ relative to the external magnetic field (Fig.~\ref{fig:SI_diode}(b)).\par


\subsection{Non-collinear hyperfine interaction: quadrupolar effects}
\begin{figure*}
    \includegraphics[scale=0.95]{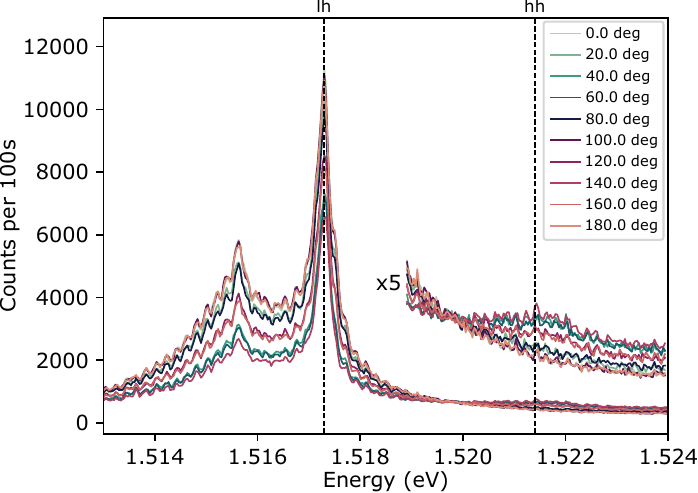}
    \caption{Photoluminescence measurement of the free exciton spectrum in bulk GaAs. This data was measured in the bulk GaAs part of the wafer in the vicinity of the QD used in this work. The legend indicates the angle of a half-wave plate placed before a linear polarizer in front of the collection fiber going to the spectrometer. The partially polarized lh free exciton peak is at 1.517293(1) eV, while the linearly-polarized hh peak is at 1.5214(1) eV. The counts in the vicinity of the hh peak are additionally plotted at 5x scale for clarity.}
    \label{fig:SI_strain}
\end{figure*}

Fig.~\ref{fig:SI_strain} shows a photoluminescence measurement of the free exciton in bulk GaAs measured on the QD device used in this work, in the vicinity of the QD. The light hole (lh) exciton peak is seen at 1.517293(1) eV and is partially polarization dependent, while the heavy-hole (hh) electron peak, centered at 1.5214(1) eV, disappears completely for polarizations where the lh peak is most visible (e.g. for half-wave plate angle $0\degree$) and appears at a perpendicular polarization (e.g. for half-wave plate angle $40\degree$). From these peaks, we extract a lh-hh splitting of 4.1(1) meV, which corresponds to a nuclear quadrupolar shift $B_{Q} = 75(2)$ kHz \cite{chekhovichCrossCalibrationDeformation2018}. For ${}^{75}\mathrm{As}$, this nuclear quadrupolar shift would give us a quadrupolar-mediated non-collinear hyperfine $a_{\mathrm{nc}}^{Q}= a B_{Q}/\omega_{n} = 0.5(1)$ kHz \cite{gangloffQuantumInterfaceElectron2019}. This is in contrast to the measured non-collinear hyperfine $a_{\mathrm{nc}} = 58\pm15$ kHz at $\omega_{e}^{0}=3\,\mathrm{GHz}$ (see Knight shift measurements in main text), roughly a factor 100 times stronger than $a_{\mathrm{nc}}^{Q}$. We thus attribute the electron-nuclear spin-flip interactions in this work to the non-collinear hyperfine mediated by a g-factor anisotropy of the electron, $a\sin(\varphi)$.\par

\subsection{Non-collinear hyperfine interaction: electronic g-factor anisotropy}

\begin{figure}[tb]
	\includegraphics[width=0.6\textwidth]{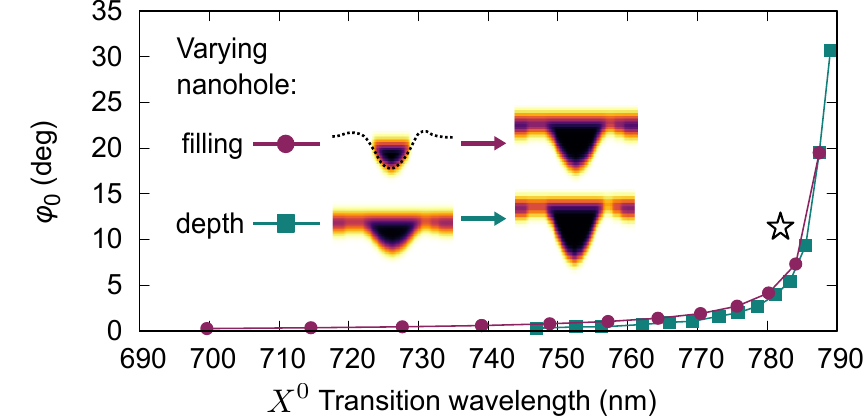}
	\caption{Simulated results for electron $g$-factor anisotropy. Angle $\varphi_0$ between the electron and nuclear spin quantization axes is plotted versus QD emission wavelength for two cases of underlying geometry change - varying nanohole filling level (circles) and varying nanohole depth (squares). Lines are to guide the eye. Plotted versus QD neutral excition ($X^0$) emission wavelength. The star symbol is the value measured in this work.}
	\label{fig:sup:g_anisotropy}
\end{figure}

Figure~\ref{fig:sup:g_anisotropy} shows the angle $\varphi_0$ between the electron and nuclear spin quantization axes for QDs emitting at different wavelengths and subject to a $B$-field along the [010] axis. The variation of wavelength is based on varying QD geometry: level of nano hole filling (dots) and nanohole depth (squares), as pictured in the legend. Both series show similar results with $\varphi_0$ ranging from below $1\degree$ to even above $\sim30\degree$ at longer wavelengths. Further (not shown here) it takes negative values after a discontinuity caused by $g$-factor sign reversal. It needs to be underlined that the highest $\varphi_0$ values at longer wavelengths concide with overall low $g$-factor values. Below we describe the simulations behind these results.

We model the studied GaAs QDs based on atomic force microscopy (AFM) measurements of a nanohole left after droplet etching on a test sample. To simulate series of QDs with different emission wavelength, we either assume different levels of GaAs filling or modify the nanohole profile by shrinking its height. The top GaAs surface is assumed to be flat. We use an axis-wise uniform numerical grid to discretize the material composition profiles. Next, we apply Gaussian averaging with a spatial extent $\sigma = 1.2$~nm to the material composition profile to mimic the normal diffusion of materials at interfaces. The lattice mismatch of the considered materials is very low, yet we still use the theory of continuous elasticity to calculate structural strain. The shear strain-induced piezoelectric polarization is calculated up to second order terms in strain-tensor elements.

For such series of simulated QDs, we calculate the lowest-energy electron eigenstates using the implementation of the multiband $\bm{k} \cdot \bm{p}$ method within the envelope-function approximation \cite{bahderEightbandModelStrained1990} described in Ref.~\citenum{gawareckiElectronStatesDouble2014}. The calculation includes the spin-orbit interaction, strain, and piezoelectric field. The explicit form of the Hamiltonian may be found in Ref.~\citenum{mielnik-pyszczorskiDominantRoleShear2018}, while material parameters used in calculations are given in Ref.~\citenum{gawelczykExcitonLifetimeEmission2017} and references therein. We find the elements of the $g$-tensor by applying low magnetic field $B$ along selected spatial directions $\widehat{\bm{e}}$ and extracting the Zeeman splitting, $g_{\widehat{\bm{e}}}=E_{\mathrm{Z}}/\mu_{\mathrm{B}}B$ for $\bm{B}\parallel\widehat{\bm{e}}$, where $\mu_{\mathrm{B}}$ is the Bohr magneton. We first verify our expectation of [110] and [$\bar{1}10$] being the in-plane $g$-factor principal axes based on the distinguished role of these axes in the zincblende material combined with almost ideal cylindrical symmetry of the QD shape. Next, we extract $g_{[110]}$ and $g_{[\bar{1}10]}$ components for the two series of QD geometries. Finally, we calculate the angle $\varphi_0$ between the electron spin quantization axis and the magnetic field (and thus nuclear spin quantization axis) for $\bm{B}\parallel$[010] as used in our work.

\section{Experimental Setup}
\begin{figure*}
    \includegraphics[scale=0.85]{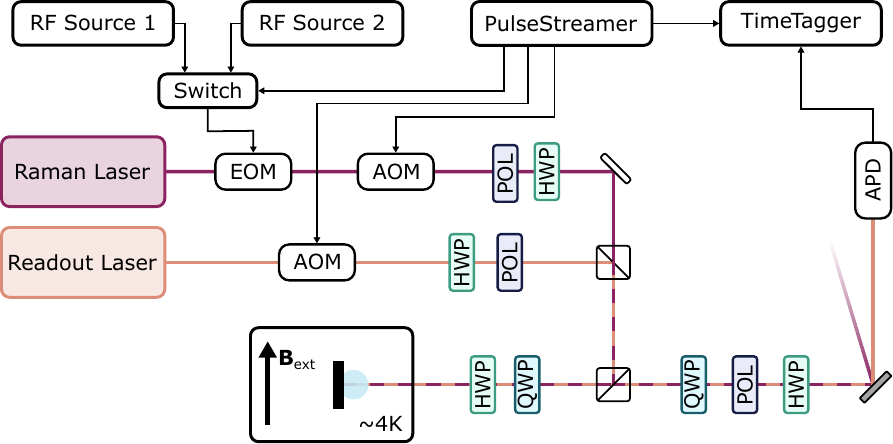}
    \caption{A schematic depiction of the experimental setup used in this work. EOM = electro-optic modulator, AOM = acousto-optic modulator, APD = avalanche photodiode; POL = polarizer, QWP = quarter-wave plate, HWP = half-wave plate.}
    \label{fig:SI_setup}
\end{figure*}

Fig.~\ref{fig:SI_setup} shows a schematic of the cryogenics, microscope, and laser control systems used in this work. The QD device is held at 4K in a closed-cycle cryostat (MyCryo OptiDry 250), while a 6T magnetic field is applied perpendicular to the crystal growth axis, i.e. in Voight geometry, but at $45\degree$ relative to the $[110]$ crystal axis. A 1mm solid immersion lens (SIL) is mounted directly onto the sample. A laser resonant with the $\ket{\downarrow}\rightarrow\ket{\downarrow\uparrow\Downarrow}$ transition provides spin-dependent optical readout and initialization \cite{xuFastSpinState2007, zaporskiIdealRefocusingOptically2023}. Due to the small Zeeman splitting between the $\ket{\uparrow}$ and $\ket{\downarrow}$ states, this laser is linearly polarized to ensure that only the $\ket{\downarrow}\rightarrow\ket{\downarrow\uparrow\Downarrow}$ transition is excited, rather than the cross-polarized $\ket{\downarrow}\rightarrow\ket{\downarrow\uparrow\Uparrow}$ transition. An acousto-optic modulator (AOM) placed in the readout laser beam path provides power stabilization and pulsing capabilities.\par
The spin state of the electron is coherently manipulated via a two-photon Raman scheme. Raman control is performed via a circularly-polarized single-frequency laser detuned from the excited states by $\Delta_{c} = 400\,\mathrm{GHz}$ \cite{bodeyOpticalSpinLocking2019,gangloffQuantumInterfaceElectron2019}. Circular polarization helps to ensure the highest available Rabi rate for a given Raman laser power \cite{pressCompleteQuantumControl2008}. The Raman laser is modulated by an electro-optic modulator (EOM) at a frequency $\omega_{\mathrm{mw}}$, producing a control field with two frequencies $\omega_c \pm \omega_{\mathrm{mw}}$ that drives a two-photon Raman transition between the electron spin states, where $\omega_{c}$ is the frequency of the Raman laser. The EOM drive frequencies are selected such that the condition $\omega_e = 2\omega_{\mathrm{mw}}$ is fulfilled. The EOM is driven by two Rohde \& Schwarz SM-line microwave sources passed through a switch to enable fast frequency switching. The Knight shift measurements were taken with an AWG (Tektronix 70002B), as it enables a wider range range of frequencies in a single experiment. An AOM placed after the EOM provides both power stabilization of the Raman laser and pulsing capability.\par 
We employ a cross-polarization scheme to reject the resonant laser excitation; this means we only collect the the V-polarized scattering from the QD, corresponding to the $\ket{\uparrow\downarrow\Downarrow}\rightarrow\ket{\uparrow}$ transition. The QD emission is deteced with an avalanche photodiode (APD), and time-tagged and histogrammed with a Swabian Timetagger. The Raman laser is filtered off with a grating placed before the APD.\par
Pulse sequences are generated via a Swabian PulseStreamer, which sends TTL control pulses to both the readout and Raman laser AOMs, as well as triggers histogram collection on the SwabianTimeTagger.

\section{System Hamiltonian}
\begin{figure*}
    \includegraphics[scale=0.8]{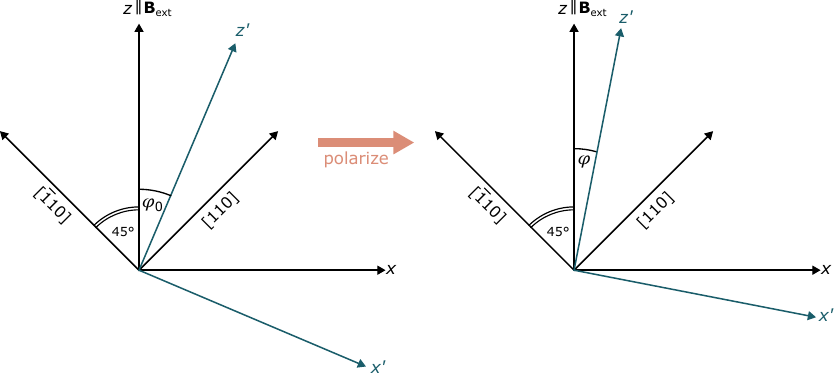}
    \caption{The axes used in the derivation of the Hamiltonian for the electron interacting with the nuclear bath (Eq.~\ref{sec_shiftH}). The $z$-axis, aligned along the external B-field, is the $z$-axis of the lab frame and constitutes the quantization axis of the nuclei. $z^{\prime}$ is the quantization axis of the electron in the QD in the diagonalized (i.e. rotated) frame. As the nuclei are polarized along $z$, $z^{\prime}$ moves closer to the $z$-axis of the lab frame.}
    \label{fig:SI_axes}
\end{figure*}
Consider the axes in Fig.~\ref{fig:SI_axes}, where $z$ is aligned along the external magnetic field, at $45\degree$ to the $[110]$ crystal axis of the wafer. Due to the anisotropy in the electron g-factor, the quantization axis of the electron is tilted from $z$ by an angle:
\begin{equation}
\varphi_{0} = \arctan\left( \frac{g_{110}-g_{\bar{1}10}}{g_{110}+g_{\bar{1}10}}\right).
\end{equation}
We have then the Hamiltonian of the electron quantized along this tilted axis interacting with the nuclear ensemble via an isotropic hyperfine interaction. Note that the nuclei are still quantized along the $z$-axis:
\begin{equation}
\label{basicH}
\hat{\mathcal{H}}= \omega_{e}^{0}\cos(\varphi_{0})\hat{S}_{z} + \omega_{e}^{0}\sin(\varphi_{0})\hat{S}_{x} + \sum_{j}\omega_{n}^{j}\hat{I}_{z}^{j} + \sum_{j}a^{j}\left( \hat{S}_{x}\hat{I}_{x}^{j} + \hat{S}_{y}\hat{I}_{y}^{j} + \hat{S}_{z}\hat{I}_{z}^{j} \right)
\end{equation}
where $j$ is summed over all nuclei, $\omega_{e}^{0}$ is the angular frequency of the electron Zeeman splitting, and $\omega_{n}^{j}$ are the nuclear Larmor frequencies of a given nuclear species $n$. As we are concerned with increasing the mean-field polarization along the $z$-axis, we split the final term of our isotropic hyperfine interaction into two parts:
\begin{equation}
\label{mf_def}
    \sum_{j}a^{j}\hat{S}_{z}\hat{I}_{z}^{j} = \delta_{\mathrm{oh}} +  \sum_{j}a^{j}\hat{S}_{z}\Delta\hat{I}_{z}^{j},
\end{equation}
where $\delta_{\mathrm{oh}}$ is the mean-field Overhauser shift seen by the electron due to the nuclear ensemble and $\Delta\hat{I}_{z}^{j}$ is the residual fluctuation of the nuclear polarization about the mean field. With this identification of the mean field Overhauser shift, we can write the Hamiltonian as:
\begin{equation}
\label{mf_ham}
\hat{\mathcal{H}}= \left( \omega_{e}^{0}\cos(\varphi_{0}) + \delta_{\mathrm{oh}} \right) \hat{S}_{z} + \omega_{e}^{0}\sin(\varphi_{0})\hat{S}_{x} + \sum_{j}\omega_{n}^{j}\hat{I}_{z}^{j} + \sum_{j}a^{j}\left( \hat{S}_{x}\hat{I}_{x}^{j} + \hat{S}_{y}\hat{I}_{y}^{j} + \hat{S}_{z}\Delta\hat{I}_{z}^{j} \right).
\end{equation}
Now we define the mean-field shifted angular frequency $\omega_{e} \equiv \sqrt{\left(\omega_{e}^{0}\cos(\varphi_{0}) + \delta_{\mathrm{oh}} \right)^{2} + \left( \omega_{e}^{0}\sin(\varphi_{0}) \right)^{2}}$ ($\approx \omega_{e}^{0}+\delta_{\mathrm{oh}}$ for small $\varphi_{0}$) and diagonalize the Hamiltonian via the unitary operator $\hat{U} = e^{-i\varphi \hat{S}_{y}}$:
\begin{equation}
\label{shift_ham}
\hat{\mathcal{H}}^{\prime} = \hat{U}^{\dag}\hat{\mathcal{H}}\,\hat{U} = \omega_{e}\hat{S}_{z^{\prime}} + \sum_{j}\omega_{n}^{j}\hat{I}_{z}^{j} + \sum_{j}a^{j}\left[\hat{S}_{x^{\prime}}\!\left( \cos(\varphi)\hat{I}_{x}^{j} - \sin(\varphi)\Delta\hat{I}_{z}^{j} \right) + \hat{S}_{y^{\prime}}\hat{I}_{y}^{j} + \hat{S}_{z^{\prime}}\!\left( \sin(\varphi)\hat{I}_{x}^{j} + \cos(\varphi)\Delta\hat{I}_{z}^{j}   \right)     \right]
\end{equation}
where the primed coordinates are the new principal axes for the electronic Hamiltonian given by $\hat{S}_{x} = \cos(\varphi)\hat{S}_{x^{\prime}} + \sin(\varphi)\hat{S}_{z^{\prime}}$, $\hat{S}_{z} = -\sin(\varphi)\hat{S}_{x^{\prime}} + \cos(\varphi)\hat{S}_{z^{\prime}}$, and $\hat{S}_{y} = \hat{S}_{y^{\prime}}$, and $\varphi$ is defined by:
\begin{equation}
\cos(\varphi) \equiv \frac{\omega_{e}^{0}\cos(\varphi_{0}) + \delta_{\mathrm{oh}}}{\omega_{e}}, \,\, \sin(\varphi) \equiv \frac{\omega_{e}^{0}\sin(\varphi_{0})}{\omega_{e}}.
\end{equation}
We make a secular approximation, noting that the energy scale of the qubit splitting is dominant in this Hamiltonian and suppresses the components of the hyperfine not parallel to $\hat{S}_{z^{\prime}}$:
\begin{equation}
\label{sec_shiftH}
\hat{\mathcal{H}}^{\prime} = \omega_{e}\hat{S}_{z^{\prime}} + \sum_{j}\omega_{n}^{j}\hat{I}_{z}^{j} + \sum_{j}a^{j}\hat{S}_{z^{\prime}}\!\left( \sin(\varphi)\hat{I}_{x}^{j} + \cos(\varphi)\Delta\hat{I}_{z}^{j}   \right)
\end{equation}
Finally, we identify the collinear $\left( a^{j}\cos(\varphi)\hat{S}_{z^{\prime}}\Delta\hat{I}_{z} \right)$ and spin-flip enabling non-collinear $\left( a^{j}\sin(\varphi)\hat{S}_{z^{\prime}}\hat{I}_{x} \right)$ hyperfine interactions in the final term of Eq.~\ref{sec_shiftH}. 

\subsection{Modeling the magnon Rabi rate}
We now transform Eq.~\ref{sec_shiftH} into the frame of the Raman drive laser rotating at drive frequency $\omega_{d}$ and define the collective nuclear operators $\hat{I}_{\alpha}\equiv \sum_{j}\hat{I}_{\alpha}^{j}$ (for $\alpha = x,y,z$) as in the main text, and can write:
\begin{equation}
    \hat{\mathcal{H}}^{\prime} = \delta\hat{S}_{z^{\prime}} + \Omega\hat{S}_{x^{\prime}} + \omega_{n}\hat{I}_{z} + a\hat{S}_{z^{\prime}}\!\left( \sin(\varphi)\hat{I}_{x} + \cos(\varphi)\Delta\hat{I}_{z}   \right),
\end{equation}
where $\Omega$ is the Rabi rate of the Raman drive and $\delta = \omega_{d}-\omega_{e}$ is the two-photon detuning of the Raman drive. We now transform to the dressed states of the electron, using the unitary transformation:
\begin{equation}
    \hat{U} = \begin{pmatrix}
        \sin\theta & \cos\theta \\
        \cos\theta & -\sin\theta
    \end{pmatrix}_{\mathrm{electron}} \otimes \mathbb{I}_{\mathrm{nuc}},
\end{equation}
where we have defined the St{\"u}ckelberg angle $\tan(2\theta) = -\Omega/\delta$. The dressed state Hamiltonian is then:
\begin{equation}
   \tilde{\mathcal{H}} = \Omega^{\prime}\tilde{S}_{z} + \omega_{n}\hat{I}_{z} - \left(1+\frac{\Omega^{2}}{\delta^{2}}\right)^{-\frac{1}{2}}\left(\tilde{S}_{z}+\frac{\Omega}{\delta}\tilde{S}_{x}\right)\left( a\hat{I}_{z}+\frac{a_{\mathrm{nc}}}{2}(\hat{I}_{+} + \hat{I}_{-} )  \right),
\end{equation}
where $\Omega^{\prime} = \sqrt{\delta^{2}+\Omega^{2}}$ is the effective Rabi frequency. To calculate the magnon activation rate $\Omega_{\mathrm{mag}}$, we must calculate the matrix element for an electro-nuclear swap:
\begin{equation}
    \frac{\Omega_{\mathrm{mag},+}}{2} = \left|\bra{I,M+1,\tilde{\downarrow}}\tilde{\mathcal{H}}\ket{I,M,\tilde{\uparrow}}  \right| = \left(1+\frac{\Omega^{2}}{\delta^{2}}\right)^{-\frac{1}{2}} \frac{a_{\mathrm{nc}}\Omega}{4\delta}\bra{I,M+1}\hat{I}_{+}\ket{I,M},
\end{equation}
where $I$ is the total nuclear spin and $M$ is the angular momentum for the addressed nuclear species, and where the collective enhancement is given by:
\begin{equation}
    \bra{I,M+1}\hat{I}_{+}\ket{I,M}=\sqrt{I(I+1)-M(M+1)}.
\end{equation}
For the case of a detuned Raman drive resonant with the magnon sideband, we have $\delta\approx\omega_{n}\gg\Omega$ and can write the magnon injection rate as:
\begin{equation}
    \Omega_{\mathrm{mag}} = \frac{a_{\mathrm{nc}}\Omega}{2\omega_{n}}\sqrt{I(I+1)-M(M+1)}.
\end{equation}
For the high number of experimental repetition rates performed to generate the histogram of the data, the experimentally measured quantity is the root-mean-square (RMS) value of the magnon Rabi rate:
\begin{equation}
\label{eq:rms_Om}
    \sqrt{\langle \Omega_{\mathrm{mag}}^{2}\rangle} =  \frac{a_{\mathrm{nc}}\Omega}{2\omega_{n}} \sqrt{\langle I(I+1)-M(M+1)\rangle}.
\end{equation}
To calculate the expectation value of the collective enhancement, first note that for the $N\sim 10^{5}$ nuclei that comprise the nuclear ensemble the total spin $I$ and angular momentum $M$ are both much greater than unity,
so that $\langle I(I+1)-M(M+1)\rangle\approx \langle I^{2}-M^{2}\rangle$. We then need to calculate the probability distribution function for a given spin and angular momentum state. We assume that the values of the spins along each axis $I_{\alpha} = \sum_{j}I_{\alpha}^{j}$ (for $\alpha = x,y,z$) are Gaussian distributions with a variance given by $\sigma^{2} = (1/3)NI(I+1)=5N/4$, where $N$ is the number of nuclear spins of the nuclear species in question. The total spin state $I=\sqrt{I_{x}^{2}+I_{y}^{2}+I_{z}^{2}}$ then has a probability distribution function given by the $\chi^{(3)}$ distribution:
\begin{equation}
    p(I) = I^{2}\exp\left(-\frac{1}{2}\left( \frac{I}{\sigma}\right)^{2}\right) = I^{2}\exp\left(-\frac{2I^{2}}{5N^{2}}\right).
\end{equation}
The probability distribution function for a given state with spin $I$ and angular momentum $M$ is then:
\begin{equation}
    p(I,M) = \frac{p(I)}{2I}
\end{equation}
where at thermal equilibrium the angular momenta are equally distributed along $-I$ to $I$. The expectation value for the collective enhancement is then given by:
\begin{align}
    \langle I^{2}-M^{2}\rangle &= \frac{\int_{0}^{\infty}dI\int_{-I}^{I}dM(I^{2}-M^{2})p(I,M)}{\int_{0}^{\infty}dI\int_{-I}^{I}dMp(I,M)}\nonumber\\
    &= \frac{5N}{2}.
\end{align}
Plugging this value back into Eq.~\ref{eq:rms_Om}, we arrive at the final value for the magnon Rabi rate:
\begin{equation}
\label{eq:Omag_SI}
    \sqrt{\langle \Omega_{\mathrm{mag}}^{2}\rangle} =  \frac{a_{\mathrm{nc}}\Omega}{2\omega_{n}}\sqrt{\frac{5N}{2}}.
\end{equation}
Note that in the main text the RMS value is denoted as $\Omega_{\mathrm{mag}} = \sqrt{\langle \Omega_{\mathrm{mag}}^{2}\rangle}$.



\section{Conversion from counts to spin $\ket{\downarrow}$ population}\label{sec:count_conv}
The magnon sideband spectra shown in the main text in Fig.~1 and Fig.~4 are displayed in units of spin $\ket{\downarrow}$ population. We convert from recorded counts in the histogram to spin $\ket{\downarrow}$ population by measuring the counts in a Rabi oscillation experiment. The peak-to-peak amplitude, in counts, of an exponentially damped sine function fit to the recorded Rabi oscillations gives us the maximum population of the $\ket{\downarrow}$ state for a given histogram duration and number of repetitions. This allows us to convert the recorded counts of a magnon sideband spectrum to the $\ket{\downarrow}$ population.\par


\section{Filter function model}

We now turn to the visibility of the Carr-Purcell (CP) sequence datasets. The measured visibility of a spin qubit in a dynamical decoupling (DD) sequence is given by \cite{cywinskiHowEnhanceDephasing2008}:
\begin{equation}
W(t) = \mathrm{exp}\left( -\int_{0}^{\infty} \frac{d\omega}{2\pi} S(\omega) \frac{\mathcal{F}(\omega t)}{\omega^{2}}  \right)
\end{equation}
where $\mathcal{F}(\omega t)$ is the filter function of the DD sequence and $S(\omega)$ is the power spectral density experienced by the electron, in this case due to the nuclear ensemble. We can calculate the noise spectrum via the Wiener-Khinchin theorem:
\begin{equation}
\label{eq:noise_spec}
S(\omega) = \int_{-\infty}^{\infty} dt \, e^{i\omega t} \sum_{i,j} a_{\mathrm{nc}}^{i} a_{\mathrm{nc}}^{j} \langle \hat{I}_{x}^{i}(t)\hat{I}_{x}^{j}(0)\rangle
\end{equation}
where $i,j$ is a sum over all nuclear spins and $a_{\mathrm{nc}}^{i} = a^{i}\sin(\varphi)$ is the non-collinear hyperfine coupling constant. We will assume from here on out that we are dealing with species-specific macrospins - this is mathematically equivalent to summing over individual nuclei, but conceptually simpler to deal with. Here we have chosen to ignore three-body interaction effects, as well as longitudinal polarization noise, as we are concerned with the coherence dips and revivals at short time scales, where the non-collinear hyperfine interaction term dominates. To determine the noise spectrum, we now need to calculate the two point nuclear correlators for $\hat{I}_{x}$:
\begin{equation}
    \langle \hat{I}_{x}^{i}(t)\hat{I}_{x}^{j}(0)\rangle = \mathrm{Tr}\left( \hat{\rho}_{n} V(t)\hat{I}_{x}^{i}V^{\dag}(t)\hat{I}_{x}^{j}\right) 
\end{equation}
where $\hat{\rho}_{n}$ is the nuclear density operator following the algorithmic Ramsey cooling and dragging of the mean-field polarization, and:
\begin{equation}
    V(t) = \exp\left(-i\sum_{j}\omega_{n}^{j}\hat{I}_{x}^{j}t\right).
\end{equation}
Note that the quadrupolar part of nuclear Hamiltonian is ignored, as this contributes to an overall decay in coherence on longer time scales, whereas here we are interested in the initial collapses and revivals of coherence caused by the non-collinear coupling. It can be verified that:
\begin{equation}
    V(t)\hat{I}_{x}^{j}V^{\dag}(t) = \frac{1}{2}\left(e^{+i\omega_{n}^{j}t}\hat{I}_{+}^{j} + e^{-i\omega_{n}^{j}}\hat{I}_{-}^{j}\right) 
\end{equation}
To calculate the nuclear correlator, we thus need to determine some of the features of the nuclear density operator $\hat{\rho}_{n}$ for our nuclear ensemble.\par

\begin{figure*}
    \includegraphics[scale = 0.8]{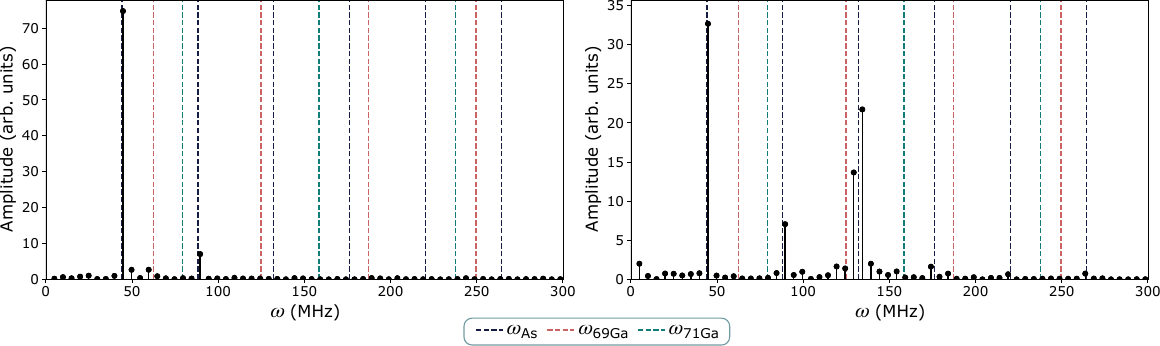}
    \caption{Fourier transform of the CP1 data (left panel) and CP2 data(right panel). Dotted lines highlight the Larmor frequencies of the nuclear species, as well as the higher-order harmonics for each Larmor frequency.}
    \label{fig:SI_FFT}
\end{figure*}

For a thermal nuclear ensemble (i.e. considering the nuclear ensemble to be at infinite temperature and fully uncorrelated), we can take $\hat{\rho}_{n} \propto \mathds{1}$. However, in this case we cannot do that formally, as the algorithmic narrowing significantly reduces the fluctuations of longitudinal nuclear polarization $\hat{I}_{z}$. Jackson et al. \cite{jacksonOptimalPurificationSpin2022} calculated $\hat{\rho}_{n}$ at zero net polarization for a single species by comparing simulation to experimental data, but here we have a non-zero mean-field polarization and are not dragging a single species selectively. Rather than calculate $\hat{\rho}_{n}$ explicitly, we will estimate some features of $\rho_{n}$ based on the experimentally measured values of $W(t)$. Fig.~\ref{fig:SI_FFT} shows the Fourier transform of the measured $W(t)$, for both CP1 and CP2 datasets. In this data, we primarily observe strong peaks at the nuclear Larmor frequency of ${}^{75}\mathrm{As}$, as well as weaker peaks at the Larmor frequencies corresponding to ${}^{69}\mathrm{Ga}$ and ${}^{71}\mathrm{Ga}$. However, the lack of frequency components $\left| \omega_{n}^{i} \pm \omega_{n}^{j} \right|$ point to an absence of correlations between transverse noise components of distinct nuclear species $i$ and $j$. Further, we have no reason to suppose that nuclear narrowing or dragging of the mean-field results in a build-up of significant net transverse polarization of any of the species. We can then write:
\begin{equation}
\label{eq:2pt_corr}
    \langle \hat{I}_{x}^{i}(t)\hat{I}_{x}^{j}(0)\rangle = \frac{\delta_{ij}}{4}\left( \langle\hat{I}_{+}^{j}\hat{I}_{-}^{j}\rangle e^{+i\omega_{n}^{j} t} + \langle\hat{I}_{-}^{j}\hat{I}_{+}^{j}\rangle e^{-i\omega_{n}^{j} t} \right) 
\end{equation}
Expanding the raising and lowering operators, we get:
\begin{equation}
    \hat{I}_{\pm}^{j}\hat{I}_{\mp}^{j} = (\hat{I}_{x}^{j})^{2} + (\hat{I}_{y}^{j})^{2} \pm \hat{I}_{z}^{j} 
\end{equation}
which, when plugged into Eq.~\ref{eq:2pt_corr}, gives:
\begin{equation}
    \langle \hat{I}_{x}^{i}(t)\hat{I}_{x}^{j}(0)\rangle = \frac{\delta_{ij}}{2}\left[ \cos(\omega_{n}^{j}t)\langle(\hat{I}_{x}^{j})^{2} + ( \hat{I}_{y}^{j})^{2}\rangle + i\sin(\omega_{n}^{j}t)\langle\hat{I}_{z}^{j}\rangle  \right].
\end{equation}\par
Gathering everything back together in the expression for the nuclear noise spectrum (Eq.~\ref{eq:noise_spec}), we have:
\begin{equation}
    \label{eq:n_spec_d}
    S(\omega>0) = \sum_{j}(a_{\mathrm{nc}}^{j})^{2}\frac{\pi}{2}\left[\langle(\hat{I}_{x}^{j})^{2} + ( \hat{I}_{y}^{j})^{2}\rangle + \langle\hat{I}_{z}^{j}\rangle  \right]\delta(\omega - \omega_{n}^{j}).
\end{equation}

\begin{figure*}
    \includegraphics[scale = 0.9]{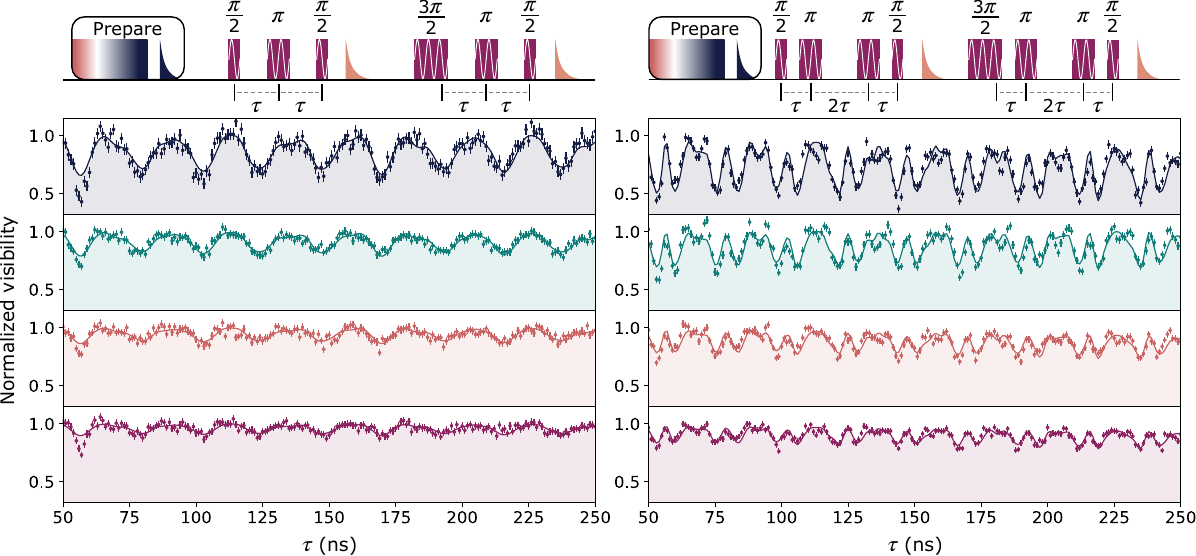}
    \caption{Rescaled CP1 data (left panel) and CP2 data(right panel), along with fits to the model described in Eq.~\ref{eq:W_fits}. A schematic of the pulse sequences is shown above each dataset.}
    \label{fig:SI_fullCP}
\end{figure*}

From this expression, we can see that the nuclear noise spectrum consists of three delta functions at the nuclear Larmor frequencies with strengths determined by the non-collinear hyperfine constants and the expectation values of the square of the transverse angular momentum $\langle(\hat{I}_{x}^{j})^{2} + ( \hat{I}_{y}^{j})^{2}\rangle$ and the polarization $\langle\hat{I}_{z}^{j}\rangle$ of each species. The visibility function is then:
\begin{equation}
    W(t)=\exp\left( -\sum_{j}\frac{(a_{\mathrm{nc}}^{j})^{2}}{4(\omega_{n}^{j})^{2}} \left[\langle(\hat{I}_{x}^{j})^{2} + ( \hat{I}_{y}^{j})^{2}\rangle + \langle\hat{I}_{z}^{j}\rangle  \right] \mathcal{F}(\omega_{n}^{j}t)      \right).
\end{equation}\par
We now want to calculate the expectation values of the nuclear operators. To do so, we will assume that in narrowing the nuclear ensemble we are only significantly reducing the fluctuation of $\sum_{j} a^{j}\hat{I}_{z}^{j}$, rather than the individual $\langle \Delta^{2}\hat{I}_{z}^{j}\rangle$, and thus the expectation values involving operators for a single species are still very close to thermal. For the expectation value of the transverse angular momentum, this assumption would give:
\begin{equation}
    \langle(\hat{I}_{x}^{j})^{2} + ( \hat{I}_{y}^{j})^{2}\rangle \approx \frac{2}{3}\langle (\mathbf{\hat{I}}^{j})^{2}\rangle = N_{j} \times \frac{5}{2} 
\end{equation}
where $N_{j}$ is the number of nuclei of species $j$. For dragging of the mean field nuclear polarization covered in this paper (up to a few GHz $\delta_{oh}$), $\langle\hat{I}_{z}^{j}\rangle \ll \frac{2}{3}\langle (\mathbf{\hat{I}}^{j})^{2}\rangle$ (by roughly a factor 100), and can be neglected.\par
Finally, recalling the definition of the per-nucleus hyperfine constant:
\begin{equation}
    a_{\mathrm{nc}}^{j}= \frac{\sin(\varphi)c_{j}A^{j}}{N_{j}} 
\end{equation}
for nuclear concentrations $c_{75\mathrm{As}}=1$, $c_{69\mathrm{Ga}}=0.604$, $c_{71\mathrm{Ga}}=0.396$, we have:
\begin{equation}
    \label{eq:W_final}
    W(t) = \mathrm{exp} \left[ -\sum_{j}\frac{5}{4}\frac{1}{N}\left(\frac{A^{j}\sin(\varphi)\sqrt{c_{j}}}{\omega_{n}^{j}}\right)^{2} \mathcal{F}
    (\omega_{n}^{j}t)    \right].
\end{equation}
Finally, the filter functions for CP1 and CP2 are given by \cite{cywinskiHowEnhanceDephasing2008,malinowskiNotchFilteringNuclear2017}:
\begin{equation}
    \mathcal{F}_{\mathrm{CP1}}(\omega t) = 8\sin^{4}\left( \frac{\pi\omega t}{2} \right),\,\, \mathcal{F}_{\mathrm{CP2}}(\omega t) = \frac{8\sin^{4}(\pi\omega t/2)\sin^{2}(2\pi\omega t) }{\cos^{2}(\pi\omega t)}. 
\end{equation}

\subsection{Fitting $W(t)$}
The visibility is fit with one free microscopic parameter ($\sin(\varphi)$) and three technical free parameters: an overall visibility $v_{0}$ that accounts for pi-pulse infidelities, a slight scaling $b$ to the magnetic field that accounts for a small deviation of the magnetic field from the stated 6T value, and a time constant $\tau_{d}$ that accounts for a slight drop in overall visibility as a function of delay time. We note that the time constant $\tau_{d}$ likely arises from the quadrupolar broadening of the nuclear noise spectrum that we did not consider in the model. This broadening was not included in the model as it has no effect on the size of the periodic visibility decays and revivals, and we lack a method to constrain the quadrupolar broadening directly. While the broadening can be measured directly via NMR techniques \cite{chekhovichCrossCalibrationDeformation2018, zaporskiIdealRefocusingOptically2023}, this was not feasible to do in this work. Rather than introduce another free microscopic parameter into the model, we thus fit the small overall decoherence over the measurement timescale with an exponential decay. In summary, the expression fit to the data is then:
\begin{equation}
    \label{eq:W_fits}
    W_{\mathrm{fit}}(t;\sin(\varphi),v_{0},b,\tau_{d}) = v_{0}*\exp\left[ -\sum_{j}\frac{5}{4}\frac{1}{N}\left(\frac{A^{j}\sin(\varphi)\sqrt{c_{j}}}{(\omega_{n}^{j}*b)}\right)^{2} \mathcal{F}
    ((\omega_{n}^{j}*b)t)    \right]*\exp(-t/\tau_{d})
\end{equation}
The microscopic free parameter $\sin(\varphi)$ is fit globally between the CP1 and CP2 data for each value of $\omega_{e}$. The magnetic scaling parameter $b=1.017700(7)$ is fit globally to all datasets (i.e. fit to the same value for all $\omega_{e}$ for both CP1 and CP2). From the fitted value for $b$, we calculate the magnetic field in the cryostat to have a strength of $B=6.10620(4)\mathrm{T}$. While this shift is less than 2\%, it allows us to accurately fit the locations of the coherence dips, which occur at the corresponding Lrmor periods. The remaining technical parameters, $v_{0}$ and $\tau_{d}$ are fit freely for each dataset. The full CP1 and CP2 datasets normalized by $v_{0}$, along with the fits to the model, are shown in Fig.~\ref{fig:SI_fullCP}.

\begin{figure*}
    \includegraphics[scale=0.87]{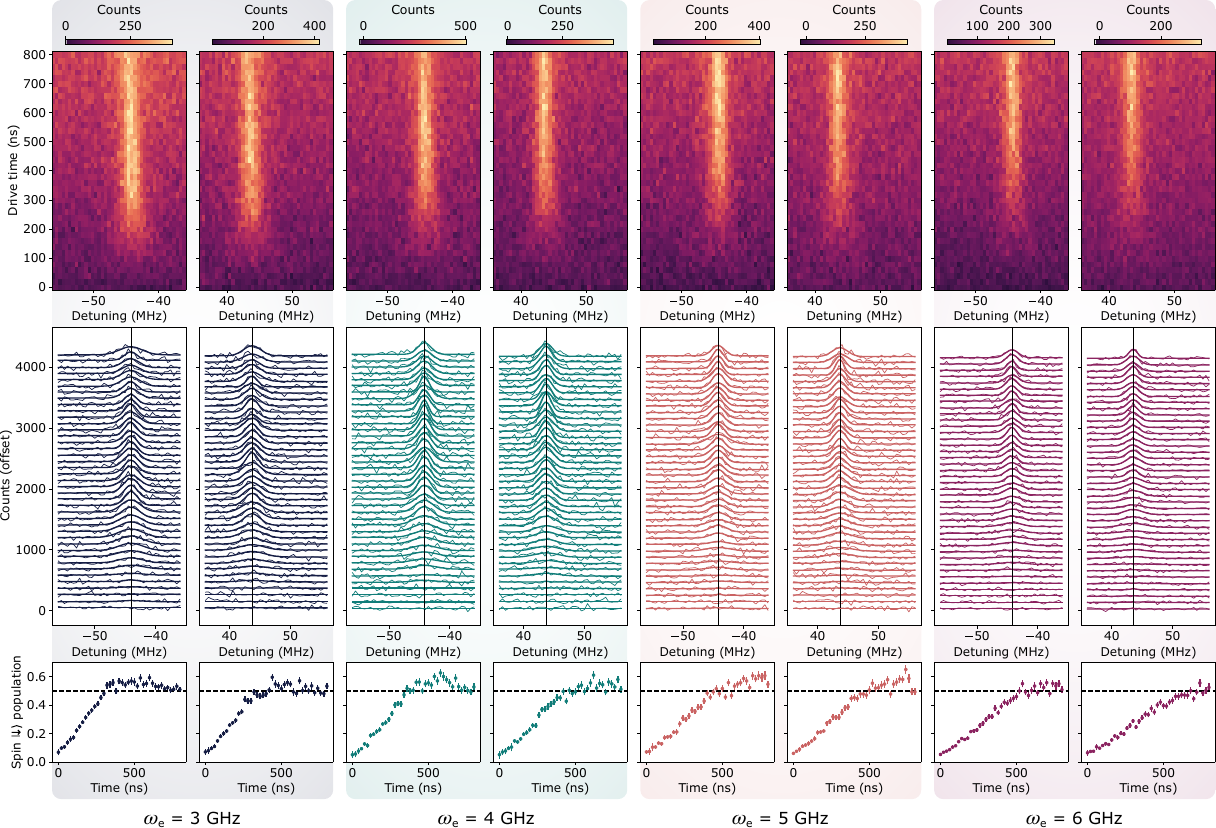}
    \caption{Extracting the amplitude of nuclear sidebands from fine step-size spectroscopy. Top row: fine resolution sideband spectra of the negative and positive ${}^{75}\mathrm{As}$ nuclear sidebands at the values of $\omega_{e}$ used in the text. Middle row: individual time steps of the sideband spectra along with a fit to a Gaussian distribution. The vertical line indicates the location of the peak of the Gaussian distribution, which was fixed across all time steps. Bottom row: the spin population at each time step extracted from the fitted Gaussian distribution. The dashed horizontal line indicates a 0.5 spin $\ket{\downarrow}$ population.}
    \label{fig:SI_magsat}
\end{figure*}

\begin{figure*}
    \includegraphics[scale=0.9]{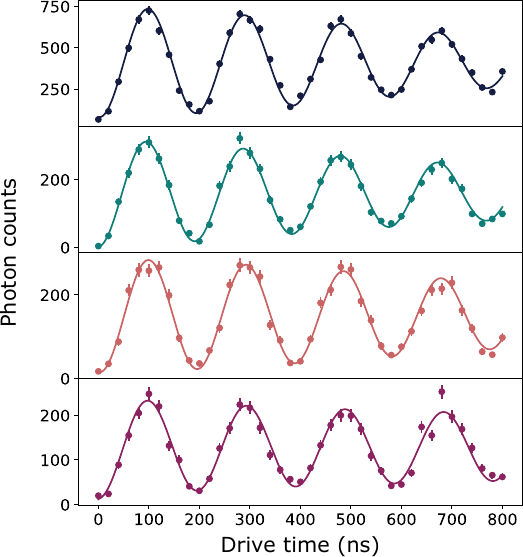}
    \caption{Measurements of Rabi oscillations at all values of $\omega_{e}$ used in the magnon Rabi fitting section, along with fits to an exponentially damped sine function (solid curves). These were measured to calibrate the $\delta=0$ electron Rabi frequency $\Omega$ for the measurements of the magnon sidebands in Fig.~\ref{fig:SI_magsat}, as well as to calculate the amplitude of counts that corresponds to a given $\ket{\downarrow}$ population. From top to bottom, the values of $\omega_{e}$ are given by $\omega_{e}=\{3,4,5,6\}\,\mathrm{GHz}$, with fitted electron Rabi frequencies of $\Omega = \{5.21(2),5.19(2), 5.17(2), 5.12(3)\}\,\mathrm{MHz}$ and amplitude $A = \{3.3(2),1.59(8),1.35(8),1.2(1)\}\times 10^{2}\,\mathrm{counts}$. The total integration time was 200 seconds, except for $\omega_{e}=3\,\mathrm{GHz}$, which was integrated for 560 seconds.}
    \label{fig:SI_calib_rabi}
\end{figure*}

\subsection{Dependence of electron $T_2$ on $\omega_\text{e}$}

In Fig.\,4 of the main text, we fit our qubit visibility at long delays under CP1 to a phenomenological function $\exp(-(\tau/T_{2})^{\alpha})$. We expect the qubit decoherence to be dominated by quadratic coupling (second-order hyperfine interaction) to the decorrelating transverse nuclear field with amplitude $a^2/\omega_\text{e}$ \cite{stockillQuantumDotSpin2016,zaporskiIdealRefocusingOptically2023}, leading to an $\omega_\text{e}^{-2}$ dependence in this interaction's power spectral density (Eq.\,19). Connecting our phenomenological fit parameters to this field amplitude leads to $T_2 \propto \omega_\text{e}^{2/\alpha}$, as used in the main text.

\section{Fitting magnon Rabi frequencies}

\begin{figure*}
    \includegraphics[scale=0.8]{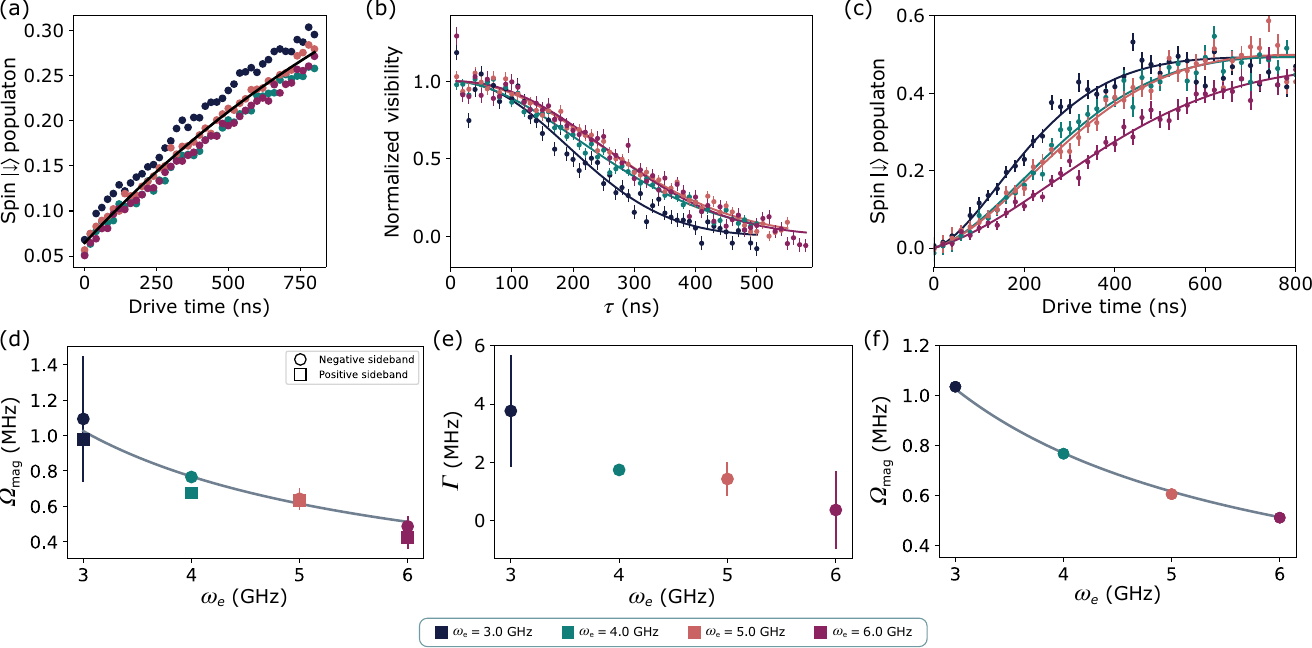}
    \caption{(a) background counts of the As sideband as a function of Rabi drive time. The black curve is a fit to an exponential saturation function $f(t) = 0.5(1-\exp(-2\gamma_{1} t) + B$, with fitted $\gamma_{1} = 3.4(1) \times 10^{-4} \,\,\mathrm{s}^{-1}$. (b) electron spin Ramsey interferometry measurement for the values of $\omega_{e}$ where magnon sideband Rabi was fit. The electron dephasing time is fit to be $T_{2}^{*} = \{253(7), 308(4), 330(4), 326 (4)\}\,\,\mathrm{ns}$ for $\omega_{e} = \{3,4,5,6\} \,\,\mathrm{GHz}$, with fitted $\alpha = \{2.2(2),1.91(8), 2.10(9),2.2(2)\}$. (c) the $\ket{\downarrow}$ spin population of the positive ${}^{75}\mathrm{As}$ sideband as a function of Rabi drive time for several values of $\omega_{e}$. The solid curves are fits to a simulation of a damped two-level system. (d) the fitted $\Omega_{\mathrm{mag}}$ values for the negative (circle markers) and positive (square markers) sidebands as a function of $\omega_{e}$. The grey curve is an interpolated prediction of an ab initio calculation of $\Omega_{\mathrm{mag}}$ using Eq.\,\ref{eq:Omag_SI}. (e) the simultaneously fitted dephasing parameter $\Gamma$ as a function of $\omega_{e}$. (f) the ab initio values of $\Omega_{\mathrm{mag}}$ for the relevant values of $\omega_{e}$, calculated using Eq.~\ref{eq:Omag_SI}. The grey line is a fit $\Omega_{\mathrm{mag}} = \Omega_{\mathrm{mag}}^{0}\times(\omega_{e}/\omega_{e}^{0})$ to the calculated values, with fitted $\Omega_{\mathrm{mag}}^{0} =1.025(5)\,\mathrm{MHz}$. The fit to the ab initio calculated values of $\Omega_{\mathrm{mag}}$ is used as the interpolated prediction of $\Omega_{\mathrm{mag}}$ in the main text and in panel (d) of this figure.}
    \label{fig:SI_magsat2}
\end{figure*}

\begin{figure*}
    \includegraphics[scale=0.87]{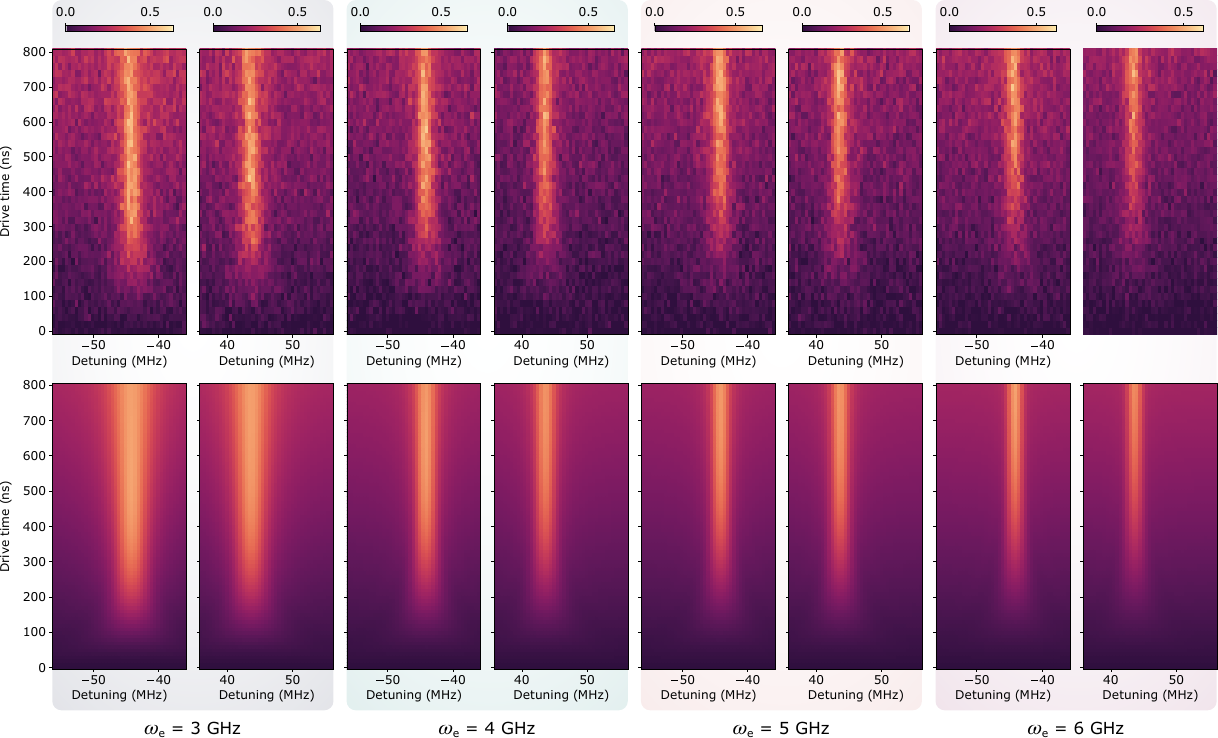}
    \caption{The fine-resolution As sidebands for all $\omega_{e}$ (top panel), along with the As sidebands simulated using the fitted $\gamma_{1}$, $\Omega_{\mathrm{mag}}$, and $\Gamma$ (bottom panel).}
    \label{fig:SI_2dsim}
\end{figure*}

We measure the positive- and negative-detuned As sidebands with fine-resolution detunings (varying $\omega_{\mathrm{mw}}$ in 0.25 MHz steps, see Fig.~\ref{fig:SI_magsat} top panel) so that the peak amplitude of the sideband can be extracted with a Gaussian fit, from $t_{\mathrm{Rabi}}=0$ to $t_{\mathrm{Rabi}}=800\,\mathrm{ns}$ in 20ns steps at $\omega_{e} = 3,4,5,6\,\mathrm{GHz}$. The recorded counts for the positive and negative sidebands are shown in the top panel of Fig.~\ref{fig:SI_magsat} for all values of $\omega_{e}$. We fit a Gaussian peak along the frequency axis at at each Rabi drive duration (Fig.~\ref{fig:SI_magsat} middle panel), fixing the frequency of the peak to the fitted peak frequency of the sidebands summed along the time axis. We extract the background and amplitude from each fitted Gaussian; the summed amplitude and background are plotted for each sideband at each value of $\omega_{e}$ in the bottom panel of Fig.~\ref{fig:SI_magsat}. Here the counts have been converted to spin $\ket{\downarrow}$ population via the method described in Sec.~\ref{sec:count_conv}.\par


\subsection{Model for magnon sideband dynamics}
We simulate the magnon dynamics with a damped two-level system, where the ground state is given by $\ket{\uparrow,\hat{I}_{z}}$ and the excited state is given by $\ket{\downarrow,\hat{I}_{z}+1}$ ($\ket{\downarrow,\hat{I}_{z}-1}$) for a positive (negative)-detuned sideband. This can be expressed in master equation form for the density operator $\rho(t)$ as \cite{steckQuantumAtomOptics2019}:
\begin{equation}
\label{eq:SI_two_level}
    \dot{\rho}(t) = -i\left[  \mathcal{H}_{0},\rho(t)\right] + \mathcal{D}[\sqrt{\gamma_{1}}\hat{S}_{+} ]\rho(t) + \mathcal{D}[ \sqrt{\gamma_{1}}\hat{S}_{-}]\rho(t)+\mathcal{D}[\sqrt{\Gamma} \hat{S}_{z}]\rho(t)
\end{equation}
where $\mathcal{D}[c]\rho \equiv c\rho c^{\dag} - (1/2)(c^{\dag}c\rho + \rho c^{\dag}c)$ denotes the Lindblad superoperator for a given operator $c$, $\gamma_{1}$ is a spin-flip rate, $\Gamma$ is a dephasing rate, and the two-level Hamiltonian $\mathcal{H}_{0}$ is given by:
\begin{equation}
    \mathcal{H}_{0} = \pi 
    \begin{pmatrix}
        \Delta & \Omega_{\mathrm{mag}} \\
        \Omega_{\mathrm{mag}} & -\Delta
    \end{pmatrix} 
\end{equation}
where $\Delta$ is the detuning of the control field from the nuclear resonance and $\Omega_{\mathrm{mag}}$ is the Rabi oscillation frequency. The evolution of the damped two-level system is simulated in Python using the QuTip simulation suite \cite{johanssonQuTiPPythonFramework2013}.\par

\subsection{Fitting the damped two-level model}
The decoherence rate $\gamma_{1}$ is fit to the background population (i.e. the background of the fit Gaussian at each time step, as in the middle row of Fig.~\ref{fig:SI_magsat}) of the spectra. This fit is shown in Fig.~\ref{fig:SI_magsat2}(a), along with the background at each drive time. The fit is performed globally over all values of $\omega_{e}$ simultaneously.\par
The master equation model accounts for the interaction of the MW control field with a single magnon. However, the electron hosted in the QD experiences a distribution of total nuclear polarization leading to small fluctuations in the Overhauser shift. This leads to a characteristic electron coherence time $T_{2}^{*}$, and we can simulate the effect of the Overhauser field fluctuations using the $T_{2}^{*}$ times at each value of $\omega_{e}$ (Fig.~\ref{fig:SI_magsat2}(b)):
\begin{equation}
    \Delta_{\mathrm{oh}} = \sigma \frac{\sqrt{2}}{2\pi T_{2}^{*}}
\end{equation}
where $\sigma\in [-2,2]$. This gives a $2\sigma$ distribution of Overhauser shifts $\Delta_{\mathrm{oh}}$. The evolution of the damped-two level system is then simulated for each value of $\Delta_{\mathrm{oh}}$, and summed up weighted by a normal probability distribution centered on $\sigma = 0$ to give the ``ensemble" damped two-level system dynamics.\par
Finally, the two free parameters in the model, $\Omega$ and $\Gamma$ are fit. The dephasing $\gamma_{1}$ is fixed as described above. $\Gamma$ is fit simultaneously to both the positive and negative magnon sidebands at each value of $\omega_{e}$, while $\Omega$ is fit freely for both sidebands. The positive sideband fits are shown in Fig.~\ref{fig:SI_magsat2}(c) with the fitted values for $\Omega_{\mathrm{mag}}$ and $\Gamma$ in Fig.~\ref{fig:SI_magsat2}(d) and Fig.~\ref{fig:SI_magsat2}(e), respectively.\par 
Figure~\ref{fig:SI_magsat2}(f) shows the values of $\Omega_{\mathrm{mag}}$ calculated ab initio for the relevant values of $\omega_{e}$ using Eq.~\ref{eq:Omag_SI}, using the value for $a$ measured via the Knight shift and the values for $\sin(\varphi)$ fit from the CP measurements. These calculated values are then fit with a power law $\Omega_{\mathrm{mag}} = \Omega_{\mathrm{mag}}^{0}\times(\omega_{e}/\omega_{e}^{0})$ to give the interpolated prediction of $\Omega_{\mathrm{mag}}$ presented in the main text. The error bars on each value of $\Omega_{\mathrm{mag}}$ calculated using Eq.~\ref{eq:Omag_SI} are generated based on the one-sigma error on $a$ derived from the fits to the Knight shift data, and lie below the thickness of the markers. The error on $a$ is propagated through the value of $N_{\mathrm{As}}$ and the fits of $\sin(\varphi)$ performed using the CP data, leading to final errors on the value of $\Omega_{\mathrm{mag}}$. The standard deviation of the values of $\Omega_{\mathrm{mag}}$ calculated for $a$, $a+\sigma_{a}$, and $a-\sigma_{a}$ (where $\sigma_{a}$ is the standard error on $a$) at each value of $\omega_{e}$ is added in quadrature with the standard error on $\Omega$ extracted from the fits to the $\delta=0$ Rabi oscillations shown in Fig.~\ref{fig:SI_calib_rabi}, and this gives the final one-sigma error on the ab initio calculated values of $\Omega_{\mathrm{mag}}$.\par
Figure~\ref{fig:SI_2dsim} shows the fine-resolution sideband spectra used to fit the magnon Rabi rates alongside a simulation of the magnon sidebands generated using the QuTip implementation of Eq.~\ref{eq:SI_two_level} with the fitted parameters $\Omega_{\mathrm{mag}}$, $\gamma_{1}$, and $\Gamma$.\par 

\bibliography{gaas_aniso}